\newif\ifarxiv%
\arxivtrue%

\ifarxiv%
\pdfoutput=1
\fi

\documentclass[TCS,biber]{nowfnt}

\usepackage{macros}

\title{Kleene Algebra}
\maintitleauthorlist{}

\author[1]{Tobias Kapp\'e}
\affil[1]{Leiden University}

\author[2]{Alexandra Silva}
\affil[2]{Cornell University}

\author[3]{Jana Wagemaker}
\affil[3]{Radboud University}

\articledatabox{\nowfntstandardcitation}

\issuesetup{%
    copyrightowner= {},
    volume        = xx,
    issue         = xx,
    pubyear       = xxxx,
    isbn          = xxx-x-xxxxx-xxx-x,
    eisbn         = xxx-x-xxxxx-xxx-x,
    doi           = 10.1561/XXXXXXXXX,
    firstpage     = 1,
    lastpage      = 18
}

\let\epsilon\varepsilon%

\ifarxiv%
\makeatletter
\renewcommand\nowfnt@prohibitiontext{%
    This is the Open Access version of a manuscript of the same name.
    Copyright is held by the authors.
    You may share this version freely under the terms of the CC-BY license: \url{https://creativecommons.org/licenses/by/4.0/}.
}
\makeatother
\fi%

\begin{document}
\renewcommand{\onlyinsubfile}[1]{}

\chapter{Introduction}%
\label{chapter:introduction}

Arithmetic has a number of laws, many of which are familiar to most people on an intuitive if not formal level.
For instance, multiplication \emph{distributes} over addition: if $x$, $y$ and $z$ are numbers, then $x \cdot (y + z)$ is \emph{equivalent} to $x \cdot y + x \cdot z$.
Using similar laws, we can reason about equivalence of arithmetic expressions by manipulating their syntax.
\emph{Kleene Algebra} is motivated by the idea that these principles are not limited to numbers --- indeed, in the words of Hoare et al.~\cite{hoare-hayes-he-etal-1987}:
\begin{quote}
    \emph{programs are mathematical expressions [\ldots] subject to a set of laws as rich and elegant as any other branch of mathematics, engineering, or natural science.}
\end{quote}
\noindent
By studying these laws, we can gain theoretical insights and practical tools to reason about program correctness.
Applications include:
\begin{enumerate}
    \setlength{\itemsep}{0em}
    \item
    When rearranging code to improve readability, we want to be certain that the new program has the same semantics, or at the very least, does not introduce any new behavior.
    \item
    We want to check whether our implementation of an algorithm conforms to its (abstract) specification, or at least refines it.
    \item
    A compiler typically transforms its code, for instance to conform to hardware constraints, and we want to check that these transformations never change the meaning of the program.
\end{enumerate}

To illustrate the first case a bit more, consider the programs below, written in a standard imperative programming language.
\[
    \begin{minipage}{52mm}
        \vspace{-0.4em}
        \begin{algorithm}[H]
            \While{\testvar{a} \textnormal{\texttt{and}} \testvar{b}}{
                \progvar{p}\;
            }
            \While{\testvar{a}}{
                \progvar{q}\;
                \While{\testvar{a} \textnormal{\texttt{and}} \testvar{b}}{
                   \progvar{p}\;
                }
            }
        \end{algorithm}
    \end{minipage}
\]

Here, the terms $\testvar{a}$ and $\testvar{b}$ stand in for any Boolean expressions (tests), whereas $\progvar{p}$ and $\progvar{q}$ represent general programs.
As you can see, this program is somewhat messy: it includes two occurrences of the program $\progvar{p}$, and no fewer than \emph{three} occurrences of Boolean expression $\testvar{a}$!

Clearly, this code could benefit from some refactoring.
After staring at it for a while, you realise that $\progvar{p}$ runs only when $\testvar{a}$ and $\testvar{b}$ hold, while $\progvar{q}$ is executed when $\testvar{a}$ is true and $\testvar{b}$ is false.
Furthermore, if $\testvar{a}$ is false after running $\progvar{p}$ or $\progvar{q}$, the program will stop; otherwise, it will continue with (some) loop.
This eventually leads you to refactor the code as follows:
\[
    \begin{minipage}{40mm}
        \vspace{-0.4em}
        \begin{algorithm}[H]
            \While{\testvar{a}}{
                \eIf{\testvar{b}}{
                    $\progvar{p}$\;
                }{
                    $\progvar{q}$\;
                }
            }
        \end{algorithm}
    \end{minipage}
\]
At this point, you may have some intuition of why these two programs are the same, or perhaps you are still skeptical.
By the end of this text, you will be able to formally \emph{prove} that these two programs have the same behavior, for any choice of $\progvar{p}$, $\progvar{q}$, $\testvar{a}$ and $\testvar{b}$.

In \Cref{\res{chapter:basics}}, we will go over the basic theory of Kleene Algebra.
\Cref{\res{chapter:coalgebra}} follows this up with some basic material on \emph{coalgebra}, which we will rely on in later chapters.
Next, in \Cref{\res{chapter:automata}}, we instantiate coalgebra to study automata, and their connection to Kleene Algebra.
This is followed by a proof of one of the most important results in Kleene Algebra: the \emph{completeness theorem}, in \Cref{\res{chapter:completeness}}.
\Cref{\res{chapter:kat}} extends the preceding material by introducing \emph{Kleene Algebra with Tests}, an extension of Kleene Algebra that can be used to reason about programs such as the ones on the previous page.
\Cref{\res{chapter:further-reading}} concludes with an overview of relevant literature that may be of interest to the reader.

\paragraph{Acknowledgements}
We express our heartfelt gratitude to Uli Fahrenberg, Walter Guttmann, and Ga\"{e}tan Staquet for their helpful and detailed comments, which substantially improved this manuscript.
We also ant to thank Mark de Jongh for managing the publication process.

T.~Kappé was partially supported by the European Union’s Horizon 2020 research and innovation programme under grant no.\ 101027412 (VERLAN), and partially by the Dutch research council (NWO) under grant no.\ VI.Veni.232.286 (ChEOpS). J.~Wagemaker was partially supported by the NWO under grant no.\ VI.Veni.242.134 (VerHyp). A.~Silva was was partially supported by ERC grant Autoprobe (no. 101002697).

\paragraph{Source material}
This text, or \emph{booklet} as we have lovingly come to call it, does not communicate new scientific knowledge.
Rather, it is based on revised and edited material excerpted from one author's PhD thesis~\cite{silva-thesis} and another's lecture notes developed for a course taught at the University of Amsterdam and ESSLLI 2023.
A.~Silva has used these notes as part of Cornell's course on Kleene Algebra (CS 6861).

\paragraph{Reading guide}
This booklet is structured as lecture notes, to be used as self-study material or as the basis for a course, with exercises at the end of each chapter.
There are three possible paths, drawn in \Cref{\res{figure:booklet-paths}}.
For a graduate-level course, you could consider following the blue dashed path, and covering all material.
For a shorter (possibly undergraduate) course it might be more appropriate to skip the completeness results and the corresponding coalgebraic theory, and instead focus on automata and soundness.
This is captured by the red dotted paths.

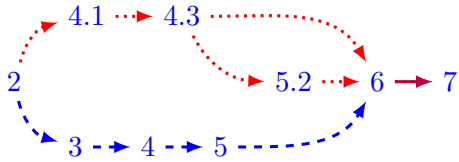
\begin{figure}
    \begin{center}
        \begin{tikzpicture}[node distance=5mm]
            \begin{scope}[every node/.style={anchor=center}]
                \node (basics) {\bref{chapter:basics}};
                \node[below right=of basics] (coalgebra) {\bref{chapter:coalgebra}};
                \node[right=of coalgebra] (automata) {\bref{chapter:automata}};
                \node[above right=of basics] (derivatives) {\bref{section:derivatives}};
                \node[right=of derivatives] (kleenes-theorem) {\bref{section:kleenes-theorem}};
                \node[right=of automata] (completeness) {\bref{chapter:completeness}};
                \node[above right=of completeness] (soundness) {\bref{section:soundness}};
                \node[right=of soundness] (kat) {\bref{chapter:kat}};
                \node[right=of kat] (related) {\bref{chapter:further-reading}};
            \end{scope}
            \begin{scope}[line width=0.4mm, -latex]
                \begin{scope}[blue,dashed]
                    \draw (basics) to[bend right] (coalgebra);
                    \draw (coalgebra) to (automata);
                    \draw (automata) to (completeness);
                    \draw (completeness) to[out=0,in=-120] (kat);
                \end{scope}
                \begin{scope}[red,dotted]
                    \draw (basics) to[bend left] (derivatives);
                    \draw (derivatives) to (kleenes-theorem);
                    \draw (kleenes-theorem) to[bend right] (soundness);
                    \draw (soundness) to (kat);
                    \draw (kat) to (related);
                    \draw (kleenes-theorem) to[->,out=0,in=120] (kat);
                \end{scope}
                \begin{scope}[purple]
                    \draw (kat) to (related);
                \end{scope}
            \end{scope}
        \end{tikzpicture}
    \end{center}
    \caption{%
        Possible paths through this booklet.
    }%
    \label{figure:booklet-paths}
\end{figure}

\paragraph{Errata}
Corrections are tracked on \url{https://ka-booklet.github.io}, where the most recent revision of this text can also be found.

\section*{Notation}
We use $\naturals$ to denote the set of natural numbers (including zero).

\paragraph{Relations}
A \emph{relation} between two sets $X$ and $Y$ is a subset of $X\times Y$.
We often use infix notation to denote membership of a relation: for a relation $R\subseteq X\times Y$, we can write $x \mathrel{R} y$ to denote $(x,y)\in R$.
Given a set $X$, we call the set $\id_X=\{(x,x)\mid x\in X\}$ the \emph{identity relation on $X$}.
If $R$ is a relation between sets $X$ and $Y$ and $R'$ a relation between sets $Y$ and $Z$, we write $R\circ R'$ for the \emph{relational composition} of $R$ and $R'$, that is: $R\circ R'=\{(x,z)\mid \exists y\in Y \text{ such that } (x,y)\in R\wedge (y,z)\in R'\}$.

A relation $R \subseteq X \times X$ is said to be \emph{reflexive} when for all $x \in X$ we have $x \mathrel{R} x$, \emph{symmetric} when $x\mathrel{R}y$ implies $y\mathrel{R}x$, \emph{transitive} when $x\mathrel{R}y$ and $y\mathrel{R}z$ imply $x\mathrel{R}z$, and \emph{antisymmetric} when $x\mathrel{R} y$ and $y \mathrel{R} x$ implies $x = y$.
Given $R \subseteq X \times X$, we write $R^*$ for the \emph{reflexive-transitive closure} of $R$, which is the smallest reflexive and transitive relation on $X$ that contains $R$; since reflexivity and transitivity are preserved by intersections, such a relation is guaranteed to exist.
If $R \subseteq X \times X$ is reflexive, antisymmetric and transitive it is called a \emph{partial order}.

\paragraph{Equivalence relations}
If $R \subseteq X \times X$ is reflexive, symmetric and transitive it is called an \emph{equivalence relation}.
Given $x \in X$, we write $[x]_R$ for the \emph{equivalence class} of $X$ under $R$, which is the set $[x]_R=\{y\in X\mid x\mathrel{R}y\}$; we may omit the subscript $R$ if the equivalence relation is clear from context.
We write $X/R =\{[x]_R\mid x\in X\}$ for the \emph{quotient} of $X$ by $R$.
Furthermore, we write $\equiv_R$ for the \emph{equivalence closure} of $R$, which is the smallest equivalence relation containing $R$.

\paragraph{Congruence relations}
Let $R$ be an equivalence relation defined on a set $X$, and let $f: X^n \to X$ be an operator on $X$.
We call $R$ \emph{congruence} w.r.t.\ $f$ if it is compatible with $f$ in the sense that if $x_1 \mathrel{R} x_1', \dots, x_n \mathrel{R} x_n'$, then $f(x_1, \dots, x_n) \mathrel{R} f(x_1', \dots, x_n')$.
For instance, if we consider a binary operator $+$, and we know that $e\mathrel{R}f$ and $g\mathrel{R}h$, then $R$ being a congruence w.r.t.\ $+$ tells us that $(e+g) \mathrel{R} (f+h)$.

\paragraph{Disjoint unions}
Given two sets $X$ and $Y$, we write $X + Y$ for their \emph{disjoint union}.
Formally, this is $\{ <x, 0> \mid x \in X \} \cup \{ <y, 1> \mid y \in Y \}$; we elide this detail, and simply treat elements of $X$ or $Y$ as elements of $X + Y$, and elements of $X + Y$ as either elements of $X$ or $Y$ (but not both).
Overlap can be avoided by renaming elements.

\paragraph{Functions and powersets}
A \emph{function} $f$ from sets $X$ to $Y$, denoted $f\colon X\rightarrow Y$, assigns to each element of $X$ exactly one element of $Y$, written $f(x)$.
We write $Y^X$ for the set of functions from $X$ to $Y$.
We may also use $\id_X$ to denote the \emph{identity} function on $A$, which assigns each element to itself (overloading earlier notation for the identity relation).

We use $1$ as an abbreviation for a one element set $\{ * \}$, and $2$ as an abbreviation for a two element set $\{ 0, 1 \}$.
Given this convention, we also have that $2^X$, the set of functions from $X$ to $2$, is in one-to-one correspondence with the \emph{powerset} of $X$, i.e., the set of subsets of $X$.
We therefore simply use $2^X$ to denote this set, and treat its members as subsets of $X$ or as their characteristic functions, interchangeably.

\onlyinsubfile{%
    \printbibliography%
}

\chapter{Basics}%
\label{chapter:basics}

In this chapter, we develop a basic algebraic theory of regular expressions, known as Kleene Algebra.
If you have a background in Computer Science, you may have encountered these expressions in undergraduate computer science courses with titles such as ``Formal Languages'', ``Theory of Computation'' or ``Foundations of Computer Science''.
You may also be familiar with them as a tool for finding patterns of characters in text using tools like GNU Grep or the Perl Compatible Regular Expressions (\texttt{pcre}) functionality available in many programming languages today.
At their core, regular expressions provide a way to describe \emph{patterns of events}, whether those are validation conditions or packet forwarding behavior in a network.
More broadly, we can think of regular expressions as an abstract syntax for imperative programs, which is the main motivation for studying them in this booklet.

We will start by defining regular expressions, along with two semantics, both of which have an intuitive meaning in connection to programming languages.
We then go on to introduce the laws of Kleene Algebra, which can be used to reason about \emph{equivalence} of regular expressions, in the sense that they have the same semantics.

\section{Syntax and Semantics}%
\label{section:syntax-semantics}

Throughout this booklet, we fix a finite set of \emph{actions} $\alphabet = \{ \ltr{a}, \ltr{b}, \ltr{c}, \ldots \}$.
You could think of these actions as representing primitive operations that can be performed by a machine; for example, the action symbol $\ltr{a}$ could mean ``make the red LED blink once'' or ``verify that the next input letter is $\ltr{a}$''.
Usually, we will abstract from the exact meaning of the actions, preferring to focus on when and how often they occur, and in what order.
Regular expressions are built from these actions.

\begin{definition}[Regular expressions~\cite{kleene-1956}]
Regular expressions (over an alphabet $\alphabet$) are generated by the grammar:
\[
    e, f\Coloneqq 0 \mid 1 \mid \ltr{a} \mid e + f \mid e \cdot f \mid e^*,
\]
where $\ltr{a}\in \alphabet$.
We will write $\Exp$ for the set of regular expressions.
\end{definition}
When we write regular expressions, we may elide ``$\cdot$'' --- just like how multiplication is commonly omitted.
We also adopt the convention that ${}^*$ takes precedence over $\cdot$, which in turn takes precedence over $+$.
Thus, $(\ltr{a} \cdot (\ltr{b}^*)) + \ltr{c}$ can also be written as $\ltr{a}\ltr{b}^* + \ltr{c}$.
Parentheses can also be used to disambiguate expressions the same way this is done in other contexts (even though they are not part of the syntax).

In regular expressions, the constant $0$ represents a program that crashes immediately, and $1$ represents a program that succeeds immediately.
The ``union'' or ``non-deterministic choice'' operator $+$ is commonly interpreted as taking the union of possible behaviors --- i.e., the expression $e + f$ represents all patterns of events described by either $e$ or $f$.
Furthermore, the ``concatenation'' or ``sequential composition'' operator $\cdot$ combines behaviors in sequence --- i.e., $ef$ describes all behaviors achieved by first completing a behavior from $e$, and then a behavior from $f$.
Finally, the ``iteration'' or ``Kleene star'' operator ${}^*$ represents behavior achieved by running an arbitrary (possibly zero) number of patterns of events from $e$ in sequence.

\begin{example}%
\label{ex:regular-expressions}\mbox{}
\begin{enumerate}
    \setlength{\itemsep}{-1mm}
    \item
    The regular expression $\ltr{a} + \ltr{b}\ltr{c}$ denotes the behavior where either the action $\ltr{a}$ runs, \emph{or} the behavior where $\ltr{b}$ is followed by $\ltr{c}$.
    \item
    We can also write the regular expression $\ltr{a}\ltr{b}^*\ltr{c}$ which describes all sequences of events that start with $\ltr{a}$ and are followed by some (possibly zero) number of $\ltr{b}$'s, ending with the action $\ltr{c}$.
    \item
    Finally, $(\ltr{a} + \ltr{b})^*$ is a regular expression that denotes all behaviors that consist of a finite number of actions, each of which is either the event $\ltr{a}$ or the event $\ltr{b}$.
\end{enumerate}
\end{example}

\begin{remark}
Some sources may use the term \emph{rational expression} to refer to regular expressions as defined above.
This is a historic distinction that is meant to separate \emph{rational} behaviors --- denoted by expressions --- from \emph{regular} behaviors --- described by automata (see \Cref{\res{chapter:automata}}).
These classes of behaviors are the same in all settings that we encounter in this booklet; we therefore do not make this distinction in our nomenclature.
\end{remark}

\subsection{Relational Semantics}\label{section:ka-relational-semantics}

One way to assign a meaning to regular expressions is to consider the effects that each action may have on the outside world.
This naturally leads to an interpretation of regular expressions in terms of \emph{relations}, which relate each ``state'' of the world to the possible states that may result from following a behavior of the expression.
To do this we are going to need to know which states are possible, and how each primitive action may change them.
This is encapsulated as follows.

\begin{definition}[Interpretation]
An \emph{interpretation} (of the alphabet $\alphabet$) is a function from $\alphabet$ to relations on a set $S$, i.e., $\sigma\colon \alphabet \to 2^{S \times S}$.
\end{definition}

When we fix an interpretation, the relational semantics can be derived inductively.
For instance, if $e$ can transform a state $s$ into $s'$, and $f$ can transform $s'$ into $s''$, then $ef$ can transform $s$ into $s''$; it follows that the relation denoted by $ef$ is the composition of the relations obtained from $e$ and $f$.
This leads to the following semantics.

\begin{definition}[Relational semantics~\cite{pratt-1976}]
Given an interpretation $\sigma\colon \alphabet \to 2^{S \times S}$, we define the $\sigma$-semantics $\sem{-}_\sigma\colon \Exp \to 2^{S \times S}$ inductively:
\begin{align*}
\sem{0}_\sigma &= \emptyset
    & \sem{1}_\sigma &= \id_S
    & \sem{\mathtt{a}}_\sigma &= \sigma(\mathtt{a}) \\
\sem{e+f}_\sigma &= \sem{e}_\sigma \cup \sem{f}_\sigma
    & \sem{e \cdot f}_\sigma &= \sem{e}_\sigma \circ \sem{f}_\sigma
    & \sem{e^*}_\sigma &= \sem{e}_\sigma^*
\end{align*}
\end{definition}

In the last case, we use ${}^*$ to denote both reflexive-transitive closure (on the right) and the Kleene star of regular expressions (on the left).

\begin{example}\label{example:program:basics}
Suppose you want to calculate the \emph{integer square root} of $n \in \naturals$ --- in other words, the largest number $i$ such that $i^2 \leq n$.
One way to arrive at this number is to start $i$ at zero, and keep incrementing it while ${(i+1)}^2 \leq n$.
A traditional program to achieve this could be:

\begin{algorithm}[H]
$i \gets 0$\;
\While{${(i + 1)}^2 \leq n$}{
    $i \gets i + 1$\;
}
\end{algorithm}

\smallskip\noindent
The behavior of this program can be captured by this regular expression:
\[
    e = \mathtt{init} \cdot {(\mathtt{guard} \cdot \mathtt{incr})}^* \cdot \mathtt{validate}
\]
Here, $\mathtt{init}$, $\mathtt{guard}$, $\mathtt{incr}$ and $\mathtt{validate}$ are primitive programs, where:
\begin{itemize}
    \setlength{\itemsep}{0em}

    \item
    $\mathtt{init}$ sets the variable $i$ to zero.

    \item
    $\mathtt{guard}$ observes that $(i+1)^2 \leq n$.

    \item
    $\mathtt{incr}$ increments the value of $i$.

    \item
    $\mathtt{validate}$ observes that $(i+1)^2 \gr n$.
\end{itemize}
It may seem odd that some actions simply observe a property of a variable, without changing any value.
But it does make sense to include them --- after all, if $(i+1)^2 \leq n$ does not hold, then the program will not start a new iteration of the loop.
Hence, the machine must at some point perform this check, which we model as an action.
We will study such \emph{tests} in more detail in \Cref{\res{chapter:kat}}.

Returning to our example program, we can give the following interpretation to the primitive programs.
First, we need to fix our machine states.
In this case, the program has two variables $i$ and $n$, both of which are natural numbers.
This means that the state of the machine is entirely described by the current values for $i$ and $n$.
Thus, we choose for $S$ the set of functions $s\colon \{ i, n \} \to \naturals$.
Next, we can fix our interpretation of the primitive programs, as follows:
\begin{align*}
\sigma(\mathtt{init}) &= \{ \angl{s, s[0/i]} \mid s \in S \}
    \\ \sigma(\mathtt{guard}) &= \{ \angl{s, s} \mid {(s(i) + 1)}^2 \leq s(n) \} \\
\sigma(\mathtt{incr}) &= \{ \angl{s, s[s(i)+1/i]} \mid s \in S \}
    \\ \sigma(\mathtt{validate}) &= \{ \angl{s, s} \mid {(s(i) + 1)}^2 \gr s(n) \}
\end{align*}
In the above, given $s\colon \{i, n\} \to \naturals$, $n \in \naturals$ and $v \in \{i, n\}$, we write $s[n/v]$ for the function that overrides the value of $v$ to $n$ in $s$, i.e., for all $v'$:
\[
    s[n/v](v') =
        \begin{cases}
        n & v = v' \\
        s(v') & \text{otherwise}
        \end{cases}
\]

If you compute $\sem{e}_\sigma$, you will find that it relates functions $s\colon \{i, n\} \to \naturals$ to functions $s'\colon \{i,n\} \to \naturals$ where $s'(n) = s(n)$ and $s'(i)^2 \leq s(n) \leq (s'(i)+1)^2$.
We will return to this program later in \Cref{\res{ex:1:basics}}.
\end{example}

In the example above, you may note that the interpretations of $\mathtt{guard}$ and $\mathtt{validate}$ are relations that are not \emph{total}, i.e., there exist $s \in S$ that do not admit an $s' \in S$ with $<s, s'> \in \mathrel{\sigma(\mathtt{guard})}$, and similarly for $\sigma(\mathtt{validate})$.
This is intentional: the intended meaning is that there is no valid way to continue a computation with $\mathtt{guard}$ for these states, and so the outcome is discarded.
Usually, this means that some non-deterministic branch of execution does not get to contribute to the overall results of the program.
Another branch may still yield a result, although regular expressions and interpretations must be written carefully to guarantee this, as in the following example.

\begin{example}%
\label{example:if-then-else}
One common pattern is to simulate an \emph{if-then-else} construct using non-determinism and opposite tests.
For instance, suppose we want to model the following piece of code using a regular expression, where $\mathtt{a}$ and $\mathtt{b}$ are placeholders for the code contained in each branch:

\begin{algorithm}[H]
\eIf{${(i + 1)}^2 \leq n$}{%
    $\mathtt{a}$\;
}{%
    $\mathtt{b}$\;
}
\end{algorithm}

\noindent
One way to accomplish this is to reuse $\mathtt{guard}$ and $\mathtt{validate}$ from the previous example and write $\mathtt{guard} \cdot \mathtt{a} + \mathtt{validate} \cdot \mathtt{b}$.
Under the interpretation $\sigma$ used earlier, the semantics will be a relation that applies $\mathtt{a}$ to states where $(i+1)^2 \leq n$, and $\mathtt{b}$ to states where $(i+1)^2 \gr n$ --- i.e.,
\begin{align*}
    \sigma(\mathtt{guard} \cdot \mathtt{a} + \mathtt{validate} \cdot \mathtt{b}) = {}
        & \{ \angl{s, t} \in \sigma(\mathtt{a}) \mid (s(i)+1)^2 \leq s(n) \} \cup {} \\
        &  \{ \angl{s, t} \in \sigma(\mathtt{b}) \mid (s(i)+1)^2 \gr s(n) \}
\end{align*}
We will discuss an extension of Kleene algebra in \Cref{\res{chapter:kat}} that can model this kind of composition more accurately.
\end{example}

\subsection{Language semantics}\label{subsec:lang}

The relational semantics is a very versatile tool for modeling programs with regular expressions, but it hinges on the interpretation for each primitive action symbol, as well as a set of states that may be affected by the actions.
This makes it possible to reason about particular programs, but equivalence of programs in general (i.e., regardless of the interpretation) is harder to describe.
This is where the \emph{language semantics} of regular expressions comes in.
Put simply, this model assigns to each expression the possible sequences of actions that may result from running the associated program.
These \emph{traces} are recorded purely using the action names, which means that this semantics is not parameterized by an interpretation; we can focus purely on the patterns of events.

To properly introduce this semantics, we first discuss some notation.
A \emph{word} or \emph{trace} (over $\alphabet$) is a finite sequence of actions (from $\alphabet$).
For instance, if $\alphabet = \{ \ltr{a}, \ltr{b} \}$, then $\ltr{a}\ltr{b}\ltr{a}\ltr{a}$ is a word.
Words can be \emph{concatenated} by juxtaposition, i.e., if $w = \ltr{a}\ltr{b}$ and $x = \ltr{b}\ltr{a}$, then $wx = \ltr{a}\ltr{b}\ltr{b}\ltr{a}$.
We write $\epsilon$ for the \emph{empty} word, and note that it is neutral for concatenation on both sides, i.e., $w\epsilon = w = \epsilon{}w$ for all words $w$.

A \emph{language} (over $\alphabet$) is a set of words (over $\alphabet$).
When $L$ and $K$ are languages, we write $L \cdot K$ for the \emph{concatenation} of $L$ and $K$, i.e., the language $\{ wx \mid w \in L, x \in K \}$, and $L^*$ for the \emph{Kleene star} of $L$, i.e., the language $\{ w_1 \cdots w_n \mid w_1, \dots, w_n \in L \}$.
This makes $\alphabet^*$ the set of all words over $\alphabet$.
The Kleene star and multiplication operators are thus overloaded to also act on languages.

\smallskip
With this in hand, we are now ready to define the language semantics of regular expressions.

\begin{definition}[Language semantics~\cite{kleene-1956}]\label{def:algsem}
We define a map $\sem{-} \colon \Exp \to 2^{\alphabet^*}$ by induction as follows:
\begin{align*}\label{eq:re_algsem}
\sem{0} &= \emptyset
    & \sem{1} &= \{\epsilon \}
    & \sem{\ltr{a}} &= \{\ltr{a}\} \\
\sem{e+f} &= \sem{e} \cup \sem{f}
    & \sem{ef} &= \sem{e} \cdot \sem{f}
    & \sem{e^*} &= \sem{e}^*
\end{align*}
\end{definition}

\begin{example}
Let us now  calculate the languages of the regular expressions from Example~\bref{ex:regular-expressions}:
\begin{mathpar}
    \sem{\ltr{a} + \ltr{b}\ltr{c}} = \{ \ltr{a}, \ltr{b}\ltr{c} \}
    \and
    \sem{\ltr{a}\ltr{b}^*\ltr{c}} = \{ \ltr{a}\ltr{c}, \ltr{a}\ltr{b}\ltr{c}, \ltr{a}\ltr{b}\ltr{b}\ltr{c}, \ldots \}
    \and
    \sem{{(\ltr{a} + \ltr{b})}^*} = \{ \epsilon, \ltr{a}, \ltr{b}, \ltr{a}\ltr{a}, \ltr{a}\ltr{b}, \ltr{b}\ltr{a}, \ltr{b}\ltr{b}, \dots \}
\end{mathpar}
\end{example}

From the programming perspective, the language of a regular expression is really an overapproximation of the behavior that may occur --- an interpretation may give rise to a wide variety of subsets of the sequences of actions in the language.
For instance, in the regular expression $\ltr{a} + \ltr{b}\ltr{c}$ the action $\ltr{b}$ may raise an error and halt the program, preventing $\ltr{c}$ from being run.
The important point, however, is that other interpretations may disagree, and so $\ltr{b}\ltr{c}$ is included in the language semantics.

\medskip
The language semantics is closely connected to the relational semantics, in the sense that for any $e \in \Exp$ and any interpretation $\sigma$, we can reconstruct $\sem{e}_\sigma$ from $\sem{e}$, by way of the function $\hat{\sigma}\colon 2^{\alphabet^*} \to 2^{S \times S}$, which accumulates the pairs of the relations represented by each word:
\[
    \hat{\sigma}(L) = \bigcup \{ \sigma(\mathtt{a}_1) \circ \cdots \circ \sigma(\mathtt{a}_n) \mid \mathtt{a}_1 \cdots \mathtt{a}_n \in L \}
\]
\begin{lemma}\label{theorem:lang-implies-rel}
For each $e \in \Exp$ and interpretation $\sigma$, $\sem{e}_\sigma = \hat{\sigma}(\sem{e})$.
\end{lemma}
\begin{proof}[Proof sketch]
The proof proceeds by induction on $e$, and relies on the fact that $\hat{\sigma}$ is compatible with the operators of language composition.
Specifically, for all $L, K \subseteq \alphabet^*$, the following hold:
\begin{mathpar}
    \hat{\sigma}(L \cup K) = \hat{\sigma}(L) \cup \hat{\sigma}(K)
    \and
    \hat{\sigma}(L \cdot K) = \hat{\sigma}(L) \circ \hat{\sigma}(K)
    \and
    \hat{\sigma}(L^*) = {\hat{\sigma}(L)}^*
\end{mathpar}
The details are left as \Cref{\res{exercise:lang-implies-rel}}.
\end{proof}

A converse correspondence is also possible, but slightly more tricky to figure out.
The idea here is to craft a particular interpretation $\sharp\colon \alphabet \to 2^{\alphabet^* \times \alphabet^*}$ that represents languages as relations, by defining
\[
    \sharp(\mathtt{a}) = \{ \angl{w, w\ltr{a}} \mid w \in \alphabet^* \}
\]
We can then relate the relational semantics under this interpretation to languages, as follows.
\begin{lemma}[{\cite[Page~24]{pratt-1980}}]\label{theorem:rel-implies-lang}
Let $e \in \Exp$.
For all $w, x \in \alphabet^*$, we have that $\angl{w, wx} \in \sem{e}_\sharp$ if and only if $x \in \sem{e}$.
\end{lemma}
\begin{proof}[Proof sketch]
This comes down to showing that $\sem{e}_\sharp = \{ \angl{w, wx} \mid w \in \alphabet^*, x \in \sem{e} \}$.
The proof again proceeds by induction on $e$; the details are left as \Cref{\res{exercise:rel-implies-lang}}.
\end{proof}

Together, these last two lemmas tell us that agreement under the language semantics is equivalent to agreement under \emph{any} relational interpretation.
This means that the language semantics fulfills its intention of abstracting over the meaning of primitive actions.

\begin{theorem}[{\cite[Page~24]{pratt-1980}}]\label{thm:rel_equiv_lang}
Let $e, f \in \Exp$.
It holds that $\sem{e} = \sem{f}$ if and only if for all interpretations $\sigma$, we have that $\sem{e}_\sigma = \sem{f}_\sigma$.
\end{theorem}

In light of the above, we will use ``the semantics of regular expressions'' to refer to the language semantics and the (universally quantified) relational semantics interchangeably in the sequel.
After all, we are mostly interested in the equivalence induced by the semantics, and we know that both perspectives equate the same expressions.

\section{Reasoning}

Writing a program as a regular expression is a nice exercise, but not an end in itself.
We want to use this representation of the program to reason about whether or not a program encoded in this way is equivalent to another program.
In this section, we get to the essence of Kleene Algebra: the laws that underpin equivalence of regular expressions.
If you favor the relational semantics, these are the pairs of regular expressions that yield the same expression under any interpretation.
Alternatively, under the language semantics, it can be seen as the study of which expressions denote the same patterns of events.

As a very basic example, take any $e, f \in \Exp$.
Now $e + f$ has the same semantics as $f + e$.
After all, we can calculate as follows:
\[
    \sem{e + f} = \sem{e} \cup \sem{f} = \sem{f} \cup \sem{e} = \sem{f + e}
\]
In other words, the operator $+$ is \emph{commutative}.
There are many correspondences like these, and we can use them when we want to prove properties of regular expressions.
Here are a number of useful ones.

\begin{lemma}\label{lem:ka-equational}
The following hold for all $e, f, g \in \Exp$:
\begin{mathpar}
\sem{e + 0} = \sem{e}
\and
\sem{e + e} = \sem{e}
\and
\sem{e + f} = \sem{f + e}
\and
\sem{e + (f + g)} = \sem{(e + f) + g}
\and
\sem{e \cdot (f \cdot g)} = \sem{(e \cdot f) \cdot g}
\and
\sem{e \cdot (f + g)} = \sem{e \cdot f + e \cdot g}
\and
\sem{(e + f) \cdot g} = \sem{e \cdot g + f \cdot g}
\and
\sem{e \cdot 1} = \sem{e} = \sem{1 \cdot e}
\and
\sem{e \cdot 0} = \sem{0} = \sem{0 \cdot e}
\and
\sem{1 + e \cdot e^*} = \sem{e^*} = \sem{1 + e^* \cdot e}
\end{mathpar}
\end{lemma}
\begin{proof}[Proof sketch]
The proofs are similar to that of the third property, shown above.
We highlight a particular case in \Cref{\res{exercise:unfolding}}.
\end{proof}

Most of these properties are fairly easy to understand intuitively.
For instance, the equation $\sem{e + e} = \sem{e}$ says that a non-deterministic choice between a sequence of events from $e$ or $e$ is really no choice at all --- the $+$ operator is \emph{idempotent}.
Similarly, choosing between $e$ and $f + g$ is the same as choosing between $e + f$ and $g$, as witnessed by the fact that $\sem{e + (f + g)} = \sem{(e + f) + g}$ --- the operator $+$ is \emph{associative}.

\begin{example}%
\label{ex:baby-equivalence}
We can use the laws from \Cref{\res{lem:ka-equational}} to reason about equivalences.
For instance, to show that $\sem{0^*} = \sem{1}$, we can derive:
\begin{align*}
\sem{0^*}
    &= \sem{1 + 0 \cdot 0^*} \tag{$\sem{e^*} = \sem{1 + e \cdot e^*}$} \\
    &= \sem{1} + \sem{0 \cdot 0^*} \tag{def.\ $\sem{-}$} \\
    &= \sem{1} + \sem{0} \tag{$\sem{0 \cdot e} = \sem{0}$} \\
    &= \sem{1 + 0} \tag{def.\ $\sem{-}$} \\
    &= \sem{1} \tag{$\sem{e + 0} = \sem{e}$}
\end{align*}
\end{example}

Some properties of the Kleene star are conditional, i.e., they require proving some other property first.
We list two particularly useful ones.
\begin{lemma}%
\label{lem:ka-quasiequational}
The following hold for all $e, f, g \in \Exp$:
\begin{enumerate}
    \item If $\sem{e + f \cdot g} \subseteq \sem{g}$ then $\sem{f^* \cdot e} \subseteq \sem{g}$.
    \item If $\sem{e + f \cdot g} \subseteq \sem{f}$ then $\sem{e \cdot g^*} \subseteq \sem{f}$.
\end{enumerate}
\end{lemma}
The proof of this Lemma appears as Exercise~\ref{ex:ka-quasiequational}.
Intuitively, the first law above can be read as saying that, if the behavior of $g$ includes both the behaviors of $f \cdot g$ and $e$, then any sequence of events described by $g$ (including, in particular, the behavior of $e$) can be prepended with a sequence of events from $f$ to obtain a new sequence of events also allowed by $g$.
Since this can be repeated any number of times, it follows that the behavior of $f^* \cdot e$ must also be included in that of $g$; the second law can be explained similarly.

As it turns out, the properties of $\sem{-}$ that we have seen thus far are so useful, it pays off to isolate them into a \emph{reasoning system}, as follows.

\begin{definition}[Provable equivalence~\cite{kozen-1994}]\label{def:ka}
We define $\equiv$ as the smallest congruence on $\Exp$ that satisfies the following for all $e, f, g \in \Exp$:
\begin{mathpar}
 e + 0 \equiv e
\and
e + e \equiv e
\and
 e + f \equiv f + e
\and
 e + (f + g) \equiv (e + f) + g
\and
 e \cdot (f \cdot g) \equiv (e \cdot f) \cdot g
\and
 e \cdot (f + g) \equiv e \cdot f + e \cdot g
\and
 (e + f) \cdot g \equiv e \cdot g + f \cdot g
\and
 e \cdot 1 \equiv e
\and
 1 \cdot e \equiv e
\and
 e \cdot 0 \equiv 0
\and
 0 \cdot e \equiv 0
\and
 1 + e \cdot e^* \equiv e^*
\and
 1 + e^* \cdot e \equiv e^*
\and
\inferrule{%
     e + f \cdot g \leqq g
}{%
     f^* \cdot e \leqq g
}
\and
\inferrule{%
     e + f \cdot g \leqq f
}{%
     e \cdot g^* \leqq f
}
\end{mathpar}
where we write $ e \leqq f$ as a shorthand for $ e + f \equiv f$.
\end{definition}

Intuitively, the way we defined $\equiv$ means that if $ e \equiv f$ holds for some $e, f \in \Exp$, then we can prove $\sem{e} = \sem{f}$ by using only the properties in \Cref{\res{lem:ka-equational},\res{lem:ka-quasiequational}}, as well as the definition of $\sem{-}$.

\begin{lemma}[Soundness]%
    \label{lem:soundness}
    Let $e, f \in \Exp$.
    If $e \equiv f$, then $\sem{e} = \sem{f}$.
    \end{lemma}

\Cref{\res{lem:soundness}} is an instance of a more general correspondence between the laws in \Cref{\res{def:ka}} and the notion of Kleene algebra.
To see this, we formally define ``a Kleene algebra'' using exactly those laws:
\begin{definition}[Kleene algebra]\label{def2:ka}
    A Kleene algebra $\mathcal K = (X, +, \cdot,-^*, 0, 1)$ consists of a set $X$ with two distinguished elements $0$ and $1$, two binary operations $+$ and $\cdot$ (the latter usually omitted when writing the expressions), and a unary operation $-^*$, that together satisfy the laws in \Cref{\res{def:ka}} (e.g., $ e+f\equiv f+e$, hence for all $x,y\in X$ $x+y=y+x$).

    Every Kleene algebra induces a partial order $\leq$ derived from the $+$ operator in the same way that $\leqq$ is, i.e., $x \leq y$ if and only if $x + y = y$.
\end{definition}

\begin{example}
\Cref{\res{lem:soundness}} essentially amounts to noting that \[(2^{\alphabet^*}, \cup, \cdot,-^*, \emptyset, \{\epsilon\})\] is a Kleene algebra.
In this model, $\leq$ is set inclusion.
Another well-known Kleene algebra is formed by the binary relations over a set $X$:
\[
    (2^{X \times X}, \cup, \circ, -^*, \emptyset, \id_X)
\]
where this time, $-^*$ denotes reflexive-transitive closure.
\end{example}

\medskip
Returning to the laws in \Cref{\res{def:ka}}, the first line of rules summarizes the properties of $+$, and essentially says that $\mathcal K$ is a join-semilattice.
Together with the second and third line, which contain the properties of $\cdot$ and its interaction with $+$, it states that $\mathcal K$ is an idempotent semiring.
The remaining rules axiomatize the star operator.

As suggested, $\leqq$ is a partial order up to $\equiv$.
Moreover, the operators are monotone with respect to $\leqq$.
We record these properties below; their proofs are left to \Cref{\res{exercise:squeeze}}.

\begin{lemma}%
\label{lem:squeeze}
Let $e, f, g \in \Exp$.
The following properties hold:
\begin{lemmaenum}[ref={\thelemma(\arabic*)}]
    \setlength{\itemsep}{0em}
    \item\label{lemma:squeeze:reflexive}
    If $e \equiv f$ then $e \leqq f$.
    \item\label{lemma:squeeze:antisymmetric}
    If $e \leqq f$ and $f \leqq e$, then $ e \equiv f$.
    \item\label{lemma:squeeze:transitive}
    If $e \leqq f$ and $f \leqq g$, then $ e \leqq g$.
    \item\label{lemma:squeeze:monotone-plus}
    If $e \leqq f$, then $e + g \leqq f + g$ and $g + e \leqq g + f$.
    \item\label{lemma:squeeze:monotone-sequence}
    If $e \leqq f$, then $e \cdot g \leqq f \cdot g$ and $g \cdot e \leqq g \cdot f$.
    \item\label{lemma:squeeze:monotone-star}
    If $e \leqq f$, then $e^* \leqq f^*$.
\end{lemmaenum}
\end{lemma}

\begin{remark}
Some authors make $\leqq$ the primary object of study, and define $e\equiv f$ as a shorthand for ``$e\leqq f$ and $f\leqq e$''.
This is a very natural way to present Kleene Algebra, which can simplify a number of proofs.
The interdefinability of these relations as well as \Cref{\res{lem:squeeze}} allows us to switch between these perspectives at will.
\end{remark}

We can now use $\equiv$ and $\leqq$ to reason cleanly about equivalence.
We will use the associativity of $+$ and $\cdot$ as an excuse to abuse notation, and drop some parentheses, writing $e \cdot f \cdot g$ to mean either $e \cdot (f \cdot g)$ or $(e \cdot f) \cdot g$ (and similarly for $+$).\footnote{%
    This convention comes about so naturally that the reader may not have noticed that we already used it earlier on, when we mentioned the expression $\ltr{a}\ltr{b}^*\ltr{c}$.
}
Essentially, these conventions mean that we can employ the usual style of equational reasoning.

\begin{example}%
    \label{ex:baby-equivalence-redux}
    Returning to \Cref{\res{ex:baby-equivalence}}, to prove that $\sem{0^*} = \sem{1}$, we can use \Cref{\res{lem:soundness}} to instead prove that $ 0^* \equiv 1$, as follows.
    \begin{align*}
         0^*
        &\equiv 1 + 0 \cdot 0^* \tag{\ensuremath{e^* \equiv 1 + e \cdot e^*}} \\
        &\equiv 1 + 0 \tag{\ensuremath{0 \cdot e \equiv 0}, congruence} \\
        &\equiv 1 \tag{\ensuremath{e + 0 \equiv e}}
    \end{align*}
    Note how the derivation is less cluttered than the one in \Cref{\res{ex:baby-equivalence}}.
\end{example}

The only unfamiliar feature of reasoning using $\equiv$ comes from the last two laws in \Cref{\res{def:ka}}, which we will refer to as the \emph{left} and \emph{right fixpoint} rule respectively, or simply as the \emph{star axioms}.
We can use these by either assuming that the conclusion holds and using that to prove the claim, followed by a proof of the premise (backward reasoning), or proving the premise and then using the consequence later (forward reasoning).
Here is an example of the backwards reasoning style.

\begin{lemma}[{\cite[Chapter~3]{conway-1971}}]\label{lem:starstuff}
    The following hold for all $e \in \Exp$:
    \begin{mathpar}
     e^* \equiv e^* + 1
     \and
     e^*e^* \equiv e^*
     \and
     e^{**} \equiv e^*
    \end{mathpar}
\end{lemma}
\begin{proof}
    The first equality follows easily:
    \[
        e^* \equiv 1+ee^* \equiv 1 +  1+ee^* \equiv 1 + e^* \equiv e^* + 1
    \]
    For the second equality, we have:
    \[
        e^*e^* \equiv (1+ee^*)e^* \equiv e^*+ee^*e^* \geqq e^*
    \]
    Furthermore, to prove that $e^*e^* \leqq e^*$, it suffices to show that $ee^* + e^* \leqq e^*$ (by the left fixpoint rule).
    To this end, we derive that:
    \[
        ee^*+e^* \leqq  1 +ee^*+e^* \equiv e^*+e^* \equiv e^*
    \]
    For the third equality, we start by proving that $e^{**} \leqq e^*$.
    By the left fixpoint rule, it suffices to show that $e^*e^* +1 \leqq e^*$.
    This follows from the first two equalities; after all, we can derive that:
    \[
        e^*e^* + 1 \equiv e^* +1 \equiv e^*
    \]
    Conversely, to show that $e^* \leqq e^{**}$, it suffices (again by the left fixpoint rule) to prove that $ee^{**} + 1 \leqq e^{**}$, which can be done as follows:
    \[
        ee^{**} +  1
            \leqq e^*e^{**} + 1
            \equiv e^{**}
            \qedhere
    \]
\end{proof}

Another example of backward reasoning follows.
\begin{lemma}[{Slide rule; cf.~\cite[Chapter~4]{conway-1971}}]%
\label{lem:slide}
$ e \cdot {(f \cdot e)}^* \equiv {(e \cdot f)}^* \cdot e$
\end{lemma}
\begin{proof}
By \Cref{\res{lem:squeeze}}, it suffices to prove $ e \cdot {(f \cdot e)}^* \leqq {(e \cdot f)}^* \cdot e$ and $ {(e \cdot f)}^* \cdot e \leqq e \cdot {(f \cdot e)}^*$.
To argue the former, we can apply the right fixpoint rule, which tells us that it is sufficient to prove
\[
     e + {(e \cdot f)}^* \cdot e \cdot f \cdot e \leqq {(e \cdot f)}^* \cdot e
\]
This is true, as we can derive the following and apply \Cref{\res{lemma:squeeze:reflexive}}.
\begin{align*}
     e + {(e \cdot f)}^* \cdot e \cdot f \cdot e
        &\equiv 1 \cdot e + {(e \cdot f)}^* \cdot e \cdot f \cdot e \\
        &\equiv (1 + {(e \cdot f)}^* \cdot e \cdot f) \cdot e \\
        &\equiv {(e \cdot f)}^* \cdot e
\end{align*}
The proof of $ {(e \cdot f)}^* \cdot e \leqq e \cdot {(f \cdot e)}^*$ is similar, and left as \Cref{\res{exercise:sliding}}.
\end{proof}

Given how useful the provability relation is, a natural question to ask is whether it is ``enough'' to prove all equivalences that are true?
More formally, is $\equiv$ \emph{complete}, in the sense that $\sem{e} = \sem{f}$ implies $ e \equiv f$?
Answering this question requires a good deal of mathematical machinery, which will be developed over the next few chapters.
By the end of \Cref{\res{chapter:completeness}}, we will be able to answer in the positive.

\section{Exercises}

\begin{exercises}
    \item\label{ex:1:basics}
    In \Cref{\res{example:program:basics}}, we have seen how you can use programs like $\mathtt{guard}$ and $\mathtt{validate}$ to abort non-deterministic branches of program execution, and create something like a \whdbare construct using the star operator.

    You can use the same technique to encode an \itebare construct, using the non-deterministic choice operator $+$ and suitably chosen programs that abort execution under certain conditions.

    Suppose you want to encode the following program as a regular expression, where $M$, $L$, $R$ and $y$ are valued as natural numbers:

    \begin{algorithm}[H]
    \eIf{$M^2 \leq y$}{%
        $L \gets M$\;
    }{%
        $R \gets M$\;
    }
    \end{algorithm}

    We can encode this as a regular expression using the technique from \Cref{\res{example:if-then-else}}, writing $\mathtt{lte} \cdot \mathtt{writeL} + \mathtt{gt} \cdot \mathtt{writeR}$, where
    \begin{itemize}
        \item
        $\mathtt{lte}$ and $\mathtt{gt}$ are programs that check whether $M^2 \leq y$ and $M^2 \gr y$ respectively, and abort if this is not the case.

        \item
        $\mathtt{writeL}$ and $\mathtt{writeR}$ write $M$ to $L$ or to $R$, respectively.
    \end{itemize}

    Your task is two-fold:
    \begin{exercises}
        \item
        Give a semantic domain $S$ that encodes the state of this four-variable program.
        \emph{Hint: look back at how we encoded a two-variable state.}

        \item
        Give an interpretation $\sigma$ to each primitive program, in $S$.
    \end{exercises}

    \item
    The program in the previous exercise is part of a larger program, which finds the integer square root of $y$ using binary search:

    \begin{algorithm}[H]
    $L \gets 0$\;
    $R \gets y + 1$\;
    \While{$L \ls R-1$}{%
        $M \gets \lfloor \frac{L+R}{2} \rfloor$\;
        \eIf{$M^2 \leq y$}{%
            $L \gets M$\;
        }{%
            $R \gets M$\;
        }
    }
    \end{algorithm}

    \begin{exercises}
        \item
        Come up with suitable primitive programs and their description to encode this program.
        You may reuse the primitive programs from the previous exercise.

        \item
        Give a regular expression to encode the program as a whole.
        You may reuse the expression for the \itebare construct.

        \item
        Give an interpretation $\sigma$ to your new primitive programs, using the same semantic domain $S$ as in the last exercise.
    \end{exercises}

    \item\label{exercise:unfolding}
    Prove that, for all $e \in \Exp$, it holds that $\sem{1 + e \cdot e^*} = \sem{e^*} $.

    \item\label{exercise:squeeze}
    Recall that $  e \leqq f$ is shorthand for $  e + f \equiv f$.
    Let us now  verify the claims in \Cref{\res{lem:squeeze}}.
    In the following, let $e, f, g \in \Exp$; prove:
    \begin{exercises}
        \item
        If $e \equiv f$ then $e \leqq f$.
        \item
        If $e \leqq f$ and $f \leqq e$, then $ e \equiv f$.
        \item
        If $e \leqq f$ and $f \leqq g$, then $ e \leqq g$.
        \item
        If $e \leqq f$, then $e + g \leqq f + g$ and $g + e \leqq g + f$.
        \item
        If $e \leqq f$, then $e \cdot g \leqq f \cdot g$ and $g \cdot e \leqq g \cdot f$.
        \item
        If $e \leqq f$, then $e^* \leqq f^*$.
    \end{exercises}

    \item\label{exercise:sliding}
    In \Cref{\res{lem:slide}}, we showed that $  e \cdot {(f \cdot e)}^* \leqq {(e \cdot f)}^* \cdot e$.
    Argue the other direction, i.e., that $ {(e \cdot f)}^* \cdot e \leqq e \cdot {(f \cdot e)}^*$.

    Together, these two laws tell us that $(e \cdot f)^* \cdot e \equiv e \cdot (f \cdot e)^*$, which is known as the \emph{slide rule}.

    \item\label{exercise:denesting}
    Prove the \emph{denesting rule}~\cite[Chapter~3]{conway-1971}: ${(e + f)}^* \equiv {(e^* \cdot f)}^*\cdot e^* $.

    \item
    Let $e\in \Exp$.
    Prove that for all $n \gr 0$, we have $ e^* \equiv (1+e)^{n-1}(e^n)^*$, where $e^n$ is defined as $e^0=1$, and $e^n=e\cdot e^{n-1}$.

    \item\label{exercise:bisimulation-law}
    Let $e,f,g\in \Exp$.
    Prove that if $ ef \equiv fg$, then $ e^*f \equiv fg^*$~\cite{kozen-2001}.

    \item\label{exercise:tightening}
    Let $e \in \Exp$.
    Show that $(e +1)^* \equiv e^*$~\cite{salomaa-1966}.

    \item\label{exercise:lang-implies-rel}
    Let us now  fill in the proof sketch of \Cref{\res{theorem:lang-implies-rel}}.
    \begin{exercises}
        \item
        Show that for all $L, K \subseteq \alphabet^*$, the following hold:
        \begin{mathpar}
            \hat{\sigma}(L \cup K) = \hat{\sigma}(L) \cup \hat{\sigma}(K)
            \\
            \hat{\sigma}(L \cdot K) = \hat{\sigma}(L) \circ \hat{\sigma}(K)
            \and
            \hat{\sigma}(L^*) = {\hat{\sigma}(L)}^*
        \end{mathpar}
        \item
        Use the above to prove that for each interpretation $\sigma$ and $e \in \Exp$, it holds that $\sem{e}_\sigma = \hat{\sigma}(\sem{e})$, by induction on $e$.
    \end{exercises}

    \item\label{exercise:rel-implies-lang}
    We now complete the sketched proof of \Cref{\res{theorem:rel-implies-lang}}.
    Let $e \in \Exp$.
    Show that $\sem{e}_\sharp = \{ \angl{w, wx} \mid w \in \alphabet^*, x \in \sem{e} \}$, by induction on $e$.

\item\label{ex:ka-quasiequational} Prove \Cref{\res{lem:ka-quasiequational}}: for all $e, f, g \in \Exp$ the following hold:
\begin{enumerate}
    \item If $\sem{e + f \cdot g} \subseteq \sem{g}$ then $\sem{f^* \cdot e} \subseteq \sem{g}$.
    \item If $\sem{e + f \cdot g} \subseteq \sem{f}$ then $\sem{e \cdot g^*} \subseteq \sem{f}$.
\end{enumerate}

    \item
    Use \Cref{\res{lem:soundness}} to prove that ${(\mathtt{a} + \mathtt{b})}^* \not\equiv \mathtt{a}^* \cdot {(\mathtt{b} \cdot \mathtt{a})}^*$.
    Be precise in your arguments --- both in how you use soundness, and what remains to be proved after that.

    \item
    Find a set $X$ and an interpretation $\sigma\colon \alphabet \to 2^{X \times X}$ such that $\sem{{(\mathtt{a} + \mathtt{b})}^*}_\sigma \neq \sem{\mathtt{a}^* \cdot {(\mathtt{b} \cdot \mathtt{a})}^*}_\sigma$.
    Then, use \Cref{\res{lem:soundness}} and \Cref{\res{thm:rel_equiv_lang}} to conclude that ${(\mathtt{a} + \mathtt{b})}^* \not\equiv \mathtt{a}^* \cdot {(\mathtt{b} \cdot \mathtt{a})}^*$.

\end{exercises}

\onlyinsubfile{%
    \printbibliography%
}

\chapter{A primer on coalgebra}\label{chapter:coalgebra}

In this chapter, we discuss the very basics of a general theory of abstract machines, known as \emph{(universal) coalgebra}~\cite{rutten-2000}.
This will give us a very natural way to define the operational semantics of a machine, as well as a sensible idea of when two machines agree on their semantics.
Coalgebra will be particularly useful when we discuss the operational semantics of languages in the next chapter.
Although we could develop the operational semantics of regular expressions concretely, the coalgebraic perspective provides a unifying framework as we will see in \Cref{\res{chapter:kat}}, where all definitions fit the same pattern as for Kleene algebra.

The material in this chapter is largely based on~\cite{rutten-2000}.
The interested reader is referred to op.\ cit., and also to~\cite{rutten-book,jacobs-book} for extensive treatments.

\section{Functors}

Suppose you encounter a machine, in the form of an opaque box with some combination of buttons and displays.
When you press these buttons, you may change its internal state; as a result of your actions, some information may also appear on one of the displays.
Universal coalgebra is a way of thinking of such a machine in terms of the interactions it allows, and how it evolves over time as a result of these interactions.

Let us  model the states of the machine as some set $X$.
In every state $x \in X$, the machine may offer some combination of output values, as well as opportunities for interaction which yield a new machine state.
We can think of the possible values of these outputs and interactions as another set; because this set may be built from the possible states of the machine, we denote it with $F(X)$.
The behavior of the machine can then be thought of as a function $t\colon X \to F(X)$ that assigns, to every state $x \in X$, a combination of outputs and interactions $t(x)$.

\begin{example}%
\label{example:infernal-machine}
Suppose our hypothetical machine has just one button.
Upon pressing this button, the machine may self-combust, leaving behind a heap of ashes (and its perplexed user).
We can model the states of the machine as some set $X$, and its set of outputs and interactions as $M(X) = X + 1$: either $t(x) \in X$, which means the machine has transitioned to some new (but unobservable) state, or $t(x) = *$, in which case the machine has gone up in flames.
\end{example}

\begin{example}%
\label{example:coffee-machine}
Suppose we are studying a coffee machine.
When you press a button $b \in B$, the machine pours out some type of coffee $c \in C$; it then transitions to a new state.
With $X$ as a set of machine states, the interface of this machine is given by $H(X) = (C \times X)^B$, and its behavior as a function $t\colon X \to H(X)$.
For any $x \in X$ and $b \in B$, $t(x)(b) = (c, x')$ means that if button $b$ is pressed in state $x$, the machine will pour coffee of type $c$ and transition to state $x'$.
\end{example}

If $F$ is either $M$ or $H$ above, we can lift $f\colon X \to Y$ to $F(f)\colon F(X) \to F(Y)$, by applying $f$ to the values from $X$ that ``live inside'' elements of $F(X)$.
This lifting is compatible with composition and the identity, and in fact an instance of a more general notion formalized below.

\begin{definition}[Functor]
A \emph{functor} $F$ associates with every set $X$ a set $F(X)$, and with every function $f\colon X \to Y$ a function $F(f)\colon F(X) \to F(Y)$, such that (1)~for every set $X$, $F$ preserves the identity function, i.e., $F(\mathsf{id}_X) = \mathsf{id}_{F(X)}$, and (2)~for all functions $f\colon X \to Y$ and $g\colon Y \to Z$, $F$ is compatible with composition, i.e., $F(g \circ f) = F(g) \circ F(f)$.
\end{definition}

\begin{example}%
\label{example:infernal-machine-functor}
Given a set $X$, let $M(X) = X + 1$, as in \Cref{\res{example:infernal-machine}}.
We can then assign to every function $f\colon X \to Y$ a function $M(f)\colon M(X) \to M(Y)$ by setting $M(f)(x) = f(x)$ when $x \in X$, and $M(f)(*) = *$; you may check that this makes $M$ a functor --- see \Cref{\res{exercise:infernal-machine-functor}}.
\end{example}

\begin{example}%
\label{example:coffee-machine-functor}
Given a set $X$, let $H(X)$ be as in \Cref{\res{example:coffee-machine}}.
For $f\colon X \to Y$, we can choose $H(f)\colon H(X) \to H(Y)$ as follows.
Given $m\colon B \to C \times X$, we define $H(f)(m)\colon B \to C \times Y$ by setting $H(f)(m)(b) = (c, f(x))$ when $m(b) = (c, x)$.
This defines a functor; see \Cref{\res{exercise:coffee-machine-functor}}.
\end{example}

In what follows, we will refer to the functors $M$ and $H$ as defined above without further cross-references.

\begin{example}
Let $E \subseteq \naturals$ be the set of even numbers.
Suppose $F(X) = 1$ when $X = E$, and $F(X) = X$ otherwise.
This \emph{cannot} be a functor.
To see why, suppose towards a contradiction that for each $f\colon X \to Y$ we can define an $F(f)\colon F(X) \to F(Y)$ satisfying both functor properties.
Consider the maps $d\colon \naturals \to E$ and $h\colon E \to \naturals$ given by $d(n) = 2n$ and $h(n) = n/2$.
Clearly $h \circ d$ is the identity on $\mathbb{N}$; since $F$ is a functor, $F(h \circ d)$ is the identity on $F(\mathbb{N}) = \mathbb{N}$ by the first property.
Now, $F(d)$ goes from $\mathbb{N}$ to $F(E) = 1$, and is therefore constant; but then so is $F(h \circ d) = F(h) \circ F(d)$ by the second property --- a contradiction.
\end{example}

\section{Coalgebras and Homomorphisms}

We are now ready to define coalgebras.
The name \emph{coalgebra} or \text{\emph{co-algebra}} comes from category theory, where it is the dual to \emph{algebra}, an abstraction of universal algebra (see for instance~\cite{burris-sankappanavar-1981}).
If an algebra tells you how to ``compose'' objects into new ones, then a coalgebra tells you how to ``decompose'' or unfold an object; this makes them suitable to model the behavior of machines, where one state evolves to the next.

\begin{definition}[$F$-coalgebra]
Let $F$ be a functor.
An \emph{$F$-coalgebra} is a pair $(X, t)$ where $X$ is a set and $t\colon X \to F(X)$ is a function.
\end{definition}

Now suppose we have two abstract machines with the same buttons and displays, modeled as $F$-coalgebras for a certain functor $F$ that describes their interface.
How would we prove that the behavior of each state in the first machine also exists in the other?
The coalgebraic way to do this is to demonstrate a function that maps a state of the first machine to a state of the second machine, in a way that preserves the observable values and next states.
This leads to the following notion.

\begin{definition}[Homomorphism]\label{def:homomorphism}
Let $F$ be a functor, and let $(X, t)$ and $(Y, u)$ be $F$-coalgebras.
A function $f\colon X \to Y$ is called an \emph{($F$-coalgebra) homomorphism} if it holds that $u \circ f = F(f) \circ t$.
\end{definition}

In the sequel, we may write $f\colon (X, t) \to (Y, u)$ to compactly represent that $f$ is an homomorphism from $(X, t)$ to $(Y, u)$.

\begin{example}%
\label{example:infernal-machine-morphism}
An $M$-coalgebra homomorphism from $(X, t)$ to $(Y, u)$ is a function $f\colon X \to Y$ such that (1)~if $t(x) = *$, then $u(f(x)) = *$ as well, and (2)~if $t(x) \in X$, then $f(t(x)) = u(f(x))$.
The first condition says that if the first machine is in state $x$ and pressing the button detonates it, then so does pressing the button on the second machine in state $f(x)$.
The second condition tells us that if pressing the button on the first machine in state $x$ but takes us to state $x'$, then pressing the button on the second machine in state $f(x)$ takes us to $f(x')$.

It is not too hard to show that if the first machine catches fire after pressing the button $n$ times starting in state $x$, then so does the second machine when repeating this experiment starting from state $f(x)$; in this sense, $f$ maps states in $(X, t)$ to equivalent ones in $(Y, u)$.
\end{example}

\begin{example}%
\label{example:coffee-machine-morphism}
An $H$-coalgebra homomorphism from $(X, t)$ to $(Y, u)$ is a function $f\colon X \to Y$ such that if $t(x)(b) = (c, x')$, then $u(f(x))(b) = (c, f(x'))$.
This means that if pressing button $b$ when the first machine is in state $x$ yields coffee of type $c$, then so does pressing  $b$ on the second machine in state $f(x)$; moreover, a transition to $x'$ in the first machine corresponds to a transition to $f(x')$ in the second machine.
\end{example}

To make sense of properties that relate compositions of functions like the one in \Cref{\res{def:homomorphism}}, it often helps to have a graphical depiction like the diagram below: the domains and codomains of $u$, $f$, $F(f)$ and $t$ are drawn as nodes, and the functions between them as arrows. 
\[
    \begin{tikzpicture}
        \node (X) {$X$};
        \node[below= of X] (FX) {$F(X)$};
        \node[right=2cm of X] (Y) {$Y$};
        \node at (FX -| Y) (FY) {$F(Y)$};
        \draw (X) edge[->] node[above] {$f$} (Y);
        \draw (X) edge[->] node[left] {$t$} (FX);
        \draw (FX) edge[->] node[above] {$F(f)$} (FY);
        \draw (Y) edge[->] node[right] {$u$} (FY);
    \end{tikzpicture}
\]
In the sequel, we will say that such a diagram \emph{commutes} when composing functions along two paths between nodes gives the same function.
For instance, the diagram above commutes precisely when $f$ is an $F$-coalgebra homomorphism from $(X,t)$ to $(Y, u)$, i.e., when $u \circ f = F(f) \circ t$.

\medskip
We introduced coalgebra homomorphisms as a way to recover the behavior of one machine inside the other.
One might wonder whether we can create an $F$-coalgebra that unambiguously contains the behavior of \emph{all} $F$-coalgebras.
The following captures this idea.

\begin{definition}[Final Coalgebra]\label{definition:final-coalgebra}
An $F$-coalgebra $(\Omega, \omega)$ is \emph{final} if for every $F$-coalgebra $(X, t)$ there exists precisely one $F$-coalgebra homomorphism ${!_X}\colon X \to \Omega$, i.e., such that the following commutes:
\[
    \begin{tikzpicture}
        \node (X) {$X$};
        \node[below= of X] (FX) {$F(X)$};
        \node[right=2cm of X] (Omega) {$\Omega$};
        \node at (FX -| Omega) (FOmega) {$F(\Omega)$};
        \draw[dashed] (X) edge[->] node[above] {${!_X}$} (Omega);
        \draw (X) edge[->] node[left] {$t$} (FX);
        \draw[dashed] (FX) edge[->] node[above] {$F({!_X})$} (FOmega);
        \draw (Omega) edge[->] node[right] {$\omega$} (FOmega);
    \end{tikzpicture}
\]
In this diagram, we have dashed the horizontal arrows to emphasize that they arise from the vertical arrows.
\end{definition}

Existence of a homomorphism into a final $F$-coalgebra guarantees that it can model the behavior of every state of every $F$-coalgebra, while uniqueness tells us that there is only one possible way to do this.
A final coalgebra may not always exist for every functor, but when it does, its state set is typically suited to act as a semantics for $F$-coalgebras, because they contain enough structure to describe the behavior of a state.
The following examples illustrate this idea further.

\begin{example}\label{example:infernal-machine-final}
We return once more to the functor $M$.
Let $\mathbb{N}^\infty$ consist of the natural numbers augmented with the symbol $\infty$.
We can define an $M$-coalgebra $(\mathbb{N}^\infty, \bullet)$, defining $\bullet\colon \mathbb{N}^\infty \to \mathbb{N}^\infty + 1$ by
\begin{mathpar}
    \bullet(n) =
        \begin{cases}
        \infty & n = \infty \\
        * & n = 0 \\
        n-1 & n \gr 0
        \end{cases}
\end{mathpar}
This is exactly the final $M$-coalgebra: given an $M$-coalgebra $(X, t)$, we can define $!_X\colon X \to \mathbb{N}^\infty$ by sending $x \in X$ to the number of button ``safe'' button presses starting from state $x$ before the machine detonates, and to $\infty$ when this is not possible.
Showing that this is an $M$-coalgebra homomorphism from $(X, t)$ to $(\mathbb{N}^\infty, \bullet)$ is left as \Cref{\res{exercise:infernal-machine-final}}.

To see that $!_X$ is the \emph{only} such $M$-coalgebra homomorphism, let $f\colon X \to \mathbb{N}^{\infty}$ be any $M$-coalgebra homomorphism from $(X, t)$ to $(\mathbb{N}^{\infty}, \bullet)$.
If the button can be safely pressed exactly $n$ times starting from $x$, then we show that $f(x) = {!_X(x)}$ by induction on $n$.
In the base, $n = 0$ implies $t(x) = *$, and hence $\bullet(f(x)) = *$ because $f$ is a homomorphism; therefore $f(x) = 0 = {!_X(x)}$.
For the inductive step, assume the button can be safely pressed exactly $n \gr 0$ times starting from $x$.
In that case $\bullet(f(x)) = f(t(x))$ because $f$ is a homomorphism, and $f(t(x)) = {!_X(t(x))}$ by induction; since ${!_X(t(x))} = n - 1$ by definition of $!_X$, we conclude that $\bullet(f(x)) = n - 1$, and hence $f(x) = n = {!_X(x)}$.

Otherwise, if pressing the button will \emph{never} detonate the device, then $t^n(x) \in X$ for all $n \gr 0$, and so $M(\bullet^n)(f(x)) \in \mathbb{N}^\infty$ for all $n \gr 0$ as well.
This means that $f(x)$ can only be $!_X(x) = \infty$.
\end{example}

\begin{example}\label{example:coffee-machine-final}
The final $H$-coalgebra has as a set of states $S$, consisting of functions $s\colon B^+ \to C$, where $B^+$ denotes the set of non-empty sequences over $B$.
Intuitively, such a function assigns to a sequence of button presses $b_1 \cdots b_n$ the types of coffee poured as a result of this sequence.\footnote{%
    The reader familiar with coalgebra might recognize that functor $H$ captures the type of {\em Mealy Machines}.
    The final $H$-coalgebra can be characterized in two equivalent ways, as we explicit in this example or as causal functions from sequences of $b$'s (button presses) to sequences of $c$'s.
    The equivalence of these characterizations is described in~\cite[Theorem 4.1.3]{silva-thesis}
}
Its structure map $\sigma\colon S \to H(S)$ is defined by $\sigma(s)(b) = (s(b), s_b)$, where $s_b\colon B^+ \to C$ is given by $s_b(w) = s(bw)$.

Given an $H$-coalgebra $(X, t)$, we associate with each $x \in X$ a function $!_X(x)\colon B^+ \to C$, which assigns to every $w \in B^+$ a value $!_X(x)(w) \in C$: when $t(x)(b) = (c, x')$, set ${!_X(x)(b)} = c$ and ${!_X(x)(bw)} = {!_X(x')(w)}$ for all $w \in B^+$.
Note how the latter is defined by induction, as $w$ is a shorter string than $bw$.
Showing that this is the only $H$-coalgebra homomorphism from $(X, t)$ to $(S, \sigma)$ is left as \Cref{\res{exercise:coffee-machine-final}}.
\end{example}

If a final coalgebra exists, it is unique ``up to isomorphism'' --- in other words, two final coalgebras must be related by a bijective homomorphism, whose inverse is also a homomorphism.
We now record this formally.

\begin{lemma}\label{lemma:final-coalgebra-unique}
Let $(\Omega, \omega)$ and $(\Omega', \omega')$ both be final $F$-coalgebras.
There exists a bijective $F$-coalgebra homomorphism $f\colon \Omega \to \Omega'$ such that the inverse $f^{-1}\colon \Omega' \to \Omega$ is also an $F$-coalgebra homomorphism.
\end{lemma}

For this reason, we prefer to speak of \emph{the} final coalgebra.

\section{Bisimulations}

Coalgebra homomorphisms can be used to show that the behavior of a state in one coalgebra is replicated by that of another state in another coalgebra.
However, they may not be sufficient, as shown below.

\begin{example}%
\label{example:no-morphism}
Consider $M$-coalgebras $(\{0,1\}, n)$ and $(\{0,1,2\}, m)$ with
\begin{mathpar}
n(0) = 1
\and
n(1) = 0
\and
m(0) = 1
\and
m(1) = 2
\and
m(2) = 0
\end{mathpar}
Clearly, all states in both coalgebras have the same behavior, as pressing the button any number of times will keep the device intact.
However, there is no $M$-coalgebra homomorphism from $(\{0,1\}, n)$ to $(\{0,1,2\}, m)$, or vice versa.
The details are left as \Cref{\res{exercise:no-morphism}}.
\end{example}
In situations such as the above, we need a more general method to relate elements in coalgebras with the same behavior, one that relaxes the assumption that the behavior of every state in the first coalgebra is replicated by a definite state in the second coalgebra.
This brings us to the notion of \emph{bisimulation}, defined as follows.

\begin{definition}\label{def:bisim_prelim}
Let $F$ be a functor, and let $(X, t)$ and $(Y, u)$ be $F$-coalgebras.
An \emph{$F$-bisimulation} (between $(X, t)$ and $(Y, u)$) is a relation $R \subseteq X \times Y$ that can be equipped with a coalgebra structure $\rho\colon R \to F(R)$ such that the projections $\pi_1\colon R \to X$ and $\pi_2\colon R \to Y$ given by $\pi_1(x, y) = x$ and $\pi_2(x, y) = y$ are $F$-coalgebra homomorphisms.

When there exists a bisimulation that relates $x \in X$ to $y \in Y$, we say that $x$ and $y$ are \emph{bisimilar}, denoted $X, x \bisim Y, y$, or simply $x \bisim y$ when the coalgebras are clear from context.
\end{definition}

This definition is somewhat abstract, but for concrete cases a bisimulation can usually be characterised rather simply in terms of the transition structure of the coalgebras.
Let us see some examples.

\begin{example}\label{example:infernal-machine-bisimulations}
A bisimulation between $M$-coalgebras $(X, t)$ and $(Y, u)$ is precisely a relation $R \subseteq X \times Y$ such that whenever $x \mathrel{R} y$, (1)~$t(x) = *$ if and only if $u(y) = *$, and (2)~if $t(x) \in X$ and $u(y) \in Y$, then $t(x) \mathrel{R} u(y)$.
Note how $t(x) \in X$ if and only if $u(y) \in Y$ by the first condition, and so only one of the premises to the second condition needs to hold.
Proving that this characterizes $M$-bisimulations is left as \Cref{\res{exercise:infernal-machine-bisimulations}}.

Using the characterization above, we can now return to the coalgebras $(\{0,1\}, n)$ and $(\{0,1,2\}, m)$ from \Cref{\res{example:no-morphism}} and
  show that $R = \{0,1\} \times \{0,1,2\}$ is a bisimulation, and so all states are bisimilar.
\end{example}

\begin{example}\label{example:coffee-machine-bisimulations}
A bisimulation between $H$-coalgebras $(X, t)$ and $(Y, u)$ is a relation $R \subseteq X \times Y$ such that if $x \mathrel{R} y$, then for each $b \in B$ there exists a $c \in C$ such that $t(x)(b) = (c, x')$ and $u(y)(b) = (c, y')$ with $x' \mathrel{R} y'$.
Validating this characterization is left as \Cref{\res{exercise:coffee-machine-bisimulations}}.
\end{example}

Given that the states related by a bisimulation are equivalent, it should stand to reason that we can shrink our coalgebra by identifying related states.
The following result records this formally.
\begin{lemma}[{\cite[Proposition~5.8]{rutten-2000}}]%
\label{lemma:quotient-coalgebra}
Let $(X, t)$ be an $F$-coalgebra, and let $R$ be a bisimulation on $(X, t)$, with $\equiv_R$ its equivalence closure.
We can equip $X/{\equiv_R}$ with an $F$-coalgebra structure $t/{\equiv_R}$ that makes $[-]\colon X \to X/R$ a homomorphism from $(X, t)$ to $(X/{\equiv_R}, t/{\equiv_R})$.
\end{lemma}
\begin{proof}
We choose $t/{\equiv_R}\colon X/{\equiv_R} \to F(X/{\equiv_R})$ as
\begin{equation}%
\label{equation:quotient-morphism}
    t/{\equiv_R}([x]) = F([-])(t(x))
\end{equation}
To see that this is well-defined, we should check that the choice of the representative $x$ does not matter, i.e., that if $x \equiv_R x'$, then $F([-])(t(x)) = F([-])(t(x'))$.
To this end, it suffices to check the case where $x \mathrel{R} x'$.
Let $r\colon R \to FR$ be the coalgebra structure on $R$ such that $t \circ \pi_i = F(\pi_i) \circ r$ for $i \in \{1,2\}$.
We then derive as follows
\begin{align*}
F([-])(t(x))
    &= F([-])(t(\pi_1(x,x'))) \\
    &= F([-])(F(\pi_1)(r(x, x'))) \\
    &= F([-] \circ \pi_1)(r(x, x')) \\
    &= F([-] \circ \pi_2)(r(x, x')) {\tag{$*$}} \\
    &= F([-])(F(\pi_2)(r(x, x'))) \\
    &= F([-])(t(\pi_2(x,x'))) \\
    &= F([-])(t(x'))
\end{align*}
In ($*$), we used the fact that $[-] \circ \pi_1 = [-] \circ \pi_2$ by construction of $\equiv_R$: after all, if $y \mathrel{R} y'$, then $y \equiv_R y'$.
The other steps apply properties of functors, coalgebra homomorphisms, and the fact that $x \mathrel{R} x'$.

Finally, as~\eqref{equation:quotient-morphism} holds by construction, we have that $[-]$ is an $F$-coalgebra homomorphism from $(X, t)$ to $(X/{\equiv_R}, t/{\equiv_R})$.
\end{proof}

\section{Relating the notions of equivalence}
We now have two ways of arguing that two states in a coalgebra have the same behavior: we can build a bisimulation relating the states, or we can show that both states are mapped to the same element in the final coalgebra.
These two kinds of equivalence coincide for many functors, as we will detail now.
In fact, one direction holds for all functors: if two states are bisimilar then they are mapped to the same element of the final coalgebra.
This latter notion is often called \emph{behavioral equivalence}.

\begin{lemma}\label{lemma:bisim-implies-behavioral-equiv}
Let $(X, t)$ be an $F$-coalgebra, and let $(\Omega, \omega)$ be the final $F$-coalgebra.
If $x_1, x_2 \in X$ with $x_1 \bisim x_2$, then ${!_X(x_1)} = {!_X(x_2)}$.
\end{lemma}
\begin{proof}
Let $R$ be a bisimulation witnessing that $x_1 \bisim x_2$, and let $\equiv_R$ be its equivalence closure.
By \Cref{\res{lemma:quotient-coalgebra}}, we can construct the coalgebra $(X/{\equiv_R}, t/{\equiv_R})$ in a way that makes $[-]\colon X \to X/{\equiv_R}$ a coalgebra homomorphism.
Since $!_X$ and $!_{X/{\equiv_R}} \circ [-]$ are both $F$-coalgebra homomorphisms from $(X, t)$ to the final coalgebra $(\Omega, \omega)$, they must be the same homomorphism, by uniqueness.
We can then conclude that
\[
    {!_X(x_1)}
        = {!_{X/{\equiv_R}}([x_1])}
        = {!_{X/{\equiv_R}}([x_2])}
        = {!_X(x_2)}
    \qedhere
\]
\end{proof}
The previous lemma tells us that bisimulations are a sound proof technique for behavioral equivalence.
That is: if we have built a bisimulation between two states, then we know they will be mapped to same state in the final coalgebra.
The opposite direction does not hold in general, but we present below a class of functors for which it does.
We need to introduce the notion of kernel and kernel compatible functors.
\begin{definition}[Kernel]
Let $f\colon X \to Y$ be a function.
The \emph{kernel} of $f$ is the set $\mathsf{Ker}(f) = \{ (x_1, x_2) \in X^2 \mid f(x_1) = f(x_2) \}$.
The \emph{projection maps} $\pi^f_1, \pi^f_2\colon \mathsf{Ker}(f) \to X$ are given by $\pi^f_i(x_1,x_2) = x_i$ for $i \in \{1,2\}$.
\end{definition}

Intuitively, the kernel of a function is nothing more than the equivalence relation induced by which elements are identified through $f$.

\begin{definition}\label{definition:kernel-compatible}
Let $F$ be a functor.
We call $F$ \emph{kernel-compatible} when for all $f\colon X \to Y$ and $p,q \in F(X)$ with $F(f)(p) = F(f)(q)$ --- i.e., $(p, q) \in \mathsf{Ker}(F(f))$ --- there exists an $r \in F(\mathsf{Ker}(f))$ s.t.\ $F(\pi^f_1)(r) = p$ and $F(\pi^f_2)(r) = q$.
Equivalently, $F$ is kernel-compatible when for all $f\colon X \to Y$ there exists an $m_f \colon \mathsf{Ker}(F(f)) \to F(\mathsf{Ker}(f))$ making the diagram below commute.
\[
    \begin{tikzpicture}
        \node (kerfunc) {$\mathsf{Ker}(F(f))$};
        \node[below=of kerfunc] (funcker) {$F(\mathsf{Ker}(f))$};
        \node[left=of funcker] (FX1) {$F(X)$};
        \node[right=of funcker] (FX2) {$F(X)$};
        \draw[->] (kerfunc) edge node[left] {$m_f$} (funcker);
        \draw[->] (funcker) edge node[above] {$F(\pi^f_1)$} (FX1);
        \draw[->] (funcker) edge node[above] {$F(\pi^f_2)$} (FX2);
        \draw[->] (kerfunc) edge[bend right] node[above left] {$\pi^{F(f)}_1$} (FX1);
        \draw[->] (kerfunc) edge[bend left] node[above right] {$\pi^{F(f)}_2$} (FX2);
    \end{tikzpicture}
\]
\end{definition}

\begin{remark}
The experienced reader may recognize kernel-compatibility as a special case of \emph{weak pullback preservation}~\cite{rutten-2000}.
We use the former because it allows us to avoid defining pullbacks, and it is all we need to relate behavioral equivalence to bisimilarity for Set-based coalgebras.
\end{remark}

Kernel-compatibility may feel rather abstract.
On an intuitive level, we can think of it as follows: if $F$ is kernel-compatible and $F(f)$ identifies two elements $u$ and $v$ of $F(X)$, then $m_f$ helps us ``factor'' $u$ and $v$ into a \emph{single} piece of structure coming from $F$ and elements $u'$ and $v'$ of $X$ that appear inside $u$ and $v$ respectively, such that $u'$ and $v'$ are already identified by $f$.
The following examples aim to clarify this.

\begin{example}\label{example:infernal-machine-kernel-compatible}
$M$ is kernel-compatible: given $f\colon X \to Y$ and $u, v \in M(X)$ with $M(f)(u) = M(f)(v)$ (i.e., $(u, v) \in \mathsf{Ker}(M(f))$), either $u = * = v$, or $u, v \in X$ with $f(u) = f(v)$.
In the former case, set $m_f(u, v) = *$, and $m_f(u, v) = (u, v)$ in the latter; either way, $m_f(u, v) \in M(\mathsf{Ker}(f))$.
This also makes the required diagram commute; cf.\ \Cref{\res{exercise:infernal-machine-kernel-compatible}}.
\end{example}

\begin{example}\label{example:coffee-machine-kernel-compatible}
$H$ is also kernel-compatible: given $f\colon X \to Y$ and $u, v \in H(X)$ with $H(f)(u) = H(f)(v)$, we can choose $m_f(u, v)$ by setting $m_f(u, v)(b) = (c_u, (x_u, x_v))$, where $u(b) = (c_u, x_u)$ and $v(b) = (c_v, x_v)$.
Showing that $m_f(u, v) \in H(\mathsf{Ker}(f))$ and the commutativity of the diagram is left as \Cref{\res{exercise:coffee-machine-kernel-compatible}}.
\end{example}

We can now prove that for kernel-compatible functors it is indeed the case that behavioral equivalence implies bisimilarity.

\begin{lemma}[{\cite[Proposition~5.9]{rutten-2000}}]%
\label{lemma:kernel-bisimulation}
Let $(X, t)$ and $(Y, u)$ be $F$-coalgebras, and let $h$ be an $F$-coalgebra homomorphism from $(X, t)$ to $(Y, u)$.
If $F$ is kernel-compatible, then $\mathsf{Ker}(h)$ is a bisimulation on $(X, t)$.
\end{lemma}

\begin{proof}
We choose $p\colon \mathsf{Ker}(h) \to \mathsf{Ker}(F(h))$ by $p(x, x') = (t(x),t(x'))$.
To see that this is well-defined, note that if $(x, x') \in \mathsf{Ker}(h)$, then $(t(x), t(x')) \in \mathsf{Ker}(F(h))$ because $h(x) = h(x')$ and we have that
\[
    F(h)(t(x))
        = u(h(x))
        = u(h(x'))
        = F(h)(t(x'))
\]
It now suffices to prove that $(\mathsf{Ker}(h), m_h \circ p)$ is a bisimulation on $(X, t)$; to see this, consider the following diagram, in which $m_h$ is the morphism witnessing that $F$ is kernel-compatible.
\[
    \begin{tikzpicture}
        \node (kerback) {$\mathsf{Ker}(h)$};
        \node[below=of kerback] (kerbackF) {$\mathsf{Ker}(F(h))$};
        \node[below=of kerbackF] (Fkerback) {$F(\mathsf{Ker}(h))$};
        \node[left=2cm of kerback] (X1) {$X$};
        \node[right=2cm of kerback] (X2) {$X$};
        \node at (X1 |- Fkerback) (FX1) {$F(X)$};
        \node at (Fkerback -| X2) (FX2) {$F(X)$};
        \draw[->] (kerback) edge node[above] {$\pi^h_1$} (X1);
        \draw[->] (kerback) edge node[above] {$\pi^h_2$} (X2);
        \draw[->] (X1) edge node[left] {$t$} (FX1);
        \draw[->] (X2) edge node[right] {$t$} (FX2);
        \draw[->] (kerback) edge node[left] {$p$} (kerbackF);
        \draw[->] (kerbackF) edge node[left] {$m_h$} (Fkerback);
        \draw[->] (Fkerback) edge node[below] {$F(\pi^h_1)$} (FX1);
        \draw[->] (Fkerback) edge node[below] {$F(\pi^h_2)$} (FX2);
        \draw[->] (kerbackF) edge node[above,sloped] {$\pi^{F(h)}_1$} (FX1);
        \draw[->] (kerbackF) edge node[above,sloped] {$\pi^{F(h)}_2$} (FX2);
    \end{tikzpicture}
\]
Here, the lower triangles commute by definition of $m_h$, and the upper quadrangles commute by construction of $p$.
\end{proof}

\begin{corollary}\label{corollary:behavioral-equiv-implies-bisim}
Let $(X, t)$ be an $F$-coalgebra for some functor $F$.
Furthermore, let $(\Omega, \omega)$ be a final $F$-coalgebra.
If $x, x' \in X$ are such that ${!_X(x)} = {!_X(x')}$, and $F$ is kernel-compatible, then $x \bisim x'$.
\end{corollary}
\begin{proof}
Choose $\mathsf{Ker}(!_X)$ as the bisimulation witnessing that $x_1 \bisim x_2$.
\end{proof}

\section{Subcoalgebras}

When treating a transition system, it often happens that there are subsets of states that only transition to one another, and not to states outside of the set.
This notion can also be formalized within coalgebra.

\begin{definition}
Let $(X, t)$ and $(Y, u)$ be $F$-coalgebras for some functor $F$.
We say $(X, t)$ is a \emph{subcoalgebra} of $(Y, u)$ if there exists a coalgebra homomorphism $f\colon (X, t) \to (Y, u)$ that is injective (as a function).
\end{definition}

One instance of the above could have $X$ be a subset of $Y$, with the injection of $X$ into $Y$ being a homomorphism.
This corresponds well to the intuition of states in $X$ only transitioning to states in $X$.

We finish this chapter with one more observation that will be useful later.
Recall that a homomorphism $f\colon (X, t) \to (Y, u)$ shows how $(Y, u)$ can ``replicate'' the behavior of $(X, t)$.
We can recover the relevant subcoalgebra of $(Y, u)$ by means of this homomorphism, as follows.

\begin{lemma}[{cf.~\cite[Theorem 7.1]{rutten-2000}}]%
\label{lemma:kernel-subcoalgebra}
Let $F$ be kernel-compatible, let $(X, t)$ and $(Y, u)$ be $F$-coalgebras, and let $f\colon (X, t) \to (Y, u)$ be a homomorphism.
Now $(X/\mathsf{Ker}(f), t/\mathsf{Ker}(f))$ (cf.\ \Cref{\res{lemma:quotient-coalgebra},\res{lemma:kernel-bisimulation}}) is a subcoalgebra of $(Y, u)$, and $f = i \circ [-]_{\mathsf{Ker}(f)}$, where $i$ is the map from $X/\mathsf{Ker}(f)$ to $Y$ given by $i([x]_{\mathsf{Ker}(f)}) = f(x)$.
\end{lemma}
\begin{proof}
First, let us check that $i$ is a well-defined function: if $[x]_{\mathsf{Ker}(f)} = [x']_{\mathsf{Ker}(f)}$, then $f(x) = f(x')$, and hence $i([x]_{\mathsf{Ker}(f)}) = i([x']_{\mathsf{Ker}(f)})$.

Note that we already know that the square below on the left commutes by \Cref{\res{lemma:quotient-coalgebra}}; it remains to show this for the other square.
\[
    \begin{tikzpicture}
        \node (X) {$X$};
        \node[right=25mm of X] (Kerf) {$X/\mathsf{Ker}(f)$};
        \node[right=15mm of Kerf] (Y) {$Y$};
        \node[below=of X] (FX) {$F(X)$};
        \node (FKerf) at (FX -| Kerf) {$F(X/\mathsf{Ker}(f))$};
        \node (FY) at (FKerf -| Y) {$FY$};
        \draw[->] (X) to node[above] {$[-]_{\mathsf{Ker}(f)}$} (Kerf);
        \draw[->] (Kerf) to node[above] {$i$} (Y);
        \draw[->] (X) to node[left] {$t$} (FX);
        \draw[->] (Kerf) to node[left] {$t/\mathsf{Ker}(f)$} (FKerf);
        \draw[->] (Y) to node[left] {$u$} (FY);
        \draw[->] (FX) to node[below] {$F([-]_{\mathsf{Ker}(f)})$} (FKerf);
        \draw[->] (FKerf) to node[below] {$F(i)$} (FY);
    \end{tikzpicture}
\]
To this end, we should prove $u(i([x]_{\mathsf{Ker}(f)})) = F(i)(t/\mathsf{Ker}(f)([x]_{\mathsf{Ker}(f)}))$ for any $x \in X$, which we do by deriving as follows:
\begin{align*}
u(i([x]_{\mathsf{Ker}(f)}))
    &= u(f(x)) \tag{def.\ $i$} \\
    &= F(f)(t(x)) \tag{$f$ is a hom.} \\ 
    &= F(i \circ [-]_{\mathsf{Ker}(f)})(t(x)) \tag{def.\ $i$} \\
    &= F(i)(F([-]_{\mathsf{Ker}(f)})(t(x))) \tag{$F$ is a functor} \\
    &= F(i)(t/\mathsf{Ker}(f)([x]_{\mathsf{Ker}(f)})) & \tag{$[-]_{\mathsf{Ker}(f)}$ is a hom.} 
\end{align*}
Finally, $i$ is injective: if $i([x]_{\mathsf{Ker}(f)}) = i([x]_{\mathsf{Ker}(f)})$, then $f(x) = f(x')$, so $[x]_{\mathsf{Ker}(f)} = [x']_{\mathsf{Ker}(f)}(x')$; this concludes the proof.
\end{proof}

\section{Exercises}%
\label{section:exercisescoalgebra}

\begin{exercises}

\item\label{exercise:infernal-machine-functor}
Confirm that $M$, as defined in \Cref{\res{example:infernal-machine-functor}}, is a functor.

\item\label{exercise:coffee-machine-functor}
Confirm that $H$, as defined in \Cref{\res{example:coffee-machine-functor}}, is a functor.

\item\label{ex:D:coalgebra}
For a set $X$, let $D(X) = 2 \times X^\alphabet$ and, for $f\colon X\to Y$, define $D(f)\colon D(X) \to D(Y)$ as follows.
Given $(o,m)\colon 2 \times X^\alphabet$, we define $D(f)(o,m)\colon2 \times Y^\alphabet$ by setting $D(f)(o,m) = (o, f \circ m)$.
Prove that this defines a functor.

\item\label{ex:D:homomorphism}
Reconsider the functor $D$ from the previous exercise.
Can you describe the conditions satisfied by a $D$-coalgebra homomorphism, along the same lines as \Cref{\res{example:infernal-machine-morphism},\res{example:coffee-machine-morphism}}?

\item\label{exercise:homomorphism-preserved}
Let $F$ be a functor, and let $(X, t_X)$, $(Y, t_Y)$ and $(Z, t_Z)$ be $F$-coalgebras, with $f\colon X \to Y$ an $F$-coalgebra homomorphism from $(X, t_X)$ to $(Y, t_Y)$, and $g\colon Y \to Z$ an $F$-coalgebra homomorphism from $(Y, t_Y)$ to $(Z, t_Z)$.
Show that $g \circ f\colon X \to Z$ is an $F$-coalgebra homomorphism from $(X, t_X)$ to $(Z, t_Z)$.

\item\label{exercise:infernal-machine-final}
Let $(X, t)$ be an $M$-coalgebra, and let $!_X\colon X \to \mathbb{N}^\infty$ be defined as in \Cref{\res{example:infernal-machine-final}}.
Show that $!_X$ is an $M$-coalgebra homomorphism from $(X, t)$ to $(\mathbb{N}^\infty, \bullet)$.

\item\label{exercise:coffee-machine-final}
Let $(X, t)$ be an $H$-coalgebra.
Furthermore, let $S$ be the set of functions from $B^+$ to $C$, with $\sigma\colon S \to H(S)$ and $!_X\colon X \to S$ defined as in \Cref{\res{example:coffee-machine-final}}.
Show that $!_X$ is the \emph{only} $H$-coalgebra homomorphism from $(X, t)$ to $(S, \sigma)$.

\item\label{exercise:final-coalgebra-unique}
Prove \Cref{\res{lemma:final-coalgebra-unique}}.
You may use the property proved in \Cref{\res{exercise:homomorphism-preserved}}, and the fact that for any $F$-coalgebra $(X, t)$, the identity $\id_X$ is an $F$-coalgebra homomorphism from $(X, t)$ to itself.

\item\label{exercise:no-morphism} Recall the two coalgebras from \Cref{\res{example:no-morphism}}.
Can you show formally that there is no homomorphism from one to the other?

\item\label{exercise:infernal-machine-bisimulations}
Verify the characterization of $M$-bisimulations from \Cref{\res{example:infernal-machine-bisimulations}}.

\item\label{exercise:coffee-machine-bisimulations}
Verify the characterization of $H$-bisimulations from \Cref{\res{example:coffee-machine-bisimulations}}.

\item\label{exercise:D-bisimulation}
Consider the functor $D$ from \Cref{\res{ex:D:coalgebra}}.
Can you characterize $D$-bisimulations, along the same lines as \Cref{\res{example:infernal-machine-bisimulations},\res{example:coffee-machine-bisimulations}}?

\item\label{exercise:infernal-machine-kernel-compatible}
Show that the choice of $m_f$ in \Cref{\res{example:infernal-machine-kernel-compatible}} makes the diagram from \Cref{\res{definition:kernel-compatible}} commute for the functor $M$.

\item\label{exercise:coffee-machine-kernel-compatible}
Finish the argument towards kernel-compatibility of $H$, outlined in \Cref{\res{example:coffee-machine-kernel-compatible}}.
In particular, show that the choice of $m_f$ is valid --- i.e., if $(u, v) \in \mathsf{Ker}(H(f))$ then $m_f(u, v) \in H(\mathsf{Ker}(f))$ --- and that the diagram from \Cref{\res{definition:kernel-compatible}} commutes.

\item\label{ex:Dkc:coalgebra}
Show that the functor $D$ from \Cref{\res{ex:D:coalgebra}} is kernel-compatible.
\end{exercises}

\onlyinsubfile{%
    \printbibliography%
}

\chapter{Automata and Kleene's Theorem}%
\label{chapter:automata}

We will now provide an \emph{operational} view of regular languages through automata and, more abstractly, coalgebra.
The tools discussed here will provide the groundwork for the completeness proof in \Cref{\res{chapter:completeness}}.
This chapter is divided in three parts: first, we present the operational semantics of regular expressions; second, we dive into the coalgebraic view on automata and languages; third, we prove Kleene's theorem, which states that the denotational and operational semantics of regular expressions coincide --- i.e., they provide equivalent semantics.

\section{Derivatives}%
\label{section:derivatives}

We start by introducing the notion of \emph{derivative}, which is central to the study of the operational semantics of regular expressions.
We will define two functions: one that assigns to each expression an \emph{output value} in $\{0,1\}$ signaling whether computation can terminate successfully; and another that given a letter $\ltr{a} \in \alphabet$ returns its \emph{$\ltr{a}$-derivative}, which is an expression that denotes all the words obtained from the first expression after \emph{deleting} an $\ltr{a}$.
These functions were first proposed by Brzozowski~\cite{brzozowski-1964}, which is why they are known as \emph{Brzozowski derivatives}.

\begin{definition}[Brzozowski derivatives~\cite{brzozowski-1964}]
The \emph{output function} $o\colon \Exp \to 2$ is defined inductively, as follows:
\begin{align*}
\ore (0) &= 0
    & \ore (1) &= 1
    & \ore (\ltr{a}) &= 0 \\
\ore (e+f) &= \ore(e) \vee \ore(f)
    & \ore(ef) &= \ore(e) \land \ore(f)
    & \ore(e^*) &= 1
\end{align*}
Here, $b_1 \vee b_2$ is understood to be $0$ when $b_1 = b_2 = 0$, and $1$ otherwise; similarly, $b_1 \wedge b_2$ is defined as $1$ when $b_1 = b_2 = 1$ and $0$ otherwise.

We furthermore define the \emph{transition function} $\tre\colon \Exp \to \Exp^\alphabet$ inductively as follows, where we use $e_\ltr{a}$ as a shorthand for $t(e)(\ltr{a})$:
\begin{mathpar}
0_\ltr{a} =  1_\ltr{a} = 0
    \and (e+f)_\ltr{a} = e_\ltr{a} + f_\ltr{a}
    \and (e^*)_\ltr{a} = e_\ltr{a}e^* \\
\ltr{b}_\ltr{a} = \begin{cases} 1 & \text{if } \ltr{a}=\ltr{b}\\  0 &  \text{if }\ltr{a}\neq \ltr{b} \end{cases}
    \and (ef)_\ltr{a} = \begin{cases} e_\ltr{a}f & \text{if }\ore(e) = 0 \\ e_\ltr{a}f + f_\ltr{a} & \text{otherwise} \end{cases}
\end{mathpar}
\end{definition}

For a regular expression $e$, we have $\ore(e)=1$ iff $\sem{e}$ contains the empty word $\epsilon$, while $e_\ltr{a}$ denotes the language containing all words $w$ such that $\ltr{a}w$ is in the language denoted by $e$.
The derivative for sequential composition can alternatively be presented more compactly as $(ef)_\ltr{a} = e_\ltr{a}f  + \ore(e)f_\ltr{a}$, by observing that $\ore(e)\in 2 \hookrightarrow \Exp$.

\begin{example}
Consider the regular expression $e = (\ltr{a}\ltr{b} + \ltr{b})^*$.
By definition, $o(e) = 1$, meaning that $\sem{e}$ should contain the empty word --- which is indeed the case.
Moreover, we can calculate $e_\ltr{a}$ as follows:
\begin{align*}
    e_\ltr{a}
        &= (\ltr{a}\ltr{b} + \ltr{b})_\ltr{a} (\ltr{a}\ltr{b} + \ltr{b})^*
         = ((\ltr{a}\ltr{b})_\ltr{a} + \ltr{b}_\ltr{a}) (\ltr{a}\ltr{b} + \ltr{b})^* \\
        &= (\ltr{a}_\ltr{a}\ltr{b} + \ltr{b}_\ltr{a}) (\ltr{a}\ltr{b} + \ltr{b})^*
         = (1 \cdot \ltr{b} + 0) (\ltr{a}\ltr{b} + \ltr{b})^* \equiv \ltr{b}(\ltr{a}\ltr{b} + \ltr{b})^*
\end{align*}
In other words, $\ltr{a}w \in \sem{e}$ should hold if and only if $w = \ltr{b}w'$ with $w' \in \sem{e}$ --- and a cursory inspection of $e$ tells us that this is true.
\end{example}

Thus, derivatives tell us how to ``deconstruct'' a regular expression in terms of its \emph{immediate} behavior (whether or not the empty word is accepted), and its \emph{later} behavior (the words denoted that begin with a given letter).
The next result, which is useful on a number of occasions, validates this intuition by telling us that this correspondence goes both ways: an expression can be ``reconstructed'' from its derivatives.

\begin{theorem}[{Fundamental theorem of Kleene Algebra~\cite[Theorem~6.4]{brzozowski-1964}}]\label{thm:fundamental-re}
Every regular expression $e \in \Exp $ satisfies
\[
 e \equiv {\ore(e)} + \sum_{\ltr{a}\in \alphabet} \ltr{a}e_\ltr{a}
\]
\end{theorem}

\begin{remark}[Generalized sum]
In the statement above, and in the sequel, we use a ``generalized sum'' operator $\sum$, interpreted intuitively the same as for numbers.
This is a slight abuse of notation, because we use it to define a \emph{term}, and it is not clear whether $\sum \{ e, f, g \}$ denotes $f + (e + g)$ or $e + (g + f)$, or any other term described as ``the sum of $\{ e, f, g \}$''.
However, associativity and commutativity of $+$ up to $\equiv$ means that order and bracketing of sums does not matter in most contexts.

In a similar vein, we will often implicitly rely on familiar properties of summation that are valid in Kleene Algebra as well (in addition to the idempotence law $ e + e \equiv e$) for the sake of brevity.
Concretely, we will freely use the following properties of sums in the proof that follows, as well as later on, for all finite $X, Y \subseteq \Exp$ and $e \in \Exp$:
\begin{mathpar}
    e + \sum_{f \in X} f \equiv \sum_{f \in X \cup \{e\}} f
    \and
    \sum_{f \in X \cup Y} f \equiv \sum_{f \in X} f + \sum_{f \in Y} f
    \and
    e \Big( \sum_{f \in X} f \Big) \equiv \sum_{f \in X} e \cdot f
    \and
    \Big( \sum_{f \in X} f \Big) e \equiv \sum_{f \in X} f \cdot e
\end{mathpar}
\end{remark}
\begin{proof}[Proof of Theorem~\ref{thm:fundamental-re}]
We proceed by induction on the structure of $e$.
The base cases are direct consequences of the laws for $0$ and $1$.
\begin{itemize}
    \item
    When $e = 0$, we derive as follows.
    \[
         0
            \equiv 0 + \sum_{\ltr{a}\in \alphabet}\ltr{a} 0
            \equiv o(0) + \sum_{\ltr{a} \in \alphabet} \ltr{a} 0_\ltr{a}
    \]
    \item
    Next, when $e = 1$, we can calculate
    \[
         1
            \equiv   1 + \sum_{\ltr{a}\in \alphabet} \ltr{a} 0
            \equiv  o(1) + \sum_{\ltr{a} \in \alphabet} \ltr{a} 1_\ltr{a}
    \]
    \item
    Finally, when $e = \ltr{b}$ for some $\ltr{b} \in \alphabet$, we find that
    \[
         \ltr{b}
            \equiv  0 + \ltr{b} 1 + \sum_{\ltr{a} \in \alphabet\setminus\{\ltr{b}\}}\ltr{a} 0
            \equiv  o(\ltr{b}) + \sum_{\ltr{a} \in \alphabet}\ltr{a} \ltr{b}_\ltr{a}
    \]
\end{itemize}

\noindent
This leaves us with three inductive cases.
For each, our induction hypothesis is that the statement is true for the proper subterms.

\begin{itemize}
    \item
    When $e = e_1 + e_2$, we find that:
    \begin{align*}
     e_1+e_2
        &\equiv \Big({\ore(e_1)} + \sum_{\ltr{a}\in \alphabet}\ltr{a}(e_1)_\ltr{a}\Big)+\Big({\ore(e_2)} + \sum_{\ltr{a}\in \alphabet}\ltr{a}(e_2)_\ltr{a} \Big) \tag{IH} \\[1.9ex]
        &\equiv  {\ore(e_1)}+{\ore(e_2)} + \sum_{\ltr{a}\in \alphabet}\ltr{a}((e_1)_\ltr{a} + (e_2)_\ltr{a}) \\
        &\equiv {\ore(e_1+e_2)} + \sum_{\ltr{a}\in \alphabet}\ltr{a}(e_1+e_2)_\ltr{a}
    \end{align*}
    \item
    When $e = e_1e_2$, we derive:
    \begin{align*}
    e_1e_2
        &\equiv  \Big({\ore(e_1)} + \sum_{\ltr{a}\in \alphabet}\ltr{a}(e_1)_\ltr{a}\Big)e_2 \tag{IH} \\
        &\equiv {\ore(e_1)}e_2 + \sum_{\ltr{a}\in \alphabet}\ltr{a}(e_1)_\ltr{a}e_2  \\
        &\equiv {\ore(e_1)}\Big( {\ore(e_2)} + \sum_{\ltr{a}\in \alphabet} \ltr{a}(e_2)_\ltr{a}\Big) + \sum_{\ltr{a}\in \alphabet}\ltr{a}(e_1)_\ltr{a}e_2 \tag{IH} \\[1.9ex]
        &\equiv {\ore(e_1e_2)}+ \sum_{\ltr{a}\in \alphabet}\ltr{a}((e_1)_\ltr{a}e_2 + {\ore(e_1)}(e_2)_\ltr{a})  \\[1.9ex]
        &\equiv {\ore(e_1e_2)}+ \sum_{\ltr{a}\in \alphabet}\ltr{a}(e_1e_2)_\ltr{a}
    \end{align*}
    Note how in this case the induction hypotheses for $e_1$ and $e_2$ are applied subsequently, and not at the same time.
    \item
    When $e = e_1^*$, the proof is a little bit more tricky.
    In the following derivation, we write $x$ as a shorthand for $\sum_{\ltr{a}\in \alphabet} \ltr{a}(e_1)_\ltr{a}$.
    We have:
    \begin{align*}
     e_1^*
        \stackrel{\text{IH}}\equiv ({\ore(e_1)} + x)^* \stackrel\dagger\equiv x^*
        \equiv 1 + xx^*  \stackrel\dagger \equiv 1 + x({\ore(e_1)} + x)^*  \stackrel{\text{IH}}\equiv 1 + x e_1^*
    \end{align*}
    where in the two steps marked with ($\dagger$), we use that for all $f \in \Exp$, $(f + 1)^* \equiv f^*$ (cf.\ \Cref{\res{exercise:tightening}}).
    We can then finish the proof:
    \begin{align*}
        1 + x e_1^* &\equiv 1 + \sum_{\ltr{a}\in \alphabet} \ltr{a}(e_1)_\ltr{a} e_1^* \tag{distributivity} \\
        &\equiv o(e_1^*) + \sum_{\ltr{a}\in \alphabet} \ltr{a}(e_1^*)_\ltr{a}
        \tag*{\qedhere}
    \end{align*}
\end{itemize}
\end{proof}

\begin{remark}
The name \emph{Fundamental Theorem} is borrowed from~\cite{rutten-2003}, where a similar property is proved for formal power series (functions $f\colon \alphabet^* \to K$, for a semiring $K$).
As explained there, the name is chosen in analogy to calculus: if we view prefixing with $\ltr{a}$ as a kind of integration, then the theorem tells us that derivation is the opposite of integration, by obtaining $e$ from the $\ltr{a}$-derivatives $e_\ltr{a}$ and the initial value $\ore(e)$.
\end{remark}

We end this section with the semantic version of the Fundamental Theorem, which follows by soundness of the axioms.
\begin{corollary}
For all $e \in \Exp$, we have that $\sem{e} = \sem{{\ore(e)} + \sum_{\ltr{a}\in \alphabet} \ltr{a}e_\ltr{a}}$.
\end{corollary}

\section{Automata (coalgebraically)}%
\label{sec:lanautco}

The semantics of regular expressions $\sem{-}$ is \emph{denotational}, in the sense that the interpretation of a regular expression arises from the that of its subexpressions.
However, there is another way to assign semantics to regular expressions, which arises from the \emph{operational} perspective.
Here, we think of an expression as an \emph{abstract machine}, which can evolve by performing certain computational steps; informally, the semantics of such an object is then taken as the set of all possible action sequences.

\begin{remarkblue}
The definitions in this section will be presented in an elementary fashion.
However, readers who have gone through \Cref{\res{chapter:coalgebra}} will be able to relate the definitions and claims here to the corresponding abstractions there, when instantiated with the appropriate functor.
The remainder of this section is presented without this abstraction; we have inserted remarks with a blue background (like this one) to make the connection explicit.
Readers who skipped \Cref{\res{chapter:coalgebra}} can disregard these.
\end{remarkblue}

More specifically, the machines studied are the well-known \emph{(deterministic) automata}; they form the operational model of regular expressions.

\begin{definition}
A \emph{(deterministic) automaton} is a pair $(X, d_X)$ where $X$ is a set, and $d_X\colon X \to 2 \times X^\alphabet$ is a function.
We will often write $d_X$ as $<o_X, t_X>$, where $o_X\colon X \to 2$ and $t_X\colon X \to X^\alphabet$ are functions such that $d_X(x) = <o_X(x), t_X(x)>$.
We say $(X, d_X)$ is finite when $X$ is.
\end{definition}

\begin{remarkblue}
Clearly, an automaton as defined above is precisely a coalgebra for the functor $D$ given by $D(X) = 2 \times X^\alphabet$ (cf.\ \Cref{\res{ex:D:coalgebra}}).
\end{remarkblue}

We often denote $t_X(x)(\ltr{a})$ with $(x)_\ltr{a}$ when $t_X$ is clear from context.
Each $x \in X$ is a \emph{state}, and when $o_X(x) = 1$ we call $x$ \emph{accepting}.
We can draw an automaton as a graph where states are nodes (with accepting states circled twice) and transitions are labelled edges.
For instance, we can depict the automaton $(X, <o_X, t_X>)$ where $X = \{ x_1, x_2 \}$, and $o_X$ and $t_X$ are as given below on the left with the graph below on the right.
\vspace{-1.5em} 
\begin{mathpar}\label{example:automaton}
    \begin{array}{rclrcl}
    o_X(x_1) &=& 0 & o_X(x_2) &=& 1 \\
    (x_1)_\ltr{a} &=& x_2 & (x_1)_\ltr{b} &=& x_1 \\
    (x_2)_\ltr{a} &=& x_1 & (x_2)_\ltr{b} &=& x_2
    \end{array}
    \and
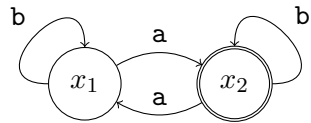
    \begin{tikzpicture}[baseline=2mm]
    \node[state] (x1) {$x_1$};
    \node[state,accepting,right=of x1] (x2) {$x_2$};
    \draw (x1) edge[->,out=180,in=90,looseness=4] node[above left] {$\ltr{b}$} (x1);
    \draw (x2) edge[->,out=0,in=90,looseness=4] node[above right] {$\ltr{b}$} (x2);
    \draw (x1) edge[bend left,->] node[above] {$\ltr{a}$} (x2);
    \draw (x2) edge[bend left,->] node[above] {$\ltr{a}$} (x1);
    \end{tikzpicture}
\end{mathpar}

\begin{remark}
The reader may be more familiar with automata presented as tuples $(Q, \delta, F)$, where $Q$ is a set of \emph{states}, $F \subseteq Q$ holds the \emph{accepting states}, and $\delta\colon Q \times \alphabet \to Q$ is called the \emph{transition function}.\footnote{%
    Many texts also choose to include an \emph{initial state} in the definition.
    The automata defined here are agnostic in this respect, because many proofs become easier when formulated as a property of all states.
    Additionally, the connection to coalgebra is clearer if we forget about the initial state.
}
We can convert between these presentations easily, by sending $(X, <o_X, t_X>)$ to $(X, \delta, F)$, where $\delta(x, \ltr{a}) = (x)_\ltr{a}$, and $F = \{ x \in X \mid o_X(x) = 1 \}$; conversely, $(Q, \delta, F)$ can be turned into $(X, <o_X,t_X>)$ by choosing $o_X(x) = 1$ when $x \in F$ and $o_X(x) = 0$ otherwise, while $(x)_{\ltr{a}} = \delta(x, \ltr{a})$.
\end{remark}

The Brzozowski derivatives $o\colon \Exp \to 2$ and $t\colon \Exp \to \Exp^\alphabet$ equip $\Exp$ with an automaton structure, making $(\Exp, <o, t>)$ the (infinite) \emph{Brzozowski automaton}; we return to this in \Cref{\res{section:brzozowski-sound}}.

Another important example of an automaton is carried by the set of languages $2^{\alphabet^*}$.
For $\ltr{a} \in \alphabet$, $l_\ltr{a}$ is defined by $l_\ltr{a}(w) = l(\ltr{a}w)$.
Whether or not $l$ is accepting is given by $l(\epsilon)$.
Together, these define an automaton $(2^{\alphabet^*}, <o_L,t_L>)$.
These maps are sometimes called \emph{semantic Brzozowski derivatives}, or \emph{simply language derivatives}.

\subsection{Morphisms and finality}
A central concept in the discussion to come is the idea of a \emph{morphism}.
Intuitively, this is a map that respects input and output derivatives.

\begin{definition}
Given two deterministic automata $(S, <o_S, t_S>)$ and $(T, <o_T,t_T>)$, a function $h\colon S\to T$ is a \emph{homomorphism} if it is compatible with the output and transition functions in the sense that $o_T(h(s)) = o_S(s)$, and $t_T(h(s))(\ltr{a}) = h(t_S(s)(\ltr{a}))$, for all $\ltr{a} \in \alphabet$.
\end{definition}

\begin{remarkblue}
Morphisms between automata as defined above are precisely morphisms between $D$-coalgebras --- that is, the equations above correspond to the commutativity of the diagram below.
\[
    \begin{tikzpicture}
        \node (S) {$S$};
        \node[right=3cm of S] (T) {$T$};
        \node[below=7mm of S] (DS) {$2 \times S^A$};
        \node at (DS -| T) (DT) {$2 \times T^A$};

        \draw[->] (S) edge node[left] {\footnotesize $<o_S, t_S>$} (DS);
        \draw[->] (T) edge node[right] {\footnotesize $<o_T, t_T>$} (DT);
        \draw[->] (S) edge node[above] {\footnotesize $h$} (T);
        \draw[->] (DS) edge node[below] {\footnotesize $D(h)$} (DT);
    \end{tikzpicture}
\]
\end{remarkblue}

Given an automaton $(S, <o_S, t_S>)$, the derivative $s_\ltr{a}$ of a state $s$ for input $\ltr{a}\in \alphabet$ can be inductively extended to the word derivative $s_w$ of $s$ for input $w\in \alphabet^*$, by defining $s_\epsilon = s$ and $s_{\ltr{a}w'} = (s_\ltr{a})_{w'}$.
This enables an easy definition of the semantics of a state $s$ of a deterministic automaton: the language $L(s) \in 2^{\alphabet^*}$ recognized by $s$ is given by:
\begin{equation}
 L(s)(w) = o_S(s_w) \label{eq:lang_dfa}
\end{equation}

\begin{example}
The language recognized  by state $s_1$ of the automaton on page~\pageref{example:automaton} is the set of all words with an odd number of $\ltr{a}$'s.
It is easy to check that, for example, $L(s_1)(\ltr{bab}) = o_S((s_1)_{\ltr{bab}}) = o_S(s_2) =1$.
\end{example}

$L$ is defined from the state space of any automaton to $2^{\alphabet^*}$.
This function takes a very special place as a particular morphism that relates every automaton to the automaton formed by language derivatives.

\begin{theorem}[Finality]\label{thm:languages-final}
For any automaton $(S,<o_S,t_S>)$, $L \colon S \to 2^{\alphabet^*}$ is the \emph{only} homomorphism from $(S, <o_S, t_S>)$ to $(2^{\alphabet^*}, <o_L, t_L>)$.
\end{theorem}

\begin{remarkblue}
Thus $(2^{\alphabet^*}, <o_L, t_L>)$ forms the final $D$-coalgebra.
\end{remarkblue}

\begin{proof}
We have to prove that the diagram commutes and that $L$ is unique.
First the commutativity with the derivatives:
\begin{align*}
o_L(L(s)) &= L(s)(\epsilon) = o_S(s_\epsilon) = o_S(s)\\
t_L(L(s))(\ltr{a})(w) &= L(s)_\ltr{a}(w) = L(s)(\ltr{a}w) \\
                      &= L (s_\ltr{a})(w) = L (t_S(s))(\ltr{a})(w)
\end{align*}
For the second-to-last step, note that, by definition of $L$ (see~\eqref{eq:lang_dfa}):
\[ L(s)(\ltr{a}w) = o_S(s_{\ltr{a}w}) = o_S((s_\ltr{a})_w) = L(s_\ltr{a})(w) \]
For the uniqueness, suppose there is a morphism $h\colon S \to 2^{\alphabet^*}$ such that, for every $s\in S$ and $\ltr{a}\in \alphabet$, $o_L (h(s)) = o_S(s)$ and $h(s)_\ltr{a} = h(s_\ltr{a})$.

We prove by induction on $w\in \alphabet^*$ that $h(s)(w) = L(s)(w)$:
\begin{align*}
h(s)(\epsilon)
    &= o_L(h(s)) = o_S(s) = L(s)(\epsilon)\\
h(s)(\ltr{a}w)
    &= (h(s)_\ltr{a})(w) = h(s_\ltr{a})(w) \stackrel{(\mathrm{IH})}= L (s_\ltr{a})(w) = L(s)(\ltr{a}w)
    \tag*{\qedhere}
\end{align*}
\end{proof}

On its own, \Cref{\res{thm:languages-final}} may seem abstract.
Its most immediate consequence is that we can relate homomorphisms to the language semantics of automata, by exploiting uniqueness.
\begin{corollary}
Given automata $(S, <o_S, t_S>)$ and $(T, <o_T, t_T>)$ with a morphism $h\colon S \to T$, it holds for $s \in S$ that $L(s) = L(h(s))$.
\end{corollary}
\begin{proof}
Follows from \Cref{\res{thm:languages-final}} and the fact that $L\colon S \to 2^{\alphabet^*}$ and $L \circ h\colon S \to 2^{\alphabet^*}$ are morphisms from $(S, <o_S, t_S>)$ to $(2^{\alphabet^*}, <o_L, t_L>)$.
\end{proof}

\subsection{Bisimulations and coinduction}

A moment ago, we saw how states related by homomorphisms must have the same language.
Unfortunately, the converse is not necessarily true: there exist automata with states that have the same language, but which are not related by any homomorphism (see \Cref{\res{exercise:no-morphism-aut}}).
To get more mileage, we need a slightly more general notion, as follows.

\begin{definition}[Bisimulation~\cite{hopcroft-karp-1971}]\label{def:automatabis}
Given $(S, <o_S, t_S>)$ and $(T, <o_T,t_T>)$, we say that a relation $R\subseteq S\times T$ is a \emph{bisimulation} if $<s,t> \in R$ implies
\[
o_S(s) = o_T(t) \ \ \ \text{ and } \ \ \ (s_\ltr{a},t_\ltr{a}) \in R\text{ for all } \ltr{a} \in \alphabet
\]
We write $s\bisim t$ whenever there exists a bisimulation $R$ containing $(s,t)$.
\end{definition}

\begin{remarkblue}\label{remark:bisimulation-coalgebra}
Unsurprisingly, bisimulations between automata as defined above are in one-to-one correspondence with bisimulations between $D$-coalgebras --- the conditions above guarantee the existence of a $D$-coalgebra structure on $R$, and vice versa; see \Cref{\res{exercise:D-bisimulation}}.
\end{remarkblue}

The next theorem guarantees that bisimilarity is a sound and complete proof principle for language equivalence of automata.
\begin{theorem}[Coinduction principle~\cite{hopcroft-karp-1971}]\label{thm:coind_dfa}
Given two deterministic automata $(S, <o_S, t_S>)$ and $(T, <o_T,t_T>)$ with $s\in S$ and $t\in T$:
\[ s\bisim t \iff L(s) = L(t) \]
\end{theorem}
\begin{remarkblue}
Coalgebraically, the claim above follows immediately from \Cref{\res{thm:languages-final}}, which tells us that $L$ is the morphism into the final $D$-coalgebra, and the fact that $D$ is kernel-compatible (\Cref{\res{ex:Dkc:coalgebra}}), which we can then use to invoke \Cref{\res{lemma:bisim-implies-behavioral-equiv}} and \Cref{\res{corollary:behavioral-equiv-implies-bisim}}.
\end{remarkblue}
\begin{proof}
For the forward implication, let $R$ be a bisimulation.
We prove, by induction on $w\in \alphabet^*$, that, for all $(s,t) \in R$, $w\in L(s)\Leftrightarrow w\in L(t)$.
\begin{align*}
L(s)(\epsilon)
    &= o_S(s) \stackrel{s\bisim t}= o_T(t) = L(t)(\epsilon)\\
L(s)(\ltr{a}w')
    &=  o_S(s_{\ltr{a}w'}) = o_S((s_\ltr{a})_{w'}) \stackrel{\text{(IH)}}{=} o_T((t_\ltr{a})_{w'}) =
L(t)(\ltr{a}w')
\end{align*}
The penultimate step follows by induction and the fact that $(s_\ltr{a}, t_\ltr{a}) \in R$.

The converse follows by the observation that the relation $\{ (s, t) \mid L(s) = L(t) \}$ is a bisimulation; we leave the details to \Cref{\res{exercise:kernel-bisimulation}}.
\end{proof}

The upshot of \Cref{\res{thm:coind_dfa}} is that we can use coinduction to determine whether two states $s$ and $t$ of two deterministic automata $(S, <o_S, t_S>)$ and $(T, <o_T,t_T>)$ recognize the same language: it is enough to construct a bisimulation containing $(s,t)$ --- indeed, if such a bisimulation does not exist, then the states \emph{cannot} have the same language.

\begin{example}[Coinduction]
Let $(S,<o_S,t_S>)$ and $(T,<o_T,t_T>)$ be the deterministic automata over the alphabet $\{\ltr{a},\ltr{b}\}$ given by
\begin{mathpar}
    \begin{tikzpicture}[every state/.style={minimum size=7mm,inner sep=0em},font=\footnotesize]
        \node[state] (s1) {$s_1$};
        \node[state,right=of s1] (s2) {$s_2$};
        \node[state,accepting,below=of s2] (s3) {$s_3$};
        \draw[->] (s1) edge node[above] {$\ltr{a}$} (s2);
        \draw[->] (s1) edge node[below left] {$\ltr{b}$} (s3);
        \draw[->] (s2) edge node[right] {$\ltr{b}$} (s3);
        \draw (s2) edge[loop right] node[right] {$\ltr{a}$} (s2);
        \draw (s3) edge[loop right] node[right] {$\ltr{a}, \ltr{b}$} (s3);
    \end{tikzpicture}
    \and
    \begin{tikzpicture}[every state/.style={minimum size=7mm,inner sep=0em},font=\footnotesize]
        \node[state] (t1) {$t_1$};
        \node[state,accepting,below=of t1] (t2) {$t_2$};
        \draw[->] (t1) edge node[right] {$\ltr{b}$} (t2);
        \draw (t1) edge[loop right] node[right] {$\ltr{a}$} (t1);
        \draw (t2) edge[loop right] node[right] {$\ltr{a}, \ltr{b}$} (t2);
    \end{tikzpicture}
\end{mathpar}
The relation $R= \{<s_1,t_1>, <s_2,t_1>, <s_3,t_2>\}$ is a
bisimulation:
\[
\begin{array}{l@{\qquad\qquad}l}
o_S(s_1) = o_S(s_2) = 0 = o_T(t_1)& o_S(s_3) = 1 = o_T(t_2)\\[1ex]
(s_1)_\ltr{a} = s_2 \mathrel{R} t_1 = (t_1)_\ltr{a} &(s_1)_\ltr{b} = s_3 \mathrel{R} t_2 = (t_1)_\ltr{b}\\[1ex]
(s_2)_\ltr{a} = s_2 \mathrel{R} t_1 = (t_1)_\ltr{a} &(s_2)_\ltr{b} = s_3 \mathrel{R} t_2 = (t_1)_\ltr{b}\\[1ex]
(s_3)_\ltr{a} = s_3 \mathrel{R} t_2 = (t_2)_\ltr{a} &
(s_3)_\ltr{b} = s_3 \mathrel{R} t_2 = (t_2)_\ltr{b}
\end{array}
\]
Thus, by \Cref{\res{thm:coind_dfa}}, $L(s_1)=L(t_1)=L(s_2)$ and $L(s_3)=L(t_3)$.
\end{example}

But how do we find a bisimulation?
Fortunately, \Cref{\res{def:automatabis}} guides us: a bisimulation relate only states that agree on their acceptance, and must be closed under $\ltr{a}$-transitions on both sides, for all $\ltr{a} \in \alphabet$.
Thus, it suffices to start with the states of interest, and add pairs reachable by simultaneous $\ltr{a}$-transitions, as long as those pairs agree on acceptance.
The algorithm in \Cref{\res{figure:bisimulation-algorithm}} implements this idea.

\begin{figure}
\begin{algorithm}[H]
    \SetKwData{R}{$R$}\SetKwData{T}{$T$}
    \KwData{An automaton $(X, <o_X, t_X>)$ and states $x_1, x_2 \in X$}
    \KwResult{A bisimulation $R$ such that $x_1 \mathrel{R} x_2$, or nothing}
    \BlankLine%
    Initialize $R$ to the empty set, and $T$ to $\{ <x_1', x_2'> \}$\;
    \While{$\T \neq \emptyset$}{
        Remove $<x_1', x_2'>$ from $T$\;
        \If{$<x_1', x_2'> \not\in R$}{
            \If{$o(x_1') \neq o(x_2')$}{%
                \tcp{No bisimulation is possible}
                \KwRet\;
            }
            Add $<x_1', x_2'>$ to $R$\;
            \ForEach{$\ltr{a} \in \alphabet$}{%
                Add $<(x_1')_\ltr{a}, (x_2')_\ltr{a}>$ to \T\;
            }
        }
    }
    \tcp{$R$ should now be a bisimulation}
    \KwRet{$R$}\;
\end{algorithm}
\caption{A simple algorithm for deciding bisimilarity.}\label{figure:bisimulation-algorithm}
\end{figure}

\begin{theorem}[{\cite{hopcroft-karp-1971}}]%
\label{theorem:algo-correct}
Suppose that the input automaton $(X, <o_X, t_X>)$ is finite.
Then the algorithm in \Cref{\res{figure:bisimulation-algorithm}} is correct, in that it constructs a bisimulation if and only if $x_1$ is bisimilar to $x_2$.
\end{theorem}
\begin{proof}[Proof sketch]
For termination, note that in each iteration either $R$ stays the same and $T$ shrinks, or $R$ grows.
Because $R \subseteq X^2$ and $X$ is finite, the latter means that $X^2 \setminus R$ shrinks, and so the loop eventually terminates.

For partial correctness, the main loop satisfies these invariants:
\begin{enumerate}
    \item
    If $R' \subseteq X^2$, $x_1 \mathrel{R'} x_2$, and $R'$ is a bisimulation, then $R \cup T \subseteq R'$.

    \item
    If $x_1' \mathrel{R} x_2'$, then $o(x_1') = o(x_2')$ and $<(x_1')_\ltr{a}, (x_2')_\ltr{a}> \in R \cup T$.
\end{enumerate}
If the algorithm decides negatively in the loop for the pair $\angl{x_1', x_2'}$, suppose towards a contradiction that $x_1$ is bisimilar to $x_2$.
Then there exists a bisimulation $R'$ on $X$ such that $x_1 \mathrel{R'} x_1$.
But by the invariant, $x_1' \mathrel{R'} x_2'$, despite $o(x_1') \neq o(x_2')$.
This is a contradiction; we must therefore conclude that $x_1$ is indeed not bisimilar to $x_2$.

Conversely, if the algorithm decides positively, then $T = \emptyset$.
The second invariant then implies that $R$ is a bisimulation.
Since $x_1 \mathrel{R} x_2$ by construction, we must conclude that $x_1$ is bisimilar to $x_2$.
\end{proof}

Of course, this algorithm leaves some room for optimization.
We refer to \Cref{\res{chapter:further-reading}} for pointers to further reading on these topics.

\subsection{Soundness of Brzozowski derivatives}\label{section:brzozowski-sound}

Recall that $(\Exp, <o, t>)$ is an automaton and thus, by Theorem~\ref{thm:languages-final}, we have a unique map $L\colon \Exp \to 2^{\alphabet^*}$ that acts as a morphism between $(\Exp, <o, t>)$ and $(2^{\alphabet^*}, <o_L, t_L>)$.
Hence, $L$ provides an alternative \emph{operational} semantics for regular expressions.
We now prove that, for any $e \in \Exp$, this semantics coincides with the one from \Cref{\res{def:algsem}}.
\begin{theorem}[{\cite{brzozowski-1964}}]
For all $e\in \Exp$ and $w\in \alphabet^*$, $\sem{e} = L(e)$.
That is,
\[
w \in \sem{e}  \Leftrightarrow  w\in L(e)
\]
\end{theorem}
\begin{proof}
By \Cref{\res{thm:languages-final}}, it suffices to show that $\sem{-}$ acts like a morphism from $(\Exp, <o, t>)$ to $(2^{\alphabet^*}, <o_L, t_L>)$.
Concretely, this comes down to proving that for all $e \in \Exp$, it holds that $o_L(\sem{e}) = o(e)$, and furthermore that for $\ltr{a} \in \alphabet$ we have that $\sem{e_\ltr{a}} = \sem{e}_{\ltr{a}}$.

For the first equality, we derive using \Cref{\res{thm:fundamental-re}} that:
\[
    o_L(\sem{e})
        = o_L(\sem{o(e) + \sum_{\ltr{a} \in \alphabet} \ltr{a}e_\ltr{a}})
        = o_L(\sem{o(e)}) = o(e)
\]
In the second-to-last step, we use that for $b \in \{ 0, 1 \}$, it holds that $o_L(\sem{b}) = b$.
For the second equality, we again use \Cref{\res{thm:fundamental-re}}:
\[
    \sem{e}_\ltr{a}
        = \sem{o(e) + \sum_{\ltr{b} \in \alphabet} \ltr{b}e_\ltr{b}}_\ltr{a}
        = \sem{\ltr{a} e_\ltr{a}}_\ltr{a}
        = \sem{e_\ltr{a}}
    \qedhere
\]
\end{proof}
Thus, the denotational semantics of regular expressions coincides with the operational semantics.
Furthermore, because bisimulations are sound and complete for deciding language equivalence of states in automata, $e$ and $f$ have the same language precisely when they are bisimilar as states of the Brzozowski automaton.
This \emph{almost} means that we can mechanize equivalence checking of expressions (through the algorithm in \Cref{\res{figure:bisimulation-algorithm}}); the only wrinkle is that the Brzozowski automaton is not finite.
We will return to this matter shortly.

\section{Kleene's theorem}%
\label{section:kleenes-theorem}

In this section, we present Kleene's theorem, which states the equivalence
between the class of languages recognized by finite deterministic
automata and the one denoted by regular expressions.
\begin{theorem}[Kleene's Theorem~\cite{kleene-1956,thompson-1968,mcnaughton-yamada-1960}]\label{thm:kleene_dfa}

For any $l \in 2^{\alphabet^*}$, $l = \sem{e}$, for some regular expression $e\in \Exp$ if and only if $l = L(s)$ for a state $s\in S$ of a finite deterministic automaton $(S,<o_S,t_S>)$.
\end{theorem}
The proof of this theorem amounts to constructing an equivalent regular expression from a state of a deterministic automaton and, conversely, to constructing a deterministic automaton which has a state that is equivalent to a given regular expression.
We will detail these constructions in the next two sections.

\subsection{Expressions to automata}%
\label{subsection:e2a}
Given a regular expression $e$, we want to construct a finite deterministic automaton with the same language.
We have shown that $\Exp$, the set of regular expressions over $\alphabet$, carries an automaton structure given by Brzozowski derivatives.
Hence, an obvious way of constructing an automaton with a state equivalent to $e$ is to carve out all states reachable from $e$ in the Brzozowski automaton $(\Exp, <\ore,\tre>)$.
However, even for very simple expressions, this may still yield an infinite automaton.
Take for example $\ltr{a}^{**}$; computing derivatives for $\ltr{a}$ yields:
\begin{mathpar}
(\ltr{a}^{**})_\ltr{a} = 1 \ltr{a}^*\ltr{a}^{**}
\and
(1 \ltr{a}^*\ltr{a}^{**})_\ltr{a} = (0\ltr{a}^* + 1 \ltr{a}^*)\ltr{a}^{**} + 1 \ltr{a}^*\ltr{a}^{**}
\and
((0\ltr{a}^* + 1 \ltr{a}^*)\ltr{a}^{**} + 1 \ltr{a}^*\ltr{a}^{**})_\ltr{a} = (0 \ltr{a}^* + 0 \ltr{a}^* +  1 \ltr{a}^*)\ltr{a}^{**} + 1 \ltr{a}^*\ltr{a}^{**} + (1 \ltr{a}^*\ltr{a}^{**})_\ltr{a}
\end{mathpar}
All derivatives above are equivalent, but they are syntactically different and therefore they will not be identified as denoting the same state.
Indeed, it is not too hard to see that every time we take a new $\ltr{a}$-step from the last expression generated, we obtain an expression that is larger than (and therefore different from) the expressions we have seen before.
Thus $\ltr{a}^{**}$ can reach infinitely many states in $(\Exp, <o, t>)$.

To achieve finiteness, it suffices to remove duplicate expressions in sums.
This can be done by leveraging the laws of associativity, commutativity and idempotence of $+$ (ACI in brief)~\cite{brzozowski-1964,conway-1971}.
In order to make this precise, we use these laws to define an equivalence.

\begin{definition}\label{definition:congruence-aci}
We define $\cong$ as the smallest congruence on $\Exp$ (w.r.t.\ $+$, $\cdot$ and $*$) satisfying the following laws, for all $e, f, g \in \Exp$:
\begin{mathpar}
e + (f + g) \cong (e + f) + g
\and
e + f \cong f + e
\and
e + e \cong e
\end{mathpar}
\end{definition}

The first thing to note about $\cong$ is that, besides relating expressions that are clearly equivalent, is also a bisimulation on $(\Exp, <o, t>)$.

\begin{proposition}\label{lemma:aci-bisimulation}
Let $e, f \in \Exp$ such that $e \cong f$.
Then $o(e) = o(f)$, and moreover $e_\ltr{a} \cong f_\ltr{a}$ for all $\ltr{a} \in \alphabet$.
\end{proposition}
\begin{proof}
It suffices to check the laws that generate $\cong$.
For instance, if $e \cong f$ because $e = g_1 + g_2$ and $f = g_2 + g_1$, then $o(e) = o(g_1) \vee o(g_2) = o(g_2) \vee o(g_1)$, and $e_\ltr{a} = (g_1)_\ltr{a} + (g_2)_\ltr{a} \cong (g_2)_\ltr{a} + (g_1)_\ltr{a} = f_\ltr{a}$.

The cases for the other rules, as well as the rules that come from the fact that $\cong$ is a congruence on $\Exp$, follow a similar pattern.
\end{proof}

The following provides a characterization of the derivative $e_w$, for a non-empty word $w\in \alphabet^*$ and $e\in \Exp$, in terms of $\cong$.
\begin{proposition}\label{prop:word_ders}
Let $w\in \alphabet^*$ be non-empty and $e,e_1,e_2\in \Exp $.
Then, the following hold for the derivatives of $(e_1+e_2)$, $e_1e_2$ and $e^*$:
\begin{align*}
(e_1+e_2)_w
    &\cong (e_1)_w + (e_2)_w\\
(e_1e_2)_w
    &\cong (e_1)_w e_2 + (e_2)_{x_1} + \cdots + (e_2)_{x_m}\\
(e^*)_w
    &\cong e_{y_1}e^* + \cdots + e_{y_n}e^*
\end{align*}
for some $m, n \geq 0$ and $w_i, y_j \in \alphabet^*$.
\end{proposition}
\begin{proof}
By induction on the length of $w$.
For the base, where $w=\ltr{a}$, the equalities hold by definition of derivatives.
For the inductive step, where $w=\ltr{a}w'$ for non-empty $w' \in \alphabet^*$, the first claim is easy to prove:
\[
(e_1+e_2)_{\ltr{a}w'}
    \stackrel{\phantom{(\mathrm{IH})}}= ((e_1)_\ltr{a} + (e_2)_\ltr{a})_{w'}
    \stackrel{(\mathrm{IH})}\cong (e_1)_{\ltr{a}w'} + (e_2)_{\ltr{a}w'}
\]
For the second claim, we note that
\[
(e_1e_2)_{\ltr{a}w'}
    = \begin{cases}
      ((e_1)_\ltr{a}e_2)_{w'} & \text{if }\ore(e_1) = 0 \\
      ((e_1)_\ltr{a}e_2)_{w'} + (e_2)_{\ltr{a}w'} & \text{otherwise}
      \end{cases}
\]
Both cases now follow by induction.
If $\ore(e_1) = 0$, we have
\[
    ((e_1)_\ltr{a}e_2)_{w'}
        \stackrel{(\mathrm{IH})}\cong (e_1)_{\ltr{a}w_1}e_2 + (e_2)_{x_1'} + \cdots + (e_2)_{x_{m'}'}
\]
Otherwise, if $o(e_1) = 1$, then we derive
\[
    ((e_1)_\ltr{a}e_2)_{w'} + (e_2)_{\ltr{a}w'}
        \stackrel{(\mathrm{IH})}\cong (e_1)_{\ltr{a}w_1}e_2 + (e_2)_{x_1'} + \cdots + (e_2)_{x_{m'}'} + (e_2)_{\ltr{a}w'}
\]
Clearly, either of these expressions fit the format of the claim.

Finally, for the third equality, we need to use the result we have just proved for sequential composition:
\begin{align*}
(e^*)_{\ltr{a}w'}
    &\stackrel{\phantom{(\mathrm{IH})}}= (e_\ltr{a}e^*)_{w'}
     \cong e_{\ltr{a}w'}e^* + (e^*)_{x_1} + \cdots + (e^*)_{x_m} \\
    &\stackrel{(\mathrm{IH})}\cong e_{y_1}e^* + \cdots + e_{y_n}e^*
\end{align*}
for appropriately chosen $n$ and $y_1, \dots, y_n \in \alphabet^*$.
\end{proof}

Using \Cref{\res{prop:word_ders}}, we can now prove the following.
\begin{proposition}[{\cite{brzozowski-1964}}]%
\label{theorem:brzozowski-finite}
Let $e\in \Exp$.
There exists a finite set of expressions $\rho(e)$ such that for $w \in \alphabet^*$, $e_w \cong e'$ for some $e' \in \rho(e)$.
\end{proposition}
\begin{proof}
By induction on the structure of the expression.
The base cases are trivial: they have either one, two or three distinct word derivatives.

For the sum $e_1 + e_2$, we can choose $\rho(e_1 + e_2) = \{ e_1' + e_2' \mid e_1' \in \rho(e_1), e_2' \in \rho(e_2) \}$.
The claim then follows by induction and \Cref{\res{prop:word_ders}}.
The cases for $e_1 \cdot e_2$ and $e_1^*$ follow similarly.
\end{proof}

This was the final insight needed to construct a finite automaton for any given expression $e$.
The results come together as follows.

\begin{theorem}[{\cite{brzozowski-1964}}]%
\label{theorem:expression-to-automaton}
Given an expression $e \in \Exp$, we can construct an automaton $(S, <o_S, t_S>)$ with a state $s \in S$ such that $L(s) = \sem{e}$.
\end{theorem}
\begin{proof}
We choose $S = \{ [e']_{\cong} \mid e' \in \rho(e) \}$, with $s = [e]_{\cong}$, while $o_S([e']) = o(e')$ and $t_S([e']_{\cong}, \ltr{a}) = [e'_\ltr{a}]_{\cong}$.
Since $\rho(e)$ is finite, so is $S$; furthermore, since there exists an $e' \in \rho(e)$ with $e' \cong e_\epsilon = e$, we know that $s \in S$.
Moreover, $o_S$ and $t_S$ are well-defined, by \Cref{\res{lemma:aci-bisimulation}}.

Now, consider the automaton $(T, <o_T, t_T>)$ where $T = \{ e_w \mid w \in \alphabet^* \}$, and $o_T(e') = o(e')$, while $t_T(e', \ltr{a}) = e'_\ltr{a}$.
Clearly, the injection of $T$ into $\Exp$ is a homomorphism, and hence the language of $e \in T$ is precisely $L(e) = \sem{e}$.
Lastly $L(e) = L(s)$, because the map that sends $e' \in T$ to $[e']_{\cong} \in S$ is a homomorphism from $(T, <o_T, t_T>)$ to $(S, <o_S, t_S>)$.
\end{proof}

We can compute $(S, <o_S, t_S>)$ as described in the proof above, by starting from $e$ and incrementally exploring the state space by computing derivatives; every time we find an expression $e'$ that is $\cong$-equivalent to an expression $e''$ seen before, we can route the transition to $e'$ instead.
This procedure is known as \emph{Brzozowski's construction}.
The example below illustrates how this algorithm proceeds in detail.

\begin{example}\label{example:big-automaton}
Let $\alphabet=\{\ltr{a},\ltr{b}\}$ and $e=(\ltr{a}\ltr{b}+\ltr{b})^*\ltr{b}\ltr{a}$.
We compute derivatives incrementally, marking with $\checkmark$ the derivatives that have been encountered before, and numbering the new ones:

{\small
\begin{longtable}{@{}r@{\ }c@{\ }p{95mm}@{\ \ \ \ }r}
$e_{\ltr{}}$
&$=$&$ (\ltr{a}\ltr{b}+\ltr{b})^*\ltr{b}\ltr{a}$
&$\mycirc{1}$ \\
$e_{\ltr{a}}$
&$=$&$ (1\ltr{b}+0)(\ltr{a}\ltr{b}+\ltr{b})^*\ltr{b}\ltr{a}+0\ltr{a}$
&$\mycirc{2}$ \\
$e_{\ltr{b}}$
&$=$&$ (0\ltr{b}+1)(\ltr{a}\ltr{b}+\ltr{b})^*\ltr{b}\ltr{a}+1\ltr{a}$
&$\mycirc{3}$ \\
$e_{\ltr{aa}}$
&$=$&$ (0\ltr{b}+0+0)(\ltr{a}\ltr{b}+\ltr{b})^*\ltr{b}\ltr{a}+0\ltr{a}$
\\
&$\cong$ & $(0\ltr{b}+0)(\ltr{a}\ltr{b}+\ltr{b})^*\ltr{b}\ltr{a}+0\ltr{a}$
&$\mycirc{4}$ \\
$e_{\ltr{ab}}$
&$=$&$ (0\ltr{b}+1+0)(\ltr{a}\ltr{b}+\ltr{b})^*\ltr{b}\ltr{a}+0\ltr{a}$
&$\mycirc{5}$ \\
$e_{\ltr{ba}}$
&$=$&$ (0\ltr{b}+0)(\ltr{a}\ltr{b}+\ltr{b})^*\ltr{b}\ltr{a}+(1\ltr{b}+0)(\ltr{a}\ltr{b}+\ltr{b})^*\ltr{b}\ltr{a}+0\ltr{a}+0\ltr{a}+1$
&$\mycirc{6}$ \\
$e_{\ltr{bb}}$
&$=$&$ (0\ltr{b}+0)(\ltr{a}\ltr{b}+\ltr{b})^*\ltr{b}\ltr{a}+(0\ltr{b}+1)(\ltr{a}\ltr{b}+\ltr{b})^*\ltr{b}\ltr{a}+1\ltr{a}+0\ltr{a}+0$
&$\mycirc{7}$ \\
$e_{\ltr{aaa}}$
&$=$&$ (0\ltr{b}+0)(\ltr{a}\ltr{b}+\ltr{b})^*\ltr{b}\ltr{a}+0\ltr{a}$
$\cong e_{\ltr{aa}}$
&$\checkmark$ \\
$e_{\ltr{aab}}$
&$=$&$ (0\ltr{b}+0)(\ltr{a}\ltr{b}+\ltr{b})^*\ltr{b}\ltr{a}+0\ltr{a}$
$\cong e_{\ltr{aa}}$
&$\checkmark$ \\
$e_{\ltr{aba}}$
&$=$&$ (0\ltr{b}+0+0)(\ltr{a}\ltr{b}+\ltr{b})^*\ltr{b}\ltr{a}+(1\ltr{b}+0)(\ltr{a}\ltr{b}+\ltr{b})^*\ltr{b}\ltr{a}+0\ltr{a}+0\ltr{a}$
\\
&$\cong$ & $(0\ltr{b}+0)(\ltr{a}\ltr{b}+\ltr{b})^*\ltr{b}\ltr{a}+(1\ltr{b}+0)(\ltr{a}\ltr{b}+\ltr{b})^*\ltr{b}\ltr{a}+0\ltr{a}$
&$\mycirc{8}$ \\
$e_{\ltr{abb}}$
&$=$&$ (0\ltr{b}+0+0)(\ltr{a}\ltr{b}+\ltr{b})^*\ltr{b}\ltr{a}+(0\ltr{b}+1)(\ltr{a}\ltr{b}+\ltr{b})^*\ltr{b}\ltr{a}+1\ltr{a}+0\ltr{a}$
\\
&$\cong$ & $(0\ltr{b}+0)(\ltr{a}\ltr{b}+\ltr{b})^*\ltr{b}\ltr{a}+(0\ltr{b}+1)(\ltr{a}\ltr{b}+\ltr{b})^*\ltr{b}\ltr{a}+1\ltr{a}+0\ltr{a}$
&$\mycirc{9}$ \\
$e_{\ltr{baa}}$
&$=$&$ (0\ltr{b}+0)(\ltr{a}\ltr{b}+\ltr{b})^*\ltr{b}\ltr{a}+(0\ltr{b}+0+0)(\ltr{a}\ltr{b}+\ltr{b})^*\ltr{b}\ltr{a}+0\ltr{a}+0$
\\
&$\cong$ & $(0\ltr{b}+0)(\ltr{a}\ltr{b}+\ltr{b})^*\ltr{b}\ltr{a}+0\ltr{a}+0$
&$\mycirc{10}$ \\
$e_{\ltr{bab}}$
&$=$&$ (0\ltr{b}+0)(\ltr{a}\ltr{b}+\ltr{b})^*\ltr{b}\ltr{a}+(0\ltr{b}+1+0)(\ltr{a}\ltr{b}+\ltr{b})^*\ltr{b}\ltr{a}+0\ltr{a}+0$
&$\mycirc{11}$ \\
$e_{\ltr{bba}}$
&$=$&$ (0\ltr{b}+0)(\ltr{a}\ltr{b}+\ltr{b})^*\ltr{b}\ltr{a}+(0\ltr{b}+0)(\ltr{a}\ltr{b}+\ltr{b})^*\ltr{b}\ltr{a}+(1\ltr{b}+0)(\ltr{a}\ltr{b}+\ltr{b})^*\ltr{b}\ltr{a}+0\ltr{a}+0\ltr{a}+1+0\ltr{a}+0$
&$\mycirc{12}$ \\
$e_{\ltr{bbb}}$
&$=$&$ (0\ltr{b}+0)(\ltr{a}\ltr{b}+\ltr{b})^*\ltr{b}\ltr{a}+(0\ltr{b}+0)(\ltr{a}\ltr{b}+\ltr{b})^*\ltr{b}\ltr{a}+(0\ltr{b}+1)(\ltr{a}\ltr{b}+\ltr{b})^*\ltr{b}\ltr{a}+1\ltr{a}+0\ltr{a}+0+0\ltr{a}+0$
$\cong e_{\ltr{bb}}$
&$\checkmark$ \\
$e_{\ltr{abaa}}$
&$=$&$ (0\ltr{b}+0)(\ltr{a}\ltr{b}+\ltr{b})^*\ltr{b}\ltr{a}+(0\ltr{b}+0+0)(\ltr{a}\ltr{b}+\ltr{b})^*\ltr{b}\ltr{a}+0\ltr{a}$
$\cong e_{\ltr{aa}}$
&$\checkmark$ \\
$e_{\ltr{abab}}$
&$=$&$ (0\ltr{b}+0)(\ltr{a}\ltr{b}+\ltr{b})^*\ltr{b}\ltr{a}+(0\ltr{b}+1+0)(\ltr{a}\ltr{b}+\ltr{b})^*\ltr{b}\ltr{a}+0\ltr{a}$
&$\mycirc{13}$ \\
$e_{\ltr{abba}}$
&$=$&$ (0\ltr{b}+0)(\ltr{a}\ltr{b}+\ltr{b})^*\ltr{b}\ltr{a}+(0\ltr{b}+0)(\ltr{a}\ltr{b}+\ltr{b})^*\ltr{b}\ltr{a}+(1\ltr{b}+0)(\ltr{a}\ltr{b}+\ltr{b})^*\ltr{b}\ltr{a}+0\ltr{a}+0\ltr{a}+1+0\ltr{a}$
$\cong e_{\ltr{ba}}$
&$\checkmark$ \\
$e_{\ltr{abbb}}$
&$=$&$ (0\ltr{b}+0)(\ltr{a}\ltr{b}+\ltr{b})^*\ltr{b}\ltr{a}+(0\ltr{b}+0)(\ltr{a}\ltr{b}+\ltr{b})^*\ltr{b}\ltr{a}+(0\ltr{b}+1)(\ltr{a}\ltr{b}+\ltr{b})^*\ltr{b}\ltr{a}+1\ltr{a}+0\ltr{a}+0+0\ltr{a}$
$\cong e_{\ltr{bb}}$
&$\checkmark$ \\
$e_{\ltr{baaa}}$
&$=$&$ (0\ltr{b}+0)(\ltr{a}\ltr{b}+\ltr{b})^*\ltr{b}\ltr{a}+0\ltr{a}+0$
$\cong e_{\ltr{baa}}$
&$\checkmark$ \\
$e_{\ltr{baab}}$
&$=$&$ (0\ltr{b}+0)(\ltr{a}\ltr{b}+\ltr{b})^*\ltr{b}\ltr{a}+0\ltr{a}+0$
$\cong e_{\ltr{baa}}$
&$\checkmark$ \\
$e_{\ltr{baba}}$
&$=$&$ (0\ltr{b}+0)(\ltr{a}\ltr{b}+\ltr{b})^*\ltr{b}\ltr{a}+(0\ltr{b}+0+0)(\ltr{a}\ltr{b}+\ltr{b})^*\ltr{b}\ltr{a}+(1\ltr{b}+0)(\ltr{a}\ltr{b}+\ltr{b})^*\ltr{b}\ltr{a}+0\ltr{a}+0\ltr{a}+0$
\\
&$\cong$ & $(0\ltr{b}+0)(\ltr{a}\ltr{b}+\ltr{b})^*\ltr{b}\ltr{a}+(1\ltr{b}+0)(\ltr{a}\ltr{b}+\ltr{b})^*\ltr{b}\ltr{a}+0\ltr{a}+0$
&$\mycirc{14}$ \\
$e_{\ltr{babb}}$
&$=$&$ (0\ltr{b}+0)(\ltr{a}\ltr{b}+\ltr{b})^*\ltr{b}\ltr{a}+(0\ltr{b}+0+0)(\ltr{a}\ltr{b}+\ltr{b})^*\ltr{b}\ltr{a}+(0\ltr{b}+1)(\ltr{a}\ltr{b}+\ltr{b})^*\ltr{b}\ltr{a}+1\ltr{a}+0\ltr{a}+0$
$\cong e_{\ltr{bb}}$
&$\checkmark$ \\
$e_{\ltr{bbaa}}$
&$=$&$ (0\ltr{b}+0)(\ltr{a}\ltr{b}+\ltr{b})^*\ltr{b}\ltr{a}+(0\ltr{b}+0+0)(\ltr{a}\ltr{b}+\ltr{b})^*\ltr{b}\ltr{a}+0\ltr{a}+0+0$
$\cong e_{\ltr{baa}}$
&$\checkmark$ \\
$e_{\ltr{bbab}}$
&$=$&$ (0\ltr{b}+0)(\ltr{a}\ltr{b}+\ltr{b})^*\ltr{b}\ltr{a}+(0\ltr{b}+1+0)(\ltr{a}\ltr{b}+\ltr{b})^*\ltr{b}\ltr{a}+0\ltr{a}+0+0$
$\cong e_{\ltr{bab}}$
&$\checkmark$ \\
$e_{\ltr{ababa}}$
&$=$&$ (0\ltr{b}+0)(\ltr{a}\ltr{b}+\ltr{b})^*\ltr{b}\ltr{a}+(0\ltr{b}+0+0)(\ltr{a}\ltr{b}+\ltr{b})^*\ltr{b}\ltr{a}+(1\ltr{b}+0)(\ltr{a}\ltr{b}+\ltr{b})^*\ltr{b}\ltr{a}+0\ltr{a}+0\ltr{a}$
$\cong e_{\ltr{aba}}$
&$\checkmark$ \\
$e_{\ltr{ababb}}$
&$=$&$ (0\ltr{b}+0)(\ltr{a}\ltr{b}+\ltr{b})^*\ltr{b}\ltr{a}+(0\ltr{b}+0+0)(\ltr{a}\ltr{b}+\ltr{b})^*\ltr{b}\ltr{a}+(0\ltr{b}+1)(\ltr{a}\ltr{b}+\ltr{b})^*\ltr{b}\ltr{a}+1\ltr{a}+0\ltr{a}$
$\cong e_{\ltr{abb}}$
&$\checkmark$ \\
$e_{\ltr{babaa}}$
&$=$&$ (0\ltr{b}+0)(\ltr{a}\ltr{b}+\ltr{b})^*\ltr{b}\ltr{a}+(0\ltr{b}+0+0)(\ltr{a}\ltr{b}+\ltr{b})^*\ltr{b}\ltr{a}+0\ltr{a}+0$
$\cong e_{\ltr{baa}}$
&$\checkmark$ \\
$e_{\ltr{babab}}$
&$=$&$ (0\ltr{b}+0)(\ltr{a}\ltr{b}+\ltr{b})^*\ltr{b}\ltr{a}+(0\ltr{b}+1+0)(\ltr{a}\ltr{b}+\ltr{b})^*\ltr{b}\ltr{a}+0\ltr{a}+0$
$\cong e_{\ltr{bab}}$
&$\checkmark$
\end{longtable}}
\noindent
The resulting automaton has $14$ states (with $e_{\ltr{ba}}$ and $e_{\ltr{bba}}$ accepting):
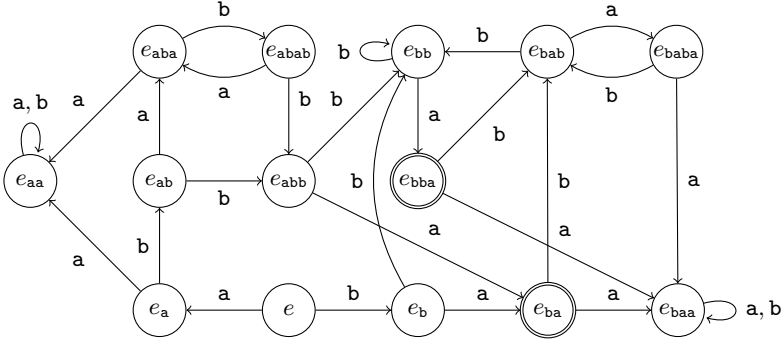
\[
    \begin{tikzpicture}[every state/.style={minimum size=7mm,inner sep=0em},font=\footnotesize]
    \node[state] (aa) {\footnotesize $e_{\ltr{a}\ltr{a}}$};
    \node[state,right=of aa] (ab) {\footnotesize $e_{\ltr{a}\ltr{b}}$};
    \node[state,below=of ab] (a) {\footnotesize $e_\ltr{a}$};
    \node[state,above=of ab] (aba) {\footnotesize $e_{\ltr{a}\ltr{b}\ltr{a}}$};
    \node[state,right=of a] (e) {\footnotesize $e$};
    \node[state,right=of ab] (abb) {\footnotesize $e_{\ltr{a}\ltr{b}\ltr{b}}$};
    \node[state,right=of aba] (abab) {\footnotesize $e_{\ltr{a}\ltr{b}\ltr{a}\ltr{b}}$};
    \node[state,right=of abab] (bb) {\footnotesize $e_{\ltr{b}\ltr{b}}$};
    \node[state,right=of e] (b) {\footnotesize $e_\ltr{b}$};
    \node[state,accepting,right=of b] (ba) {\footnotesize $e_{\ltr{b}\ltr{a}}$};
    \node[state,right=of bb] (bab) {\footnotesize $e_{\ltr{b}\ltr{a}\ltr{b}}$};
    \node[state,accepting] at (abb -| bb) (bba) {\footnotesize $e_{\ltr{b}\ltr{b}\ltr{a}}$};
    \node[state,right=of ba] (baa) {\footnotesize $e_{\ltr{b}\ltr{a}\ltr{a}}$};
    \node[state,right=of bab] (baba) {\footnotesize $e_{\ltr{b}\ltr{a}\ltr{b}\ltr{a}}$};
    \draw (aa) edge[loop above] node {$\ltr{a}, \ltr{b}$} (aa);
    \draw[->] (a) to node[below left] {$\ltr{a}$} (aa);
    \draw[->] (aba) to node[above left] {$\ltr{a}$} (aa);
    \draw[->] (a) to node[left] {$\ltr{b}$} (ab);
    \draw[->] (ab) to node[left] {$\ltr{a}$} (aba);
    \draw[->] (e) to node[above] {$\ltr{a}$} (a);
    \draw (bb) edge[loop left] node {$\ltr{b}$} (bb);
    \draw[->] (aba) to[bend left] node[above] {$\ltr{b}$} (abab);
    \draw[->] (abab) to[bend left] node[below] {$\ltr{a}$} (aba);
    \draw[->] (e) to node[above] {$\ltr{b}$} (b);
    \draw[->] (ab) to node[below] {$\ltr{b}$} (abb);
    \draw[->] (b) to[bend left] node[left] {$\ltr{b}$} (bb);
    \draw[->] (abb) to node[above left] {$\ltr{b}$} (bb);
    \draw[->] (abab) to node[above right] {$\ltr{b}$} (abb);
    \draw[->] (abb) to node[above right] {$\ltr{a}$} (ba);
    \draw[->] (bb) to node[right] {$\ltr{a}$} (bba);
    \draw[->] (b) to node[above] {$\ltr{a}$} (ba);
    \draw[->] (ba) to node[right] {$\ltr{b}$} (bab);
    \draw[->] (bba) to node[below right] {$\ltr{b}$} (bab);
    \draw[->] (bba) to node[above right] {$\ltr{a}$} (baa);
    \draw[->] (ba) to node[above] {$\ltr{a}$} (baa);
    \draw[->] (bab) to[bend left] node[above] {$\ltr{a}$} (baba);
    \draw[->] (baba) to[bend left] node[below] {$\ltr{b}$} (bab);
    \draw[->] (baba) to node[right] {$\ltr{a}$} (baa);
    \draw (baa) edge[loop right] node {$\ltr{a}, \ltr{b}$} (baa);
    \draw[->] (bab) to node[above] {$\ltr{b}$} (bb);
    \end{tikzpicture}
\]
\end{example}
In the last example, we strictly followed the procedure described above.
The goal was to construct a finite automaton with a state equivalent to $e$ and we did not worry about its size.
If we allow expressions to be simplified even further, the resulting automaton could be much smaller.
Formally, we can do this by defining $\simeq$ as the smallest congruence that contains $\cong$ but also satisfies that $0e \simeq 0$, $1e \simeq e$, and $e + 0 \simeq e$ for all $e \in \Exp$.\label{page:simeq}
It is not too hard to verify that the properties of $\cong$ can be transferred to $\simeq$, namely that (1)~it is a bisimulation and (2)~only finitely expressions can be reached from any expression, up to $\simeq$ (since $\simeq$ contains $\cong$, the latter follows immediately from \Cref{\res{theorem:brzozowski-finite}}).

\begin{example}
With $\simeq$, the calculations from \Cref{\res{example:big-automaton}} become:
{\small
\begin{longtable}{@{}r@{\ }c@{\ }p{95mm}@{\ \ \ \ }r}
$e_{\ltr{}}$
&$=$&$ (\ltr{a}\ltr{b}+\ltr{b})^*\ltr{b}\ltr{a}$
&$\mycirc{1}$ \\
$e_{\ltr{a}}$
&$=$&$ (1\ltr{b}+0)(\ltr{a}\ltr{b}+\ltr{b})^*\ltr{b}\ltr{a}+0\ltr{a}$
$\simeq \ltr{b}(\ltr{a}\ltr{b}+\ltr{b})^*\ltr{b}\ltr{a}$
&$\mycirc{2}$ \\
$e_{\ltr{b}}$
&$=$&$ (0\ltr{b}+1)(\ltr{a}\ltr{b}+\ltr{b})^*\ltr{b}\ltr{a}+1\ltr{a}$
$\simeq (\ltr{a}\ltr{b}+\ltr{b})^*\ltr{b}\ltr{a}+\ltr{a}$
&$\mycirc{3}$ \\
$e_{\ltr{aa}}$
&$=$&$ 0(\ltr{a}\ltr{b}+\ltr{b})^*\ltr{b}\ltr{a}$
$\simeq 0$
&$\mycirc{4}$ \\
$e_{\ltr{ab}}$
&$=$&$ 1(\ltr{a}\ltr{b}+\ltr{b})^*\ltr{b}\ltr{a}$
$\simeq e_{\ltr{}}$
&$\checkmark$ \\
$e_{\ltr{ba}}$
&$=$&$ (1\ltr{b}+0)(\ltr{a}\ltr{b}+\ltr{b})^*\ltr{b}\ltr{a}+0\ltr{a}+1$
$\simeq \ltr{b}(\ltr{a}\ltr{b}+\ltr{b})^*\ltr{b}\ltr{a}+1$
&$\mycirc{5}$ \\
$e_{\ltr{bb}}$
&$=$&$ (0\ltr{b}+1)(\ltr{a}\ltr{b}+\ltr{b})^*\ltr{b}\ltr{a}+1\ltr{a}+0$
$\simeq e_{\ltr{b}}$
&$\checkmark$ \\
$e_{\ltr{aaa}}$
&$=$&$ 0$
$\simeq e_{\ltr{aa}}$
&$\checkmark$ \\
$e_{\ltr{aab}}$
&$=$&$ 0$
$\simeq e_{\ltr{aa}}$
&$\checkmark$ \\
$e_{\ltr{baa}}$
&$=$&$ 0(\ltr{a}\ltr{b}+\ltr{b})^*\ltr{b}\ltr{a}+0$
$\simeq e_{\ltr{aa}}$
&$\checkmark$ \\
$e_{\ltr{bab}}$
&$=$&$ 1(\ltr{a}\ltr{b}+\ltr{b})^*\ltr{b}\ltr{a}+0$
$\simeq e_{\ltr{}}$
&$\checkmark$
\end{longtable}
}
\noindent
The resulting automaton would then have $5$ states:
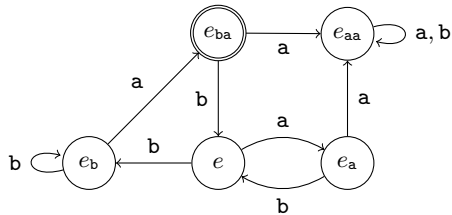
\[
    \begin{tikzpicture}[every state/.style={minimum size=7mm,inner sep=0em},font=\footnotesize]
        \node[state] (e) {$e$};
        \node[state,right=of e] (a) {$e_{\ltr{a}}$};
        \node[state,accepting,above=of e] (ba) {$e_{\ltr{ba}}$};
        \node[state,above=of a] (aa) {$e_{\ltr{aa}}$};
        \node[state,left=of e] (b) {$e_\ltr{b}$};
        \draw[->] (e) edge[bend left] node[above] {$\ltr{a}$} (a);
        \draw[->] (a) edge[bend left] node[below] {$\ltr{b}$} (e);
        \draw[->] (e) edge node[above] {$\ltr{b}$} (b);
        \draw[->] (a) edge node[right] {$\ltr{a}$} (aa);
        \draw[->] (ba) edge node[below] {$\ltr{a}$} (aa);
        \draw[->] (ba) edge node[left] {$\ltr{b}$} (e);
        \draw[->] (b) edge node[above left] {$\ltr{a}$} (ba);
        \draw (aa) edge[loop right] node {$\ltr{a}, \ltr{b}$} (aa);
        \draw (b) edge[loop left] node {$\ltr{b}$} (b);
    \end{tikzpicture}
\]
\end{example}

The automaton in the last example is in fact minimal.
This is not always the case; for that one would have to consider the complete set of axioms from \Cref{\res{def:ka}}.
However, in many cases the automaton computed using just a subset of the axioms is minimal: an empirical study about this phenomenon appears in~\cite{JFP09}.

\begin{remark}
There are, of course, many more ways to construct an equivalent automaton from a regular expression.
We focused on Brzozowski's construction here because it works well with the coalgebraic perspective, but you may be familiar with the syntax-directed translation proposed by Thompson~\cite{thompson-1968}.
Other constructions can be found in~\cite{antimirov-1996,glushkov-1961,berry-sethi-1986,mcnaughton-yamada-1960}; we refer to~\cite{watson-1993} for an excellent overview.
\end{remark}

\subsection{Automata to expressions}%
\label{subsection:a2e}

To present the construction of a regular expression from a
state of a finite deterministic automaton we build a system of equations on expressions and then solve it.
Let $(S, <o_S,t_S>)$ be a finite deterministic automaton.
We associate with each $s\in S$ a variable $x_s$ and an equation
\begin{align}\label{eq:dfa2re}
x_s \equiv \sum_{\ltr{a}\in \alphabet} \ltr{a} x_{s_\ltr{a}} +  {o_S(s)}
\end{align}
The sum is well defined because $\alphabet$ is finite.
In this way, we build a system with $n$ equations and $n$ variables, where $n$ is the size of $S$.

\begin{definition}[Solutions to systems of equations]%
\label{def:solutions}
Consider a system of equations  $x_s \equiv \sum_{\ltr{a}\in \alphabet} \ltr{a} x_{s_\ltr{a}} +  {o_S(s)}, s\in S$ as in
\Cref{\res{eq:dfa2re}}.
We define a \emph{solution} to this system as a mapping that assigns to each variable  $x_s$ an  expression $e_s\in\Exp$ (over $\alphabet$) satisfying $ e_s \equiv \sum_{\ltr{a}\in \alphabet} \ltr{a} e_{s_\ltr{a}} +  {o_S(s)}$.
\end{definition}

Observe that if $e_s$ is a solution to $x_s\equiv t_{s}$ and if $ t_{s} \equiv t_{s}'$, then $e_s$ is also a solution to $x_s\equiv t_{s}'$.
Moreover, if the equation is of the shape $x\equiv tx+c$, then $t^*c$ is a solution to the equation~\cite{kozen-1994}.
This can be used to iteratively solve a system, see below for an example.

Our definition of solution is sound, in the sense that every solution is necessarily language equivalent to the corresponding state.
This can be shown directly (see \Cref{\res{exercise:solution-sound}}), but a rather elegant proof by coinduction (cf.~\Cref{\res{sec:lanautco}} and \Cref{\res{exercise:solution-sound-coalgebra}}) is also possible.

We will now illustrate the construction of solutions with an example, and then we
will present an alternative definition of the solution of a system of
equations of the same form as~(\ref{eq:dfa2re}) using matrices.
\begin{example}\label{example:a2e}
Consider the following deterministic automaton with two states, over the alphabet $\alphabet=\{\ltr{a},\ltr{b}\}$:
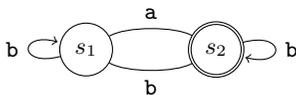
\[
    \begin{tikzpicture}[every state/.style={minimum size=7mm,inner sep=0em},font=\footnotesize]
        \node[state] (s1) {$s_1$};
        \node[state,accepting,right=of s1] (s2) {$s_2$};
        \draw (s1) edge[bend left,looseness=0.6] node[above] {$\ltr{a}$} (s2);
        \draw (s2) edge[bend left,looseness=0.6] node[below] {$\ltr{b}$} (s1);
        \draw (s1) edge[loop left] node[left] {$\ltr{b}$} (s1);
        \draw (s2) edge[loop right] node[right] {$\ltr{b}$} (s2);
    \end{tikzpicture}
\]
Let $e_1$ and $e_2$ denote $e_{s_1}$ and $e_{s_2}$, respectively.
The system of equations arising from the automaton is as follows:
\begin{mathpar}
    e_1 \equiv \ltr{a}e_2 + \ltr{b}e_1 + 0
    \and
    e_2 \equiv \ltr{a}e_1 + \ltr{b}e_2 + 1
\end{mathpar}
To solve for $e_1$ and $e_2$, we first apply the left fixpoint rule to the second equation to find that $\ltr{b}^*(\ltr{a}e_1 +  1) \leqq e_2$.
Indeed, if we fill in $\ltr{b}^*(\ltr{a}e_1 + 1)$ for $e_2$ in the second equation, it is not too hard to see that this constraint is still satisfied, regardless of the expression we end up choosing for $e_1$:
\[
    \ltr{b}^*(\ltr{a}e_1 + 1)
        \equiv (1 + \ltr{b}\ltr{b}^*)(\ltr{a}e_1 +1)
        \equiv \ltr{a}e_1 + \ltr{b}\ltr{b}^*(\ltr{a}e_1 + 1) + 1
\]
Thus, if we find a closed expression for $e_1$ satisfying the first equation, we can plug this into the above to find a closed expression for $e_2$ such that the second equation holds.
Now, given that we expressed $e_2$ in terms of $e_1$, we can replace $e_2$ with $\ltr{b}^*(\ltr{a}e_1 + 1)$ in the first equation:
\[
e_1 \geqq \ltr{a}\ltr{b}^*(\ltr{a}e_1 + 1) + \ltr{b}e_1 + 0 \equiv (\ltr{a}\ltr{b}^*\ltr{a} + \ltr{b})e_1 +\ltr{a}\ltr{b}^*
\]
By the left fixpoint rule, this tells us that $(\ltr{a}\ltr{b}^*\ltr{a} + \ltr{b})^*\ltr{a}\ltr{b}^* \leqq e_1$.
If we take the left-hand side of this as our closed expression for $e_1$, we can plug it into $\ltr{b}^*(\ltr{a}e_1 + 1)$ to find the following closed expression for $e_2$:
\[
e_2 \equiv \ltr{b}^*(\ltr{a}(\ltr{a}\ltr{b}^*\ltr{a} + \ltr{b})^*\ltr{a}\ltr{b}^* + 1) \equiv \ltr{b}^*\ltr{a}(\ltr{a}\ltr{b}^*\ltr{a} + \ltr{b})^*\ltr{a}\ltr{b}^*+\ltr{b}^*
\]
Together, these choices for $e_1$ and $e_2$ satisfy the equation for $e_1$; our previous arguments show that they should also satisfy the equation for $e_2$.
Lastly, and crucially, one can show that $\sem{e_1} = L(s_1)$ and $\sem{e_2} = L(s_2)$.
\end{example}

The methodology above generalizes to larger automata, but can become tedious to work out by hand (we will return to it in \Cref{\res{chapter:completeness}}).
For now, we briefly present an alternative general method~\cite{kozen:97:kat} to construct a solution to a system of $n$ equations of the form
\[
x_i \equiv e_{i1}x_1+ \cdots + e_{in}x_n +  f_i \ \ \ \ \ 1\leq i \leq n
\]
Given a system of $n$ equations as above, construct an $n\times n$ matrix $M$ with all the $e_{ij}$, an $n$-dimensional vector $O$ with all the $f_i$, and an $n$-dimensional vector $X$ with all the $x_i$.
The constraints of the above form can then collectively be written as the matrix-vector equation below, where matrix-vector multiplication and vector addition works just like in linear algebra, and vector equivalence is interpreted pointwise:
\begin{equation}\label{eq:sys_eqs}
X \equiv MX + O
\end{equation}
Next, we define, for an $n\times n$ matrix $R$ with entries in $\Exp$, a matrix $R^*$, by induction on $n$.
For $n=1$, the matrix $R$ has a single entry $e$ and $R^*= \left\lbrack e^* \right\rbrack$.
If $n \gr 1$, we divide the matrix into four submatrices
\[
R = \begin{pmatrix}
S & T\\
U & V
\end{pmatrix}
\]
where $S$ and $V$ are square with dimensions, respectively, $m\times m$ (for some $0 \ls m\ls n$) and $(n-m)\times (n-m)$.
By the induction hypothesis, we can compute the ``star'' of square matrices strictly smaller than $R$; this includes the submatrices $S$ and $V$, but also other $m$-by-$m$ matrices.
This allows us to define $R^*$ as follows:\label{star-matrices}
\[
R^* = \begin{pmatrix}
(S+TV^*U)^* & (S+TV^*U)^*TV^*\\
(V+US^*T)^*US^* & (V+US^*T)^*
\end{pmatrix}
\]

\begin{remark}
The choice of expressions in $R^*$ may seem magical at first, but can be explained intuitively by thinking of $R$ as encoding the transitions of a two-state automaton, with $S$ the label of a transition from the first state to itself, $T$ as the label of the transition from the first to the second state, and so on.
Then, $R^*$ contains expressions for the words read along paths between the state --- e.g., $(S + TV^*U)^*$ describes the paths from the first state to itself, which consist of either self-transitions ($S$), or transitions to the second state ($T$), a number of loops there ($V^*$), and a transition back ($U$).
See also \Cref{\res{example:matrix-two-state-solution}}.
\end{remark}

Now we can compute $M^*$, and show that the vector $M^*O$ is the least (with respect to the
pointwise order $\leqq$ of regular expressions) solution to the system
in~(\ref{eq:sys_eqs})~\cite[Theorem A.3]{kozen-97-book}.

\begin{example}\label{example:matrix-two-state-solution}
Let us revisit \Cref{\res{example:a2e}} using this method.
First, we construct the matrices $X$, $M$ and $O$:
\[
X = \begin{pmatrix}
e_1\\
e_2
\end{pmatrix}\ \ \ \ \
M = \begin{pmatrix}
\ltr{b} & \ltr{a} \\
\ltr{a} & \ltr{b}
\end{pmatrix}\ \ \ \ \ O = \begin{pmatrix}
 0\\
 1
\end{pmatrix}\ \ \ \ \
\]
The two equations are now captured by $M X + O \equiv X$.
We can now solve for $X = M^* O$ by computing $M^*$, as follows:
\begin{mathpar}
M^* = \begin{pmatrix}
(\ltr{b} + \ltr{a}\ltr{b}^*\ltr{a})^* & (\ltr{b} + \ltr{a}\ltr{b}^*\ltr{a})^*\ltr{a}\ltr{b}^*\\
(\ltr{b} + \ltr{a}\ltr{b}^*\ltr{a})^*\ltr{a}\ltr{b}^* &  (\ltr{b} + \ltr{a}\ltr{b}^*\ltr{a})^*
\end{pmatrix}
\\
 M^* O = \begin{pmatrix}
((\ltr{b} + \ltr{a}\ltr{b}^*\ltr{a})^*) 0 + ((\ltr{b} + \ltr{a}\ltr{b}^*\ltr{a})^*\ltr{a}\ltr{b}^*) 1\\
((\ltr{b} + \ltr{a}\ltr{b}^*\ltr{a})^*\ltr{a}\ltr{b}^*) 0 +((\ltr{b} + \ltr{a}\ltr{b}^*\ltr{a})^*) 1
\end{pmatrix} \equiv \begin{pmatrix}
(\ltr{b} + \ltr{a}\ltr{b}^*\ltr{a})^*\ltr{a}\ltr{b}^*\\
(\ltr{b} + \ltr{a}\ltr{b}^*\ltr{a})^*
\end{pmatrix}
\end{mathpar}
In the last step we used some of the Kleene algebra equations to simplify regular expressions.
Now note that the expression computed for $e_1$ is almost the same as the one we obtained in \Cref{\res{example:a2e}}, whereas the one for $e_2$ is quite different (syntactically), but still equivalent.
\end{example}

With the constructions from deterministic automata to regular expressions, and vice versa, proving Kleene's theorem is straightforward.
\begin{proof}[Proof of Theorem~\ref{thm:kleene_dfa} (Kleene's Theorem)]
Suppose $l=L(e)$ for some regular expression $e\in \Exp $.
We can construct an automaton $(S,<o_S,t_S>)$ and accompanying state that accepts the same language as $e$, as described in \Cref{\res{theorem:expression-to-automaton}}.
For the converse, suppose $l = L(s)$, for state $s\in S$ of a finite deterministic automaton $(S,<o_S,t_S>)$.
We solve a system of equations as described in \Cref{\res{def:solutions}} and the result is an expression $e_s$ equivalent to $s$, that is $\sem{e_s}=L(s)$.
\end{proof}

To summarize, in this chapter we instantiated the coalgebraic framework to deterministic automata, obtaining a decidable notion of \emph{bisimulation} that characterizes language equivalence (\Cref{\res{sec:lanautco}}).
We then showed that every regular expression can be converted into an equivalent deterministic automaton, meaning equivalence of regular expressions can be checked (\Cref{\res{subsection:e2a}}).
Finally, we argued that every automaton can be converted back to an expression (\Cref{\res{subsection:a2e}}).

\section{Exercises}%
\label{section:exercisesautomata}

\begin{exercises}
    \item\label{exercise:automata-bisimilarity}
    Create an automaton for the following languages over $\alphabet = \{ \mathtt{a}, \mathtt{b} \}$:
    \begin{exercises}
        \item
        The language $L_0$ of words where every third letter is an $\mathtt{a}$, unless it is preceded by a $\mathtt{b}$.
        Words in $L_0$ include $\mathtt{a}\mathtt{a}\mathtt{a}$, $\epsilon$, $\mathtt{b}$, and $\mathtt{b}\mathtt{b}\mathtt{b}$.
        Words \emph{not} in $L_0$ include $\mathtt{a}\mathtt{a}\mathtt{b}$ and $\mathtt{a}\mathtt{a}\mathtt{a}\mathtt{b}\mathtt{a}\mathtt{b}$.

        \item
        The language $L_1$ of words that do not contain three subsequent $\mathtt{a}$'s or $\mathtt{b}$'s.
        Words in $L_1$ include $\epsilon$, $\mathtt{a}\mathtt{b}$ and $\mathtt{a}\mathtt{b}\mathtt{a}\mathtt{a}$.
        Words \emph{not} in $L_1$ include $\mathtt{a}\mathtt{a}\mathtt{a}$ and $\mathtt{a}\mathtt{b}\mathtt{b}\mathtt{b}\mathtt{a}$.
    \end{exercises}

    \item\label{exercise:no-morphism-aut}
    Demonstrate two automata $(S, <o_S, t_S>)$ and $(T, <o_T, t_T>)$ with states $s \in S$ and $t \in T$ such that $L(s) = L(t)$, but there \emph{cannot} be a morphism between these automata that relates $s$ to $t$.
    Argue in detail \emph{why} such a morphism cannot exist.

    \item
    Consider the automata drawn below:
    
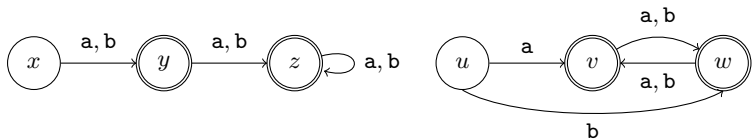
\[
        \begin{tikzpicture}[every state/.style={minimum size=7mm,inner sep=0em},font=\footnotesize]
            \node[state] (x) {$x$};
            \node[state,accepting,right=of x] (y) {$y$};
            \node[state,accepting,right=of y] (z) {$z$};
            \node[state,right=15mm of z] (u) {$u$};
            \node[state,accepting,right=of u] (v) {$v$};
            \node[state,accepting,right=of v] (w) {$w$};

            \draw[->] (x) edge node[above] {$\mathtt{a}, \mathtt{b}$} (y);
            \draw[->] (y) edge node[above] {$\mathtt{a}, \mathtt{b}$} (z);
            \draw[->] (z) edge[loop right] node[right] {$\mathtt{a}, \mathtt{b}$} (z);
            \draw[->] (u) edge node[above] {$\mathtt{a}$} (v);
            \draw (u.south) edge[looseness=0.6,bend right,->] node[below] {$\mathtt{b}$} (w.south);
            \draw[->] (v) edge[bend left] node[above] {$\mathtt{a}, \mathtt{b}$} (w);
            \draw[->] (w) edge node[below] {$\mathtt{a}, \mathtt{b}$} (v);
        \end{tikzpicture}
    \]
    Prove that $x$ and $u$ are bisimilar.

    \item\label{exercise:kernel-bisimulation}
    We return to the claim at the end the proof for \Cref{\res{thm:coind_dfa}}.
    \begin{exercises}
        \item
        Given two automata $(S, <o_S, t_S>)$ and $(T, <o_T, t_T>)$ and a morphism $h\colon S \to T$.
        Prove that the relation $\{ (s, s') \in S \times S \mid h(s) = h(s') \}$ is a bisimulation.
        \item
        Given two automata $(S, <o_S, t_S>)$ and $(T, <o_T, t_T>)$.
        Prove that the relation $\{ (s, t) \in S \times T \mid L(s) = L(t) \}$ is a bisimulation.
    \end{exercises}

    \item
    In the proof of \Cref{\res{theorem:algo-correct}}, we claimed that the following two properties were \emph{invariants} of the loop in the algorithm from \Cref{\res{figure:bisimulation-algorithm}} --- that is, if they are true before the loop body starts, then they are also true when the loop body ends:
    \begin{enumerate}
        \item
        If $R' \subseteq X^2$, $x_1 \mathrel{R'} x_2$, and $R'$ is a bisim., then $R \cup T \subseteq R'$.

        \item
        If $x_1' \mathrel{R} x_2'$, then $o(x_1') = o(x_2')$ and $<(x_1')_\ltr{a}, (x_2')_\ltr{a}> \in R \cup T$.
    \end{enumerate}

    Complete the following tasks regarding these claims:
    \begin{exercises}
        \item
        Prove that the first property is truly an invariant.
        \item
        Prove that the second property is truly an invariant.
        \item
        Check that, by the second property, when the loop ends (i.e., by returning $R$), $R$ is a bisimulation relating $x_1$ and $x_2$.
        \item
        Prove that, when the loop ends successfully, $R$ is the \emph{smallest} bisimulation relating $x_1$ and $x_2$ --- that is, if $R'$ is any bisimulation relating $x_1$ and $x_2$, then $R \subseteq R'$.
    \end{exercises}

    \item
    Use Brzozowski's construction to build a finite automaton for the expression $\mathtt{b}^* \mathtt{a}$, based on the congruence $\cong$; next, repeat this procedure with the congruence $\simeq$.

    \item
    Consider the automata from \Cref{\res{exercise:automata-bisimilarity}}.
    Derive the corresponding systems of equations, and solve them manually using the approach from \Cref{\res{example:a2e}}.
    Next, do the same thing, but use the alternative (matrix-based) approach.

    \item
    Construct the Brzozowski automaton for $e=\mathtt{a} \mathtt{b}^*$ and $f = \mathtt{a} + \mathtt{a} \mathtt{b} + \mathtt{a} \mathtt{b} \mathtt{b}^*$, using the congruence $\cong$.
    Next, show that the states for $e$ and $f$ are bisimilar.

    \item\label{exercise:solution-sound}
    Let $(S, <o_S, t_S>)$ be an automaton, and suppose that for each $s \in S$ we have an expression $e_s$, such that these expressions together form a solution to this automaton (cf.\ \Cref{\res{def:solutions}}).
    \begin{exercises}
        \item
        Show that for $s \in S$, $o_S(s) = 1$ if and only if $o(e_s) = 1$.
        \item
        Show that for $s \in S$, $w \in \alphabet^*$ and $\ltr{a} \in \alphabet$, we have that $\ltr{a}w \in \sem{e_s}$ if and only if $w \in \sem{e_{s_a}}$.
        \item
        Use the two facts above to show that, for all $w \in \alphabet^*$ and $s \in S$, we have that $w \in L(s)$ if and only if $w \in \sem{e_s}$.
        In your proof, make sure you state clearly what claim you are proving by induction (and pay attention to the quantifiers).
    \end{exercises}

    \item\label{exercise:solution-sound-coalgebra}
    We revisit \Cref{\res{exercise:solution-sound}} using a coalgebraic lens.
    Let $(S, <o_S, t_S>)$ be an automaton, and suppose that for $s \in S$ we have $e_s \in \Exp$, forming a solution.
    Now consider the relation
    \[
        R = \{ <e, s> \in \Exp \times S \mid \sem{e} = \sem{e_s} \}
    \]
    Show that $R$ is a bisimulation between $(S, <o_S, t_S>)$ and the Brzozowski coalgebra, and conclude that $\sem{e_s} = L(s)$ for all $s \in S$.
\end{exercises}

\onlyinsubfile{%
    \printbibliography%
}

\chapter{Completeness}%
\label{chapter:completeness}

In the previous chapters, we presented regular expressions, their denotational and operational semantics, and a set of laws to reason about their equivalence.
In this chapter, we want to prove that the laws of Kleene Algebra are enough to reason about any equivalence of regular expressions.
That is, given two regular expressions $e$ and $f$ that denote the same regular language, we can prove that $e \equiv f$.
This property is called \emph{completeness} with respect to the semantic model.
The goal of this chapter is therefore to show that the laws of Kleene Algebra are \emph{complete} for the language model.
We will use the fact that language equivalence and bisimilarity coincide for regular expressions, and show:
\begin{theorem}[Soundness and completeness]\label{thm:sc}
Let $e, f \in \Exp$; now
\[
    e \equiv f \iff \text{$e \bisim f$ in $(\Exp, <o, t>)$}
\]
\end{theorem}
The forward implication, called \emph{soundness}, says that all equivalences proved using the laws of \KA are in fact between bisimilar expressions.
The converse, called \emph{completeness}, says that all true equations between regular expressions can be proved using only the laws of \KA.

Many of the results presented in this chapter have appeared in the literature (e.g.,~\cite{conway-1971,kozen-2001,jacobs-2006}).
We will include proofs inspired by these works, and in some cases we will present them differently.
In particular, we give a new account using \emph{locally finite matrices} which fills in a gap in previous coalgebraic proofs.

\section{Matrices over Kleene algebras}%

We will start with what is perhaps the single most useful observation about Kleene algebra, which comes down to the following slogan:

\begin{quote}
\emph{Matrices over a Kleene algebra form a Kleene algebra!}
\end{quote}
We will first develop this idea for finite matrices.
Matrices indexed by infinite sets are a bit more subtle, and we will spend some time working on that in the latter half of this section.
For the remainder of this section, we fix a Kleene algebra $\mathcal{K} = (X, +, \cdot,-^*, 0, 1)$.

\subsection{Finite matrices}
To get an idea of how our slogan works for finite matrices, consider $2 \times 2$-matrices formed using elements of a Kleene algebra $\mathcal{K}$.
We can define addition and multiplication of these matrices in the usual way:
\begin{mathpar}
\begin{pmatrix}e_1&e_2\\e_3&e_4\end{pmatrix} +
\begin{pmatrix}e_5&e_6\\e_7&e_8\end{pmatrix} =
\begin{pmatrix}e_1+e_5&e_2+e_6\\e_3+e_7&e_4+e_8\end{pmatrix},
\\
\begin{pmatrix}e_1&e_2\\e_3&e_4\end{pmatrix} \cdot
\begin{pmatrix}e_5&e_6\\e_7&e_8\end{pmatrix} =
\begin{pmatrix}e_1e_5 + e_2e_7&e_1e_6 +
e_2e_8\\e_3e_5+e_4e_7&e_3e_6+e_4e_8\end{pmatrix}.
\end{mathpar}
It should be clear that the neutral elements for $+$ and $\cdot$ are, respectively:
\begin{mathpar}
    \begin{pmatrix}0&0\\0&0\end{pmatrix},
    \and
    \begin{pmatrix}1&0\\0&1\end{pmatrix}.
\end{mathpar}

The ``only'' question that remains is: how do we define an operation on matrices that satisfies the axioms of the Kleene star?
As it turns out, for $n = 2$, the following is a valid choice:
\[
    \begin{pmatrix}
    e_1 & e_2 \\
    e_3 & e_4
    \end{pmatrix}^* =
        \begin{pmatrix}
        (e_1 + e_2e_4^*e_3)^* & (e_1 + e_2e_4^*e_3)^*e_2e_4^*\\
        (e_4 + e_3e_1^*e_2)^*e_3e_1^* &(e_4 + e_3e_1^*e_2)^*
        \end{pmatrix},
\]
This may look like an arbitrary soup of symbols right now, although the eagle-eyed reader might recognize the operation on matrices from page~\pageref{\res{star-matrices}}, defined to solve systems of equations.
We will derive a way to compute the star of $n \times n$-matrices for a given $n$ in a moment; for now, let us pause and formalize what we already know.

\begin{definition}
    Let $Q$ be a set.
    A \emph{$Q$-vector (over $\mathcal{K}$)} is a function $v\colon Q \to \mathcal{K}$, and a \emph{$Q$-matrix (over $\mathcal{K}$)} is a function $M\colon Q \times Q \to \mathcal{K}$.

    Addition of matrices and vectors is defined pointwise, e.g., for $Q$-matrices $M$ and $N$, we define $M + N$ as the $Q$-matrix defined by:
    \[
        (M + N)(q, q') = M(q, q') + N(q, q')
    \]

    When $Q$ is finite, matrix-matrix and matrix-vector multiplication works as usual: for $Q$-matrices $M$ and $N$ and $Q$-vectors $v$, we have that $M \cdot N$ and $M \cdot v$ are respectively the $Q$-matrix and $Q$-vector defined by:
    \begin{mathpar}
    (M \cdot N)(q, q') = \sum_{q''} M(q, q'') \cdot N(q'', q')
    \and
    (M \cdot v)(q) = \sum_{q'} M(q, q') \cdot v(q')
    \end{mathpar}
    Note how the sums on the right-hand side are finite, because $Q$ is finite.

    Finally, given a $Q$-vector $v$ and $x \in \mathcal{K}$, we write $v \fatsemi x$ for the \emph{(right) scalar multiplication} of $v$ by $x$, which is simply the $Q$-vector where
    \[
        (v \fatsemi x)(q) = v(q) \cdot x
    \]
\end{definition}

\begin{lemma}%
\label{lemma:matrix-semiring}
Let $X$, $Y$ and $Z$ be finite $Q$-matrices.
We write $\mathbf{1}$ for the identity $Q$-matrix, given by $\mathbf{1}(q, q) = [q = q']$, where we use $[q = q']$ as a shorthand for $1$ when $q = q'$, and $0$ otherwise.\footnote{%
    We will use this \emph{Iverson bracket} in the sequel, too.
}
Furthermore, we write $\mathbf{0}$ for the $Q$-matrix where $\mathbf{0}(q, q') = 0$.
The following hold:
\begin{mathpar}
X + \mathbf{0} = X
\and
X + X = X
\and
X + Y = Y + X
\\
X + (Y + Z) = (X + Y) + Z
\and
X \cdot (Y \cdot Z) = (X \cdot Y) \cdot Z
\\
X \cdot (Y + Z) = X \cdot Y + X \cdot Z
\and
(X + Y) \cdot Z = X \cdot Z + Y \cdot Z
\\
X \cdot \mathbf{1} = X = X \cdot \mathbf{1}
\and
X \cdot \mathbf{0} = \mathbf{0} = \mathbf{0} \cdot X
\end{mathpar}
\end{lemma}
\begin{proof}
Each law can be proved by unfolding the definitions of the operators, and applying the corresponding law on the level of $\mathcal{K}$.
\Cref{\res{exercise:left-distributivity-matrix}} highlights the left distributivity law.
\end{proof}

Because the $+$ operation on matrices satisfies the same laws as expected from the $+$ in a Kleene algebra, we automatically get an order defined on $Q$-matrices in the same way: $M \leq N$ when $M + N = N$.

Next we show that finite matrices over a Kleene algebra satisfy the star axioms.
We gave a direct definition of the star of a matrix on page~\pageref{\res{star-matrices}}.
For our purposes it is more practical to define the star of a matrix using least solutions.
These two definitions can be shown to be equivalent~\cite[Theorem~A.3]{kozen-97-book}.
We start with a technical construction.

\begin{lemma}%
\label{lemma:least-fixpoint-matrix}
Let $Q$ be a finite set, with $M$ a $Q$-matrix and $b$ a $Q$-vector.
We can construct a $Q$-vector $s$ such that $b + M \cdot s \leq s$, and furthermore if $t$ is a $Q$-vector and $z \in \mathcal{K}$ with $b \fatsemi z + M \cdot t \leq t$, then $s \fatsemi z \leq t$.
\end{lemma}
\begin{proof}
    Let us first look at the special case where $Q$ is a singleton set.
    Here, $M$ and $b$ contain just one element of $\mathcal{K}$ --- so we treat them as such.
    Moreover, addition and multiplication comes down to adding and multiplying single elements.
    Thus, we are really looking for $s \in \mathcal{K}$ where $b + M \cdot s \leq s$, and if $t \in \mathcal{K}$ such that $b \cdot z + M \cdot t \leq t$, then $s \cdot z \leq t$.
    But we already know exactly such an expression: $M^* \cdot b$.
    After all,
    \begin{mathpar}
    b + M \cdot M^* \cdot b = (1 + M \cdot M^*) \cdot b = M^* \cdot b
    \and
    b \cdot z + M \cdot t \leq t \implies M^* \cdot b \cdot z \leq t
    \end{mathpar}
    Our job is to generalize this approach to sets $Q$ that contain \emph{more} elements.
    To this end, we proceed by induction on (the size of) $Q$.

    In the base, where $Q = \emptyset$, we can simply choose $s$ to be the unique $Q$-vector $s\colon \emptyset \to \mathcal{K}$, which satisfies both conditions vacuously.
    For the inductive step, let $p \in Q$ and choose $Q' = Q \setminus \{ p \}$.
    We are going to create a $Q'$-vector and $Q'$-matrix, and then use the induction hypothesis to obtain a $Q'$-vector; next, we will extend this $Q'$-vector into a $Q$-vector satisfying our objectives.

    Let $M'$ and $b'$ respectively be the $Q'$-matrix and $Q'$-vector given by
    \begin{mathpar}
        M'(q, q') = M(q, q') + M(q, p) \cdot {M(p, p)}^* \cdot M(p, q')
        \and
        b'(q) = b(q) + M(q, p) \cdot {M(p, p)}^* \cdot b(p)
    \end{mathpar}

    The intuition behind these choices is as follows.
    Let's think of $M$ and $b$ as representing a generalized ``automaton'', where $M(q, q')$ holds regular expressions for the words that can be read while moving from $q$ to $q'$ in one ``step'', and $b(q)$ holds regular expressions for the words that can be read upon halting in $q$.
    From this perspective, $M'$ and $b'$ represent an automaton meant to read the same words, just with $p$ removed.
    To achieve this, $M'(q, q')$ holds all of the words that could already be read when stepping from $q$ to $q'$ --- $M(q, q')$ --- plus the words that could be read by moving from $q$ to $p$, looping at $p$ some number of times, and then moving to $q'$ --- $M(q, p) \cdot M(p, p)^* \cdot M(p, q')$.
    Similarly, $b'(q)$ is a regular expression for the words that could already be accepted at $q$ --- $b(q)$ --- in addition to the words that could be accepted by transitioning to $p$, looping there for a while, and ultimately accepting --- $M(q, p) \cdot M(p, p)^* \cdot b(p)$.
    Through this lens, the construction mirrors precisely the classical ``state elimination'' method to convert a finite automaton to a regular expression~\cite{mcnaughton-yamada-1960}.

    By induction we obtain a $Q'$-vector $s'$ such that $b' + M' \cdot s' \leq s'$, and moreover if $t'$ is a $Q'$-vector such that $b' \fatsemi z + M' \cdot t' \leq t'$, then $s' \fatsemi z \leq t'$.

    We extend $s$ to a $Q$-vector as follows:
    \[
        s(q) =
            \begin{cases}
            s'(q) & q \in Q' \\
            {M(p, p)}^* \cdot \left( b(p) + \sum_{q' \in Q'} M(p, q') \cdot s'(q') \right) & q = p
            \end{cases}
    \]
    We leave the proof that $s$ satisfies our goal as \Cref{\res{exercise:matrix-fixpoint-vector}}.
\end{proof}

This construction then immediately gives us a way to define the star of a matrix, which we record in the following lemma.

\begin{lemma}[Star of a matrix]%
\label{lemma:matrix-star}
Let $Q$ be a finite set, and let $M$ be a $Q$-matrix.
We can construct a matrix $M^*$ such that the following hold:
\begin{enumerate}[label={(\roman*)}]
    \item
    if $s$ and $b$ are $Q$-vectors s.t.\ $b + M \cdot s \leq s$, then $M^* \cdot b \leq s$; and

    \item
    $\mathbf{1} + M \cdot M^* = M^*$, where $\mathbf{1}$ is as in \Cref{\res{lemma:matrix-semiring}}.
\end{enumerate}
\end{lemma}
\begin{proof}

For each $q \in Q$, we write $u_q$ for the $Q$-vector where $u_q(q') = [q = q']$.
Furthermore, we write $s_q$ for the $Q$-vector such that $u_q + M \cdot s_q \leq s_q$, and if $z \in \mathcal{K}$ and $t$ is a $Q$-vector such that $u_q \fatsemi z + M \cdot t \leq t$, then $s_q \fatsemi z \leq t$; note that we can construct this $Q$-vector, per \Cref{\res{lemma:least-fixpoint-matrix}}.

We now choose the $Q$-matrix $M^*$ as follows:
\[
    M^*(q, q') = s_{q'}(q)
\]
It remains to show that $M^*$ satisfies the two requirements above.
\begin{enumerate}[label={(\roman*)}]
    \item
    Suppose that $s$ and $b$ are $Q$-vectors such that $b + M \cdot s \leq s$.
    We need to show that $M^* \cdot b \leq s$, that is show that for all $q \in Q$:
    \[
        \sum_{q' \in Q} M^*(q, q') \cdot b(q') \leq s(q)
    \]
    It suffices to show that, for all $q, q' \in Q$, we have $s_{q'}(q) \cdot b(q') \leqq s(q)$, that is $s_{q'} \fatsemi b(q') \leq s$.
    By construction of $s_{q'}$, it then suffices to show that $u_{q'} \fatsemi b(q') + M \cdot s \leq s$.
    To this end, we derive:
    \[
        (u_{q'} \fatsemi b(q') + M \cdot s)(q)
            \leq (b + M \cdot s)(q)
            \leq s(q)
    \]
    which completes this part of the proof.

    \item
    For the second claim, we first note that for $q \in Q$, we have
    \[
        u_q + M \cdot (u_q + M \cdot s_q)
            \leq u_q + M \cdot s_q
    \]
    and hence $s_q \leq u_q + M \cdot s_q$, meaning that $u_q + M \cdot s_q = s_q$.

    Now, let $q, q' \in Q$; we derive as follows:
    \begin{align*}
    &\phantom{=}  (\mathbf{1} + M \cdot M^*)(q, q')\\
        &= \mathbf{1}(q, q') + \sum_{q'' \in Q} M(q, q'') \cdot M^*(q'', q') \tag{by def.} \\
        &= [q = q'] + \sum_{q'' \in Q} M(q, q'') \cdot s_{q'}(q'') \tag{def.\ $u_{q'}, M^*$} \\
        &= (u_{q'} + M \cdot s_{q'})(q) \tag{by def.} \\
        &= s_{q'}(q) \tag{derivation above} \\
        &= M^*(q, q') \tag*{(def.\ $M^*$) \qedhere}
    \end{align*}
\end{enumerate}
\end{proof}

The left fixpoint rule now follows (cf.\ \Cref{\res{exercise:matrix-fixpoint-matrix}}).
\begin{corollary}%
    \label{corollary:matrix-star-fixpoint}
    Let $Q$ be a finite set, and let $M$ and $B$ be $Q$-matrices.
    If $S$ is a $Q$-matrix such that $B + M \cdot S \leq S$, then $M^* \cdot B \leq S$.
\end{corollary}

A proof of the remaining axioms --- i.e., that (1)~$\mathbf{1} + M^* \cdot M = M^*$ and (2)~if $B + S \cdot M \leq S$ then $B \cdot M^* \leq S$ --- requires some thought.
We leave these to the reader in \Cref{\res{exercise:other-star}}, and record the following:
\begin{theorem}
The addition, multiplication and star operations on matrices satisfy all of the laws of Kleene algebra.
In particular, in addition to the laws listed in \Cref{\res{lemma:matrix-semiring}}, if $X$, $Y$ and $Z$ are $Q$-matrices for some finite set $Q$, then
\begin{mathpar}
    \mathbf{1} + X \cdot X^* = X^*
    \and
    X + Y \cdot Z \leq Z \implies Y^* \cdot Z \leq Z
    \\
    \mathbf{1} + X^* \cdot X = X^*
    \and
    X + Y \cdot Z \leq Y \implies X \cdot Z^* \leq Y
\end{mathpar}
\end{theorem}

\subsection{Infinite matrices}

In the proof of completeness to come, we will model finite automata using finite matrices, and their corresponding expressions using the Kleene star of those matrices.
However, we will also need to talk about potentially infinite automata.
It is not at all clear that the operations we defined on finite matrices still make sense for infinite ones, and if they do, whether the same laws apply.
In this subsection, we will study infinite matrices over a Kleene algebra, and argue that the same equations hold when the operations can be extended to infinite matrices.

The first operator to look at is addition.
Fortunately, there is no problem here at all: because addition of matrices is defined pointwise, it can be extended immediately to infinite matrices.
Furthermore, the laws that involve only addition, such as associativity, still work for the same reason that they do for finite matrices.

Next, we should look at multiplication.
Here, we encounter a problem when we try to extend the old definition.
Specifically, let $Q$ be infinite, and let $M, N\colon Q \times Q \to \mathcal{K}$.
If we try to define $M \cdot N$ as usual, i.e.:
\[
    (M \cdot N)(q, q') = \sum_{q'' \in Q} M(q, q'') \cdot N(q'', q')
\]
we risk the possibility that the sum on the right is infinite!
This would be ill-defined, because we only have finitary addition in $\mathcal{K}$.
To work around this problem, we introduce the following notion.

\begin{definition}[Locally-finite matrix]
Let $M\colon Q \times Q \to \mathcal{K}$ be a matrix for a possibly infinite set $Q$.
When $q \in Q$, we write $R_M(q)$ for the \emph{reach} of $q$, which is the smallest set satisfying the following rules:
\begin{mathpar}
    \inferrule{~}{%
        q \in R_M(q)
    }
    \and
    \inferrule{%
        q' \in R_M(q) \\
        M(q', q'') \neq 0
    }{%
        q'' \in R_M(q)
    }
\end{mathpar}
That is, if we regard $M$ as the adjacency matrix of a directed graph, then $R_M(q)$ is the set of vertices that can be reached from the vertex $q$.

We say that $M$ is \emph{locally-finite} when for each $q \in Q$, $R_M(q)$ is finite.
\end{definition}

\begin{example}
Every finite matrix is locally finite.
Examples of infinite but locally-finite matrices include the units $\mathbf{0}, \mathbf{1}\colon Q \times Q \to \mathcal{K}$; after all, $R_\mathbf{0}(q) = \emptyset$ for all $q$, and $R_\mathbf{1}(q) = \{ q \}$ for all $q$.
\end{example}

\begin{example}
Consider the matrix $M_{\Exp/_\equiv}$ over $\equiv$-equivalence classes of expressions given by $M_{\Exp/_\equiv}([e], [f]) = \sum \{ \ltr{a} \mid e_\ltr{a} \equiv f \}$.
This matrix is locally finite by \Cref{\res{theorem:brzozowski-finite}} and the observation that $R_{M_{\Exp/_\equiv}}([e]) = \{ [e_w] \mid w \in \alphabet^* \}$; it will play a role in the completeness proof to follow.
\end{example}

Matrix multiplication makes sense when the left operand is locally finite.
Specifically, if $M(q, q'') \cdot N(q'', q') \neq 0$, then certainly $M(q, q'') \neq 0$, and if this is the case then $q'' \in R_M(q)$; because there are only finitely many such $q''$ when $M$ is locally-finite, there are only finitely many non-zero terms that contribute to the sum on the right.

Because matrix multiplication extends to some infinite matrices with the same definition, the same laws apply when both sides of the equation are defined.
For instance, if $X, Y, Z\colon Q \times Q \to \mathcal{K}$ are matrices such that $X$, $Y$ and $X \cdot Y$ are locally finite, then $X \cdot (Y \cdot Z) = (X \cdot Y) \cdot Z$.

\begin{remark}
As can be seen from the above, some care is necessary when stating equations involving operations on infinite matrices.
For one thing, it is not sufficient to assume that $X$ and $Y$ are locally finite, because $X \cdot Y$ may not be.
Take for instance $Q = \mathbb{N}$ and set
\begin{mathpar}
X(2k, 2k+1) = 1
\and
Y(2k+1, 2k+2) = 1
\end{mathpar}
with all other entries of $X$ and $Y$ being zero.
In that case, $X$ and $Y$ are locally finite but $X \cdot Y$ is not (nor is $X + Y$, for that matter).

For the sake of brevity, we will omit such qualifications from equalities involving infinite matrices going forward, although we will remark on when certain expressions are defined, just to be rigorous.
\end{remark}

For the Kleene star, we also limit ourselves to locally finite matrices.
Intuitively, this makes sense because the Kleene star of a matrix corresponding to an automaton can be used to calculate a regular expression for the language of each state in that automaton.
If the Kleene star of a matrix could be extended to infinite matrices in general, it would mean that we could calculate regular expressions for \emph{all} infinite automata --- which would be problematic.
On the other hand, locally finite matrices correspond to (potentially infinite) automata in which every state can reach only finitely many states; it should be clear that in that setting, it is still possible to compute a regular expression for every state.

\begin{definition}
Let $M\colon Q \times Q \to \mathcal{K}$ be a matrix, with $Q$ potentially infinite.
When $P \subseteq Q$, we write $M_P$ for the $P$-indexed submatrix of $M$, i.e., $M_P\colon P \times P \to \mathcal{K}$ with $M_P(q, q') = M(q, q')$ for $q, q' \in P$.
This notation also extends to vectors, i.e., given a $Q$-vector $b\colon Q \to \mathcal{K}$, we write $b_P$ for the $P$-vector given by $b_P(q) = b(q)$.

If $M$ is locally finite, we write $M^*$ for the $Q$-indexed matrix where
\[
M^*(q, q') =
    \begin{cases}
    {(M_{R_M(q)})}^*  (q,q') & q' \in R_M(q)\\
    0 & q' \not\in R_M(q)
    \end{cases}
\]
Note that when $M$ is locally finite, $R_M(q)$ is finite for every $q \in Q$, which means that $M_{R_M(q)}$ is a finite matrix, hence ${(M_{R_M(q)})}^*$ is defined.
\end{definition}

It remains to verify the laws involving the Kleene star.
Let us start with $\mathbf{1} + M \cdot M^* = M^*$.
We will need the following technical property, which says that, under certain conditions on $P$, its the star of its $P$-indexed submatrix is the $P$-indexed submatrix of its star.
\begin{lemma}%
\label{lemma:star-commute-restrict}
Let $M\colon Q \times Q \to \mathcal{K}$ be a finite matrix, and let $P \subseteq Q$ such that if $q \in P$ and $M(q, q') \neq 0$, then $q' \in P$.
Now $(M_P)^* = (M^*)_P$.
\end{lemma}
\begin{proof}
To see that $(M_P)^* \leq (M^*)_P$, unfolding the star on the left, it suffices to show that $M_P \cdot (M^*)_P + \mathbf{1} \leq (M^*)_P$.
To this end, we derive as follows, for $q, q' \in P$:
\begin{align*}
(M_P \cdot (M^*)_P + \mathbf{1})(q, q')
    &= \sum_{q'' \in P} M_P(q, q'') \cdot (M^*)_P(q'', q') + [q = q'] \\
    &= \sum_{q'' \in P} M(q, q'') \cdot M^*(q'', q') + [q = q'] \\
    &\leq \sum_{q'' \in Q} M(q, q'') \cdot M^*(q'', q') + [q = q'] \\
    &= (M \cdot M^* + \mathbf{1})(q, q') \\
    &= M^*(q, q') = (M^*)_P(q, q')
\end{align*}
To show that $(M^*)_P \leq (M_P)^*$, we need a slightly more involved argument.
Let $N\colon Q \times Q \to \mathcal{K}$ be the matrix defined by
\[
    N(q, q') =
        \begin{cases}
        (M_P)^*(q, q') & q, q' \in P \\
        M^*(q, q') & \text{otherwise}
        \end{cases}
\]
It now suffices to show that $M^* \leq N$, because $N_P = (M_P)^*$ by construction.
To this end, we need only show that $M \cdot N + \mathbf{1} \leq N$.
There are two cases; first, if $q, q' \in P$, then we derive
\begin{align*}
(M \cdot N + \mathbf{1})(q, q')
    &= \sum_{q'' \in Q} M(q, q'') \cdot N(q'', q') + [q = q'] \\
    &= \sum_{q'' \in P} M_P(q, q'') \cdot (M_P)^*(q'', q') + [q = q'] \\
    &= (M_P \cdot (M_P)^* + \mathbf{1})(q, q') \\
    &= (M_P)^*(q, q') = N(q, q')
\end{align*}
Here, the second step is sound because if $q \in P$ and $M(q, q'') \neq 0$, then $q'' \in P$ as well.
For the other case, where $q \not\in P$ or $q' \not\in P$, we find that
\begin{align*}
(M \cdot N + \mathbf{1})(q, q')
    &= \sum_{q'' \in Q} M(q, q'') \cdot N(q'', q') + [q = q'] \\
    &\leq \sum_{q'' \in Q} M(q, q'') \cdot M^*(q'', q') + [q = q'] \\
    &= (M \cdot M^* + \mathbf{1})(q, q') \\
    & = M^*(q, q') = N(q, q')
\end{align*}
In the second step, we use $N \leq M^*$; after all, if $q, q' \in P$ then $N(q, q') = (M_P)^*(q, q') \leq (M^*)_P(q, q') = M^*(q, q')$ (by the derivation above), and if $q \not\in P$ or $q' \not\in P$, then $N(q, q') = M^*(q, q')$ by definition.
\end{proof}

Because of \Cref{\res{lemma:star-commute-restrict}}, we can write $M_P^*$ instead of $(M_P)^*$ or $(M^*)_P$ without ambiguity, provided that $P$ satisfies the condition specified.
With this in mind, we can now verify the first equation.

\begin{lemma}%
\label{lemma:matrix-unroll}
If $M\colon Q \times Q \to \mathcal{K}$ is locally finite, then $\mathbf{1} + M\cdot M^* = M^*$.
\end{lemma}
\begin{proof}
We prove only that $\mathbf{1} + M \cdot M^* \leq M^*$; the converse inequality follows from the fixpoint rule for matrices, proved in \Cref{\res{lemma:matrix-fixpoint}} without relying on the claim at hand.
It thus remains to show that for $q, q' \in Q$, we have that $(\mathbf{1} + M \cdot M^*)(q, q') \leqq M^*(q, q')$.

When $q' \not\in R_M(q)$, we have that $M^*(q, q') = 0$ by definition.
Fortunately, this also means that $q' \neq q$, and that $q' \not\in R_M(q'')$ for any $q'' \in R_M(q)$ --- after all, if either were the case, then $q' \in R_M(q)$ by definition of $R_M$.
This then means that $(\mathbf{1} + M \cdot M^*)(q, q') = 0$ as well.

We thus focus on the case where $q' \in R_M(q)$, and derive:
\begin{align*}
    & (\mathbf{1} + M \cdot M^*)(q, q') \\
    &= [q = q'] + \sum_{q'' \in Q} M(q, q'') \cdot M^*(q'', q') \\
    &= [q = q'] + \sum_{q'' \in R_M(q)} M_{R_M(q)}(q, q'') \cdot M^*(q'', q') \\
    &= [q = q'] + \sum_{\substack{q'' \in R_M(q) \\ \text{s.t. $q' \in R_M(q'')$}}} M_{R_M(q)}(q, q'') \cdot (M_{R_M(q'')})^*(q'', q')
\\
    &\leq [q = q'] + \sum_{q'' \in R_M(q)} M_{R_M(q)}(q, q'') \cdot (M_{R_M(q)})^*(q'', q') \tag{\textdagger} \\
    &= (\mathbf{1} + M_{R_M(q)} \cdot (M_{R_M(q)})^*)(q, q')
     = (M_{R_M(q)})^*(q, q') = M^*(q, q')
\end{align*}
In the step marked with (\textdagger), we use that because $q'' \in R_M(q)$, it holds that $R_M(q'') \subseteq R_M(q)$, so if $q'' \in R_M(q)$ with $q' \in R_M(q'')$, then
\begin{align*}
    (M_{R_M(q'')})^*(q'', q')
        &= ((M_{R_M(q)})_{R_M(q'')})^*(q'', q') \\
        &= ((M_{R_M(q)})^*)_{R_M(q'')}(q'', q') \tag{\Cref{\res{lemma:star-commute-restrict}}} \\
        &= (M_{R_M(q)})^*(q'', q') \tag*{\qedhere}
\end{align*}
\end{proof}

We move on to the proof of the left fixpoint rule: i.e., that $B+MS\leq S$ implies $M^*B\leq S$.
As before, we instead prove a variant of the claim for vectors.
The generalisation to matrices then follows.

\begin{lemma}%
\label{lemma:matrix-fixpoint}
Let $M\colon Q \times Q \to \mathcal{K}$ be locally finite, and let $b, s\colon Q \to \mathcal{K}$.
If $b + M \cdot s \leq s$, then $M^* \cdot b \leq s$ as well.
\end{lemma}
\begin{proof}
To prove that $M^* \cdot b \leq s$, we should show that for every $q \in Q$,
\[
    \sum_{q' \in Q} M^*(q,q') \cdot b(q')\leq s(q)
\]
For all $q' \not\in R_M(q)$ we have $M^*(q,q') = 0 \leq s(q)$.
So the above is equivalent to showing that for all $q \in Q$, we have that
\[
    \sum_{q'\in R_M(q)} (M_{R_M(q)})^*(q,q') \cdot b_{R_M(q)}(q') \leq s_{R_M(q)}(q)
\]
which is true if for $q \in Q$, we have that $(M_{R_M(q)})^* \cdot b_{R_M(q)} \leq s_{R_M(q)}$.
Since this is an inequation of \emph{finite} vectors, the latter holds when $b_{R_M(q)} + M_{R_M(q)} \cdot s_{R_M(q)} \leq s_{R_M(q)}$ (cf.\ \Cref{\res{lemma:matrix-star}}).
This follows from our assumption that $b + M \cdot s \leq s$ by restricting both sides to $R_M(q)$.
\end{proof}

The right-handed laws for $M^*$, i.e., that (1)~$M^* \cdot M + \textbf{1} = M^*$, (2)~and if $B + S \cdot M \leq S$ then $B \cdot M^* \leq S$, can be derived in the same way we derived these for the finite case --- see \Cref{\res{exercise:other-star}}.

This concludes our study of the laws of Kleene algebra for matrices indexed by infinite sets; we will leverage these in the sections to come.

\section{Soundness}%
\label{section:soundness}

For the left to right implication of \Cref{\res{thm:sc}} (soundness), it is enough to prove that $\equiv$ is a bisimulation relation --- after all, since $\bisim$ is the largest bisimulation, this would imply that $\equiv$ is a subset of $\bisim$.
This will have as consequence that the Kleene algebra axiomatization of regular
expressions is sound: for all $e_1,e_2\in \Exp $, if $ e_1 \equiv e_2$ then
$e_1 \bisim e_2$.
\begin{theorem}[Soundness]\label{thm:ra_sound}
Provable equivalence is a bisimulation on $(\Exp, <o, t>)$, that is, for every $e_1,e_2\in \Exp $, if $ e_1 \equiv e_2$ then
\[
\ore(e_1) = \ore(e_2) \text{ and for all $\ltr{a} \in \alphabet$, }  (e_1)_{\ltr{a}} \equiv (e_2)_{\ltr{a}}
\]
\end{theorem}
\begin{proof}
By induction on the length of derivations for $ e_1 \equiv e_2$.

For the base cases, i.e., all but the last two equations in \Cref{\res{def:ka}}, everything follows from unwinding the definitions of $\ore$ and $\tre$.
We show the proof for the equation $e_1+e_2 \equiv e_2+e_1$:
\begin{align*}
\ore(e_1+e_2)
    &= \ore(e_1) \vee \ore(e_2)
     = \ore(e_2+e_1)\\
(e_1 + e_2)_{\ltr{a}}
    &\equiv (e_1)_{\ltr{a}} + (e_2)_{\ltr{a}}
     \equiv (e_2)_{\ltr{a}} + (e_1)_{\ltr{a}}
     \equiv (e_2 + e_1)_{\ltr{a}}
\end{align*}
For good measure, here is the proof for $ee^* +  1 \equiv e^*$:
\begin{align*}
     \ore(ee^* +  1)
    &= \ore(ee^*) \vee \ore(1)
     =  \ore(ee^*) \vee 1
     = 1
     = \ore(e^*)\\
(ee^* +  1)_{\ltr{a}}
    &\equiv (ee^*)_{\ltr{a}} +  1_{\ltr{a}}\\
    &\equiv e_{\ltr{a}} e^* +  {\ore(e)}(e^*)_{\ltr{a}} + 0 \\
    &\equiv
        \begin{cases}
            e_{\ltr{a}} e^* +  0 (e^*)_{\ltr{a}} + 0 & \text{if }\ore(e) = 0 \\
            e_{\ltr{a}} e^* + 1 (e^*)_{\ltr{a}} + 0 & \text{otherwise}
        \end{cases}\\
        &\equiv
        \begin{cases}
            (e^*)_{\ltr{a}} + 0 (e^*)_{\ltr{a}} + 0& \text{if }\ore(e) = 0 \\
            (e^*)_{\ltr{a}} + 1 (e^*)_{\ltr{a}} + 0 & \text{otherwise}
        \end{cases}\\
    &\equiv (e^*)_{\ltr{a}}
\end{align*}
In the inductive step, we illustrate the case where $e_2^*e_1 \leqq f$ given $e_1 + e_2 f \leqq f$; the other case is proved precisely in the same way.
Suppose we have just derived $e_2^*e_1 \leqq f$, using $e_1 + e_2 f \leqq f$ as a premise.
Our induction hypothesis is the claim applied to this premise, so $o(e_1 + e_2f) \leq o(f)$ and $(e_1 + e_2f)_{\ltr{a}} \leqq f_{\ltr{a}}$.
We want to prove that $\ore(e_2^*e_1) \leq \ore(f)$ and $(e_2^*e_1)_{\ltr{a}} \leqq f_{\ltr{a}}$.
The first inequality simplifies to $\ore(e_1) \leq \ore(f)$, which follows from the first part of the induction hypothesis.
For the second inequality, we derive as follows:
\begin{align*}
 (e_2^*e_1)_{\ltr{a}} &= (e_2)_{\ltr{a}} e_2^*e_1 +(e_1)_{\ltr{a}}\\
&\leqq (e_2)_{\ltr{a}} f + (e_1)_{\ltr{a}} & \tag{\ensuremath{e_2^*e_1\leqq f}}\\
&\leqq (e_2)_{\ltr{a}} f + {\ore(e_2)}f_{\ltr{a}} + (e_1)_{\ltr{a}} & \\
&\equiv (e_2f)_{\ltr{a}} +  (e_1)_{\ltr{a}}\\
&\equiv (e_1+e_2f)_{\ltr{a}} \\
&\leqq f_{\ltr{a}} & \tag{induction hypothesis}
\end{align*}
This completes the proof.
\end{proof}

\section{The road to completeness}

We now turn our attention to the central result of this chapter, namely \emph{completeness} --- the converse direction of \Cref{\res{thm:sc}}.
Our proof of this fact will involve notions from coalgebra, discussed in \Cref{\res{chapter:coalgebra}} and the blue remarks in \Cref{\res{chapter:automata}}.
Readers who skipped those parts are invited read them first, before returning back here.

Soundness (Theorem~\ref{thm:ra_sound}) tells us that $\equiv$ is a bisimulation on $\Exp$.
\Cref{\res{lemma:quotient-coalgebra}} then gives us a coalgebra
\(
    <\overline\ore,\overline\tre> \colon \Exp /_\equiv \to 2\times (\Exp /_\equiv)^\alphabet
\)
on $\Exp/_\equiv$ such that the quotient map $[-]\colon \Exp \to \Exp/_\equiv$ given by $[e] = \{e' \mid e \equiv e'\}$ is a coalgebra homomorphism.
In concrete terms, $\overline\ore$ and $\overline\tre$ work as expected: given $e \in \Exp$, $\overline\ore([e]) = \ore(e)$ and $\overline\tre([e])(\ltr{a}) = [e_{\ltr{a}}]$.

The proof of completeness that follows is inspired by a proof due to Jacobs~\cite{jacobs-2006}, which can be seen as a coalgebraic
review of Kozen's proof~\cite{kozen-2001}, though we present some lemmas (and respective proofs) slightly differently.
We start with an overview of all the necessary steps.

\medskip
\noindent\textbf{Step }\mycirc{1}
First, let $L\colon \Exp/_\equiv \to 2^{\alphabet^*}$ be the final homomorphism, and let $(I, <o_I, t_I>)$ be the $D$-coalgebra obtained by quotienting $(\Exp/_\equiv, <\overline\ore,\overline\tre>)$ using $\mathsf{Ker}(L)$, per \Cref{\res{lemma:quotient-coalgebra}}.
By \Cref{\res{lemma:kernel-subcoalgebra}}, this brings us the situation below, with $q\colon \Exp/_\equiv \to I$ being the quotient map, and $m$ being the injective homomorphism from $(I, <o_I, t_I>)$ to $(2^{\alphabet^*}, <o_L, t_L>)$.
\[
    \begin{tikzpicture}
        \node (QExp) {$\Exp /_\equiv$};
        \node[right=2cm of QExp] (I) {$I$};
        \node[right=2cm of I] (lang) {$2^{\alphabet^*}$};
        \node[below=4mm of QExp] (FQExp) {$2\times (\Exp/_\equiv )^\alphabet$};
        \node at (FQExp -| I) (FI) {$2\times I^\alphabet$};
        \node at (FI -| lang) (Flang) {$2\times (2^{\alphabet^*})^\alphabet$};

        \draw[->] (QExp) edge node[above] {\footnotesize $q$} (I);
        \draw[->] (QExp) edge node[left] {\footnotesize $<\overline\ore,\overline\tre>$} (FQExp);
        \draw[->] (FQExp) edge node[below] {\footnotesize $D(q)$} (FI);
        \draw[->] (I) edge node[right] {\footnotesize $<o_I,t_I>$} (FI);
        \draw[->] (I) edge node[above] {\footnotesize $m$} (lang);
        \draw[->] (FI) edge node[below] {\footnotesize $D(m)$} (Flang);
        \draw[->] (lang) edge node[right] {\footnotesize $<o_L, t_L>$} (Flang);
        \draw[->] (QExp) edge[bend left,looseness=0.4] node[above] {\footnotesize $L$} (lang);
    \end{tikzpicture}
\]
\noindent\textbf{Step }\mycirc{2} We prove that $q$ is a bijection.
The key idea is that $(\Exp /_\equiv,<\overline\ore,\overline\tre>)$ and $(I,<o_I,t_I>)$ are final among automata where the reach of each state is finite.
By \Cref{\res{theorem:brzozowski-finite}}, $(\Exp/_\equiv, <\overline\ore, \overline\tre>)$ is such a coalgebra.

We prove that $(\Exp /_\equiv
,<\overline\ore,\overline\tre>)$ is final using three intermediate steps:
\begin{itemize}[align=left]
\item[(\Cref{\res{lemma:exist_sol}})] For any locally finite automaton $(S,<o_S,t_S>)$, there exists a homomorphism $\ceil{-}_S \colon (S,<o_S,t_S>) \to
( \Exp /_\equiv, <\overline\ore,\overline\tre>)$.
\item[(\Cref{\res{lemma:sol_hom}})] If $f$ is a homomorphism from $(S,<o_S,t_S>)$ to $(T,<o_T,t_T>)$ and both are locally finite, then it holds that $\ceil{f(s)}_T = \ceil{s}_S$.
\item [(\Cref{\res{lemma:sol_exp_id}})] The homomorphism $\ceil{-}_{\Exp /_\equiv}$ is the
identity.
\end{itemize}
These three lemmas together imply that for any locally finite automaton $(S,<o_S,t_S>)$, there exists a \emph{unique} coalgebra homomorphism\label{claim:locally-finite-unique-morphism}
 \[
    \ceil{-}_S \colon (S,<o_S,t_S>) \to ( \Exp /_\equiv, <\overline\ore,\overline\tre>)
\]
Hence, $(\Exp /_\equiv ,<\overline\ore,\overline\tre>)$ is final among locally finite coalgebras.

With these facts in hand, the finality of $(I, <o_I, t_I>)$ is easy by comparison.
We just need to observe the following two properties:
\begin{itemize}
\item[(i)] For any locally finite automaton $(S,<o_S,t_S>)$, we can construct a coalgebra homomorphism into $(I, <o_I, t_I>)$ as follows:
\[
    (S, <o_S, t_S>) \xrightarrow{\ceil{-}_S} (\Exp/_\equiv, <\overline\ore, \overline\tre>) \xrightarrow{q} (I, o_I, t_I>)
\]
Here, $\ceil{-}_S$ is as in the statement of \Cref{\res{lemma:exist_sol}}, and $q$ is the quotient morphism from the diagram above.
\item[(ii)]
If there exist two homomorphisms $f,g\colon  (S,<o_S,t_S>) \to (I, <o_I, t_I>)$, then $f = g$.
After all, by finality of $2^{\alphabet^*}$, we know that, for any $s\in S$, $L(s) = m\circ g (s)$ and $L(s) = m\circ f (s)$.
Hence, we have that $m \circ f= m\circ g$ and, since $m$ is injective, $f=g$.
\end{itemize}
\noindent\textbf{Step }\mycirc{3}
From \mycirc{1} and \mycirc{2}, $q\colon (\Exp/_\equiv, <\overline\ore, \overline\tre>) \to (I, <o_I,t_I>)$ is an isomorphism, and hence injective.
This makes $L = m \circ q \colon \Exp /_\equiv \to 2^{\alphabet^*}$ injective, which is the key to prove completeness.

\begin{theorem}[Completeness]
If $e_1 \bisim e_2$ then $ e_1 \equiv e_2$.
\end{theorem}
\begin{proof}
For all $e_1,e_2\in \Exp $, if $e_1\bisim e_2$ then they are mapped into the same element in the final coalgebra: $L(e_1)=L(e_2)$.
Since $[-]$ is a homomorphism, $L([e_1]) = L([e_2])$, where we use $L$ to denote the map into the final coalgebra both from $\Exp $ and $\mathcal
\Exp/_\equiv$.
By \mycirc{3}, $L$ is injective and thus $[e_1] = [e_2]$, that is, by definition of $[-]$, $ e_1 \equiv e_2$.
\end{proof}

\section{Filling in the details}

It remains to prove the properties of $\Exp/_\equiv$ alluded to above.
We start by constructing the desired map, and showing that it is a homomorphism.
In the following, we work with $Q$-matrices over $\Exp$; it should be clear that the Kleene star of a finite $Q$-matrix can still be defined syntactically; moreover, we write $M \equiv N$ for matrices over $\Exp$ to mean that $M$ and $N$ are pointwise equivalent (and similarly for vectors).

\begin{lemma}\label{lemma:exist_sol}
Let $(S, <o_S, t_S>)$ be a locally finite automaton.
There exists a coalgebra homomorphism $\ceil{-}_S\colon (S, <o_S, t_S>) \to (\Exp/_\equiv, <\overline{o}, \overline{t}>)$.
\end{lemma}
\begin{proof}
Let $M_S\colon Q \times Q \to \Exp$ be the transition matrix of $S$, given by
\[
    M_S(q, q') = \sum \{ \ltr{a} \in \alphabet \mid t_S(q)(\ltr{a}) = q' \}
\]
Because $(S, <o_S, t_S>)$ is locally finite, so is $M_S$.

If we view $o_S\colon Q \to 2$ as a vector of expressions, then $M_S^* o_S$ is another vector of expressions.
We choose $\ceil{q}_S = [(M_S^* \cdot o_S)(q)]$.
To see that $\ceil{-}_S$ is a homomorphism, we must prove two equalities:
\begin{mathpar}
    \overline{o}(\ceil{q}_S) = o_S(q)
    \and
    \overline{t}(\ceil{q}_S)(\ltr{a}) = \ceil{t_S(q)(\ltr{a})}_S
\end{mathpar}
For the first equality, we derive as follows:
\begin{align*}
    \overline{o}(\ceil{q}_S)
        &= o((M_S^* \cdot o_S)(q)) \\
        &= o(o_S(q) + (M_S \cdot M_S^* \cdot o_S)(q)) \\
        &= o(o_S(q)) \vee o((M_S \cdot M_S^* \cdot o_S)(q)) \\
        &= o(o_S(q)) \vee \bigvee_{q'} o(M_S(q, q')) \wedge o((M_S^* \cdot o_S)(q')) \\
        &= o(o_S(q)) \\
        &= o_S(q)
\end{align*}
where the second-to-last step holds because $o(M_S(q, q')) = 0$.

For the second equality, we start by noting the following:
\begin{align*}
    (M_S^*o_S(q))_\ltr{a}
        &\equiv (o_S(q) + (M_S \cdot M_S^* \cdot o_S)(q))_\ltr{a} \tag{\Cref{\res{lemma:matrix-unroll}}} \\
        &\equiv \Bigl( \sum_{q'} M_S(q, q') (M_S^* \cdot o_S)(q')\Bigr)_\ltr{a} \tag{\text{$o_S(q)_\ltr{a} = 0$}} \\
        &\equiv \sum_{q'} \Bigl( \Bigl( \sum \{ \ltr{b} \in \alphabet \mid t_S(q)(\ltr{b}) = q' \} \Bigr) \cdot (M_S^* \cdot o_S)(q')\Bigr)_\ltr{a}  \\
        &\equiv \sum \{ ( \ltr{b} \cdot (M_S^* \cdot o_S)(t_S(q)(\ltr{b})))_\ltr{a} \mid \ltr{b} \in \alphabet \} \\
        &\equiv (M_S^* \cdot o_S)(t_S(q)(\ltr{a}))
\end{align*}
In the last step, we use that for $\ltr{a}, \ltr{b} \in \alphabet$ and $e \in \Exp$, we have that $(\ltr{b} \cdot e)_\ltr{a} \equiv e$ when $\ltr{a} = \ltr{b}$, and $(\ltr{b} \cdot e)_\ltr{a} \equiv 0$ otherwise.
We then conclude:
\[
    \overline{t}(\ceil{q}_S)(\ltr{a})
        = [((M_S^* \cdot o_S)(q))_\ltr{a}]
        = [(M_S^* \cdot o_S)(t_S(q)(\ltr{a}))]
        = \ceil{t_S(q)(\ltr{a})}
    \qedhere
\]
\end{proof}

Also, homomorphisms from locally finite automata to $(\Exp/_\equiv, \overline\ore, \overline\tre)$ commute with homomorphisms among locally finite automata.

\begin{lemma}\label{lemma:sol_hom}
Let $(S, <o_S, t_S>)$ and $(T, <o_T, t_T>)$ be locally finite.
For any homomorphism $f\colon (S,<o_S,t_S>) \to  (T,<o_T,t_T>)$, $\ceil{f(s)}_T = \ceil{s}_S$.
\end{lemma}
\begin{proof}
The idea behind this proof is to encode $f$ in an $S$-by-$T$ (rectangular) matrix, and exploit properties of matrix multiplication~\cite{jacobs-2006}.
In the proof below, we therefore expand matrix multiplication to rectangular matrices in the obvious way; it is routine to verify that the laws of Kleene algebra still apply to well-defined expressions.

Let $M_f\colon S \times T \to \Exp$ be given by $M_f(q, q') = [f(q) = q']$.
We start by claiming that $M_f \cdot M_T \equiv M_S \cdot M_f$.
To see this, derive as follows:
\begin{align*}
    (M_f \cdot M_T)(q, q')
        &\equiv \sum_{q''} M_f(q, q'') \cdot M_T(q'', q') \\
        &\equiv M_T(f(q), q') \tag{def.\ $M_f$} \\
        &\equiv \sum \{ \ltr{a} \in \alphabet \mid t_T(f(q))(\ltr{a}) = q' \} \\
        &\equiv \sum \{ \ltr{a} \in \alphabet \mid f(t_S(q)(\ltr{a})) = q' \} \tag{homomorphism} \\
        &\equiv \sum_{q''} \Bigl\{ \sum \{ \ltr{a} \mid t_S(q)(\ltr{a}) = q'' \} \mid f(q'') = q' \Bigr\} \\
        &\equiv \sum_{q''} M_S(q, q'') \cdot M_f(q'', q') \\
        &\equiv (M_S \cdot M_f)(q, q')
\end{align*}
We can thus conclude from the fact that $M_f \cdot M_T \equiv M_S \cdot M_f$ that $M_f \cdot M_T^* \equiv M_S^* \cdot M_f$ (see \Cref{\res{exercise:bisimulation-matrices}}).
Next, we claim that $M_f \cdot o_T \equiv o_S$; this is easier to show:
\[
    (M_f \cdot o_T)(q)
        \equiv \sum_{q'} M_f(q, q') o_T(q')
        \equiv o_T(f(q))
        = o_S(q)
\]

Putting this together, we can derive that
\begin{align*}
    \ceil{f(s)}_T
        &= [(M_T^* \cdot o_T)(f(s))] \\
        &= [(M_f \cdot M_T^* \cdot o_T)(s)] \\
        &= [(M_S^* \cdot M_f \cdot o_T)(s)] \\
        &= [(M_S^* \cdot o_S)(s)] \\
        &= \ceil{s}_S
            \tag*{\qedhere}
\end{align*}
\end{proof}

To place the final piece of the puzzle, we need the following property of $\ceil{-}$; the proof is left as \Cref{\res{exercise:ceil-monotone}}.

\begin{lemma}%
\label{lemma:ceil-monotone}
Let $e, f \in \Exp$.
If $e \leqq f$, then $\ceil{[e]}_{\Exp/_\equiv} \leq \ceil{[f]}_{\Exp/_\equiv}$.
\end{lemma}

With this in hand, we can now prove the last lemma, which says that $\ceil{-}_{\Exp/_\equiv}\colon (\Exp/_\equiv, <\overline\ore, \overline\tre>) \to (\Exp/_\equiv, <\overline\ore, \overline\tre>)$ acts as the identity.

\begin{lemma}\label{lemma:sol_exp_id}
For all $e \in \Exp$, it holds that $\ceil{[e]}_{\Exp/_\equiv} = [e]$.
\end{lemma}
\begin{proof}
In the proof below, we treat $\Exp/_\equiv$ as a Kleene algebra, extending the operators on $\Exp$ to the equivalence classes; for instance, $[f] + [g] = [f + g]$.
This is well-defined because $\equiv$ is a congruence with respect to all operators.
To prevent some clutter, we will treat $0, 1 \in 2$ as equivalence classes of expressions.
We will also write $[f] \leq [g]$ when $f \leqq g$.
Note that this makes $\leq$ a partial order on $\Exp/_\equiv$.
To prove the claim, it now suffices to show that $\ceil{[e]}_{\Exp/_\equiv}([e]) \leq [e]$ and $[e] \leq \ceil{[e]}_{\Exp/_\equiv}([e])$

To show that $\ceil{[e]}_{\Exp/_\equiv} \leq [e]$, we can use $\id\colon \Exp/_\equiv \to \Exp/_\equiv$ as a vector.
If $M_{\Exp/_\equiv}$ is the matrix corresponding to $\overline{t}\colon \Exp/_\equiv \to {(\Exp/_\equiv)}^\alphabet$, and we think of $\overline{o}\colon \Exp/_\equiv \to 2$ as a vector, then we can derive as follows:
\begin{align*}
    (M_{\Exp/_\equiv} \cdot \id + \overline{o})([e])
        &= \sum_{[f]} M_{\Exp/_\equiv}([e], [f]) \cdot [f] + \overline{o}([e]) \\
        &= \sum_{[f]} \sum_{e_\ltr{a} \equiv f} [\ltr{a} \cdot f] + o(e) \\
        &= \sum_{\ltr{a}} [\ltr{a} \cdot e_\ltr{a}] + o(e) \\
        &= [e] \tag{\Cref{\res{thm:fundamental-re}}}
\end{align*}
It then follows that $\ceil{[e]}_{\Exp/_\equiv} = [(M_{\Exp/_\equiv}^* \cdot \overline{o})([e])] \leq [e]$.

In the other direction, we prove that for $e, f \in \Exp$, we have that $[e] \cdot \ceil{[f]}_{\Exp/_\equiv} \leq \ceil{[e \cdot f]}_{\Exp/_\equiv}$.
This implies the main claim, because
\[
    [e]
        = [e] \cdot [1]
        \leq [e] \cdot \ceil{[1]}_{\Exp/_\equiv}
        \leq \ceil{[e \cdot 1]}_{\Exp/_\equiv}
        = \ceil{[e]}_{\Exp/_\equiv}
\]

The proof proceeds by induction on $e$, while still quantifying over all $f \in \Exp$.
In the base, there are three cases to consider:
\begin{itemize}
    \item
    If $e = 0$, then the claim is trivial, because we have
    \[
        [0] \cdot \ceil{[f]}_{\Exp/_\equiv}
            = [0]
            \leq \ceil{[0 \cdot f]}_{\Exp/_\equiv}
    \]
    \item
    If $e = 1$, then the claim also trivializes, because
    \[
        [1] \cdot \ceil{[f]}_{\Exp/_\equiv}
            = \ceil{[f]}_{\Exp/_\equiv}
            = \ceil{[1 \cdot f]}_{\Exp/_\equiv}
    \]
    \item
    If $e = \ltr{a}$ for some $\ltr{a} \in \alphabet$, then note that $[(\ltr{a} \cdot f)_\ltr{a}] = [f]$, and so $[\ltr{a}] \leq M_{\Exp/_\equiv}([\ltr{a} \cdot f], [f])$.
    With this in mind, we can derive
    \begin{align*}
    [\ltr{a}] \cdot \ceil{[f]}_{\Exp/_\equiv}
        &= [\ltr{a}] \cdot (M_{\Exp/_\equiv}^* \cdot \overline{o})([f]) \\
        &\leq M_{\Exp/_\equiv}([\ltr{a} \cdot f], [f]) \cdot (M_{\Exp/_\equiv}^* \cdot \overline{o})([f]) \\
        &\leq (M_{\Exp/_\equiv}^* \cdot \overline{o})([\ltr{a} \cdot f]) \\
        &= \ceil{[\ltr{a} \cdot f]}_{\Exp/_\equiv}
    \end{align*}
\end{itemize}
For the inductive step, there are three more cases.
\begin{itemize}
    \item
    If $e = e_1 + e_2$, then by induction we know that for $i \in \{1,2\}$:
    \[
        [e_i] \cdot \ceil{[f]}_{\Exp/_\equiv} \leq \ceil{[e_i \cdot f]}_{\Exp/_\equiv}
    \]
    We can then derive that
    \begin{align*}
    [e] \cdot \ceil{[f]}_{\Exp/_\equiv}
        &= [e_1] \cdot \ceil{[f]}_{\Exp/_\equiv} + [e_2] \cdot \ceil{[f]}_{\Exp/_\equiv} \\
        &\leq \ceil{[e_1 \cdot f]}_{\Exp/_\equiv} + \ceil{[e_2 \cdot f]}_{\Exp/_\equiv} \tag{IH} \\
        &\leq \ceil{[e \cdot f]}_{\Exp/_\equiv} \tag{\Cref{\res{lemma:ceil-monotone}}}
    \end{align*}
    \item
    If $e = e_1 \cdot e_2$, then the following hold for all $f \in \Exp$ and $i \in \{1,2\}$:
    \[
        [e_i] \cdot \ceil{[f]}_{\Exp/_\equiv} \leq \ceil{[e_i \cdot f]}_{\Exp/_\equiv}
    \]
    With these facts in hand, we can then derive for any $f \in \Exp$ that
    \begin{align*}
    [e] \cdot \ceil{[f]}_{\Exp/_\equiv}
        &= [e_1] \cdot ([e_2] \cdot \ceil{[f]}_{\Exp/_\equiv}) \\
        &\leq [e_1] \cdot \ceil{[e_2 \cdot f]}_{\Exp/_\equiv} \tag{IH} \\
        &\leq \ceil{[e_1 \cdot (e_2 \cdot f)]}_{\Exp/_\equiv} \tag{IH} \\
        &= \ceil{[e \cdot f]}_{\Exp/_\equiv}
    \end{align*}
    \item
    If $e = e_1^*$, then we derive as follows:
    \begin{align*}
    & [e_1] \cdot \ceil{[e \cdot f]}_{\Exp/_\equiv} + \ceil{[f]}_{\Exp/_\equiv} \\
        &\quad \leq \ceil{[e_1 \cdot (e \cdot f)]}_{\Exp/_\equiv} + \ceil{[f]}_{\Exp/_\equiv} \tag{IH} \\
        &\quad \leq \ceil{[e \cdot f]}_{\Exp/_\equiv} \tag{\Cref{\res{lemma:ceil-monotone}}}
    \end{align*}
    By the fixpoint rule, the above then lets us derive that
    \[
        [e] \cdot \ceil{[f]}_{\Exp/_\equiv} = [e_1]^* \cdot \ceil{[f]}_{\Exp/_\equiv} \leq \ceil{[e \cdot f]}_{\Exp/_\equiv}
    \qedhere
    \]
\end{itemize}
\end{proof}

\section{Leveraging completeness}\label{sec:leveraging-completeness}

Completeness now enables the use of coinduction to prove that expressions are provably equivalent using the axioms of Kleene algebra.
All we have to do to show that $e \equiv f$ is argue that that $L(e) = L(f)$.
Fortunately, the latter is decidable, by \Cref{\res{theorem:algo-correct},\res{theorem:expression-to-automaton}}.

We could alternatively try to construct a bisimulation relating $e$ to $f$ on $(\Exp, <o, t>)$ directly.
Unfortunately, since the latter automaton is infinite (and not even locally finite!) such a bisimulation may become cumbersome to describe.
However, it turns out that we do not always need to explicitly specify the whole bisimulation --- in many cases, describing a fragment is enough.
Such a fragment is commonly called a ``bisimulation up-to $X$'', where $X$ is some technique that completes the fragment into a true bisimulation.
We will now briefly discuss one instance of this technique, called \emph{bisimulation up-to congruence}~\cite{DBLP:conf/popl/BonchiP13}.

\begin{definition}
Let $R \subseteq \Exp^2$.
We write $\equiv_R$ for the \emph{congruence closure} of $R$, which is the smallest congruence on $\Exp$ that contains $R$ and $\equiv$.
We say $R$ is a bisimulation \emph{up-to congruence} when, for all $e \mathrel{R} f$, we have that $o(e) = o(f)$, and moreover $e_\ltr{a} \equiv_R f_\ltr{a}$ for all $\ltr{a} \in \alphabet$.
\end{definition}

Note how a bisimulation up-to congruence is not \emph{quite} a bisimulation in $(\Exp, <o, t>)$: the second condition says that $e_\ltr{a}$ and $f_\ltr{a}$ are related by $\equiv_R$ rather than $R$ itself.
However, bisimulations up-to congruence give rise to bisimulations rather directly.
\begin{lemma}\label{lemma:bisimulation-up-to-bisimulation}
Let $R \subseteq \Exp^2$.
If $R$ is a bisimulation up-to congruence, then $\equiv_R$ is a bisimulation on $(\Exp, <o, t>)$.
\end{lemma}
\begin{proof}
We proceed by induction on the construction of $\equiv_R$.
In the base, $e \equiv_R f$ because either $e \mathrel{R} f$ or $e \equiv f$.
In either case, we have that $o(e) = o(f)$ and $e_\ltr{a} \equiv_R f_\ltr{a}$ --- either by the premise that $R$ is a bisimulation up-to congruence, or the arguments from \Cref{\res{thm:ra_sound}}.

For the inductive step, there are several cases to consider.
Since they are all similar, we treat the one for $+$; here, $e \equiv_R f$ because $e = e_1 + e_2$ and $f = f_1 + f_2$, where $e_i \equiv_R f_i$ for $i \in \{1,2\}$.
In that case $o(e_i) = o(f_i)$ and $(e_i)_\ltr{a} \equiv_R (f_i)_\ltr{a}$ for $i \in \{1,2\}$ by induction.
This then tells us that $o(e) = o(f)$ and $e_\ltr{a} \equiv_R f_\ltr{a}$, thus completing the proof.
\end{proof}

The upshot is that, to demonstrate bisimilarity, all we need to do is come up with a bisimulation up-to congruence.

\begin{example}\label{example:bisim-up-to-involved}
Let us illustrate the power of bisimulations up-to congruence by proving the denesting rule $(\ltr{a}+\ltr{b})^* \equiv (\ltr{a}^*\ltr{b})^*\ltr{a}^*$ (\Cref{\res{exercise:denesting}}), by proposing the following relation as a bisimulation up-to congruence:
\[
R = \{<(\ltr{a}+\ltr{b})^*, (\ltr{a}^*\ltr{b})^*\ltr{a}^*>, <(\ltr{a}+\ltr{b})^*,(\ltr{a}^*\ltr{b})(\ltr{a}^*\ltr{b})^*\ltr{a}^* +\ltr{a}^*>\}
\]
We can verify this claim for the first pair, as follows:
\begin{align*}
o((\ltr{a}+\ltr{b})^*)
    &= 1
     = o((\ltr{a}^*\ltr{b})^*\ltr{a}^*) \\
((\ltr{a}+\ltr{b})^*)_{\ltr{a}}
    &\equiv (\ltr{a}+\ltr{b})^*
     \mathrel{R} (\ltr{a}^*\ltr{b})(\ltr{a}^*\ltr{b})^*\ltr{a}^* +\ltr{a}^*
     \equiv ((\ltr{a}^*\ltr{b})^*\ltr{a}^*)_{\ltr{a}} \\
((\ltr{a}+\ltr{b})^*)_{\ltr{b}}
    &\equiv (\ltr{a}+\ltr{b})^*
     \mathrel{R} (\ltr{a}^*\ltr{b})^*\ltr{a}^*
     \equiv ((\ltr{a}^*\ltr{b})^*\ltr{a}^*)_{\ltr{b}}
\intertext{
    For the second pair, our derivation goes like this:
}
o((\ltr{a}+\ltr{b})^*)
    &= 1
     = o((\ltr{a}^*\ltr{b})(\ltr{a}^*\ltr{b})^*\ltr{a}^* +\ltr{a}^*) \\
((\ltr{a}+\ltr{b})^*)_{\ltr{a}}
    &\equiv (\ltr{a}+\ltr{b})^*
     \mathrel{R} (\ltr{a}^*\ltr{b})(\ltr{a}^*\ltr{b})^*\ltr{a}^* +\ltr{a}^*
     \equiv ((\ltr{a}^*\ltr{b})(\ltr{a}^*\ltr{b})^*\ltr{a}^* +\ltr{a}^*)_{\ltr{a}} \\
((\ltr{a}+\ltr{b})^*)_{\ltr{b}}
    &\equiv (\ltr{a}+\ltr{b})^*
     \mathrel{R} (\ltr{a}^*\ltr{b})^*\ltr{a}^*
     \equiv ((\ltr{a}^*\ltr{b})(\ltr{a}^*\ltr{b})^*\ltr{a}^* +\ltr{a}^*)_{\ltr{b}}
\end{align*}
Together, this shows that $R$ is a bisimulation up-to congruence.
\end{example}

\begin{example}
Consider the following relation:
\[
R = \{ <(\ltr{b}+\ltr{a}\ltr{b}^*\ltr{a})^*, \ltr{b}^*\ltr{a}(\ltr{a}\ltr{b}^*\ltr{a}+\ltr{b})^*\ltr{a}\ltr{b}^* + \ltr{b}^*> , <\ltr{b}^*\ltr{a}(\ltr{b}+\ltr{a}\ltr{b}^*\ltr{a})^*,(\ltr{a}\ltr{b}^*\ltr{a}+\ltr{b})^*\ltr{a}\ltr{b}^*>\}
\]
This is a bisimulation up-to congruence witnessing that $(\ltr{b}+\ltr{a}\ltr{b}^*\ltr{a})^* \equiv \ltr{b}^*\ltr{a}(\ltr{a}\ltr{b}^*\ltr{a}+\ltr{b})^*\ltr{a}\ltr{b}^* + \ltr{a}^*$.
We leave checking this as \Cref{\res{exercise:check-bisim-up-to}}.

In comparison, the algebraic proof of the same equivalence requires a bit more ingenuity, using the denesting rule (left as \Cref{\res{exercise:denesting}}) and the slide rule (\Cref{\res{lem:slide}}):
\begin{align*}
 (\ltr{b}+\ltr{a}\ltr{b}^*\ltr{a})^*
&\equiv (\ltr{b}^*\ltr{a}\ltr{b}^*\ltr{a})^*\ltr{b}^*  \tag{denesting rule}\\
&\equiv \ltr{b}^*(\ltr{a}\ltr{b}^*\ltr{a}\ltr{b}^*)^* \tag{slide rule}\\
&\equiv \ltr{b}^*( \ltr{a}\ltr{b}^*\ltr{a}\ltr{b}^*(\ltr{a}\ltr{b}^*\ltr{a}\ltr{b}^*)^*+  1) \tag{\ensuremath{ e^* \equiv 1 + ee^*}}\\
&\equiv \ltr{b}^* \ltr{a}\ltr{b}^*\ltr{a}\ltr{b}^*(\ltr{a}\ltr{b}^*\ltr{a}\ltr{b}^*)^* + \ltr{b}^* \tag{distributivity and \ensuremath{ e 1 \equiv e}}\\
&\equiv \ltr{b}^* \ltr{a}(\ltr{b}^*\ltr{a}\ltr{b}^*\ltr{a})^*\ltr{b}^*\ltr{a}\ltr{b}^* + \ltr{b}^* \tag{slide rule}\\
&\equiv \ltr{b}^* \ltr{a}(\ltr{b} + \ltr{a}\ltr{b}^*\ltr{a})^*\ltr{a}\ltr{b}^* + \ltr{b}^* \tag{denesting rule}\\
&\equiv \ltr{b}^* \ltr{a}(\ltr{a}\ltr{b}^*\ltr{a} +\ltr{b})^*\ltr{a}\ltr{b}^* + \ltr{b}^* \tag{commutativity of $+$}
\end{align*}
\end{example}

\section{Exercises}

\begin{exercises}
    \item\label{exercise:left-distributivity-matrix}
    Let $Q$ be a finite set, and let $X$, $Y$ and $Z$ be $Q$-matrices.
    Prove the left-distributivity claimed in \Cref{\res{lemma:matrix-semiring}}, namely that
    \[
        X \cdot (Y + Z) = X \cdot Y + X \cdot Z
    \]

    \item
    Let $\mathcal{K}$ be a Kleene algebra, and let $M\colon Q \times Q \to \mathcal{K}$.
    \begin{exercises}
        \item\label{exercise:reach-star-finite}
        Show that if $M$ is finite and $q \in Q$, then $R_M(q) = R_{M^*}(q)$.

        \emph{Hint: use that $M^* = \mathbf{1} + M \cdot M^*$.}
        \item\label{exercise:reach-star-infinite}
        Show the same property, assuming that $M$ is locally finite.
    \end{exercises}

    \item\label{exercise:bisimulation-matrices}
    Let $Q_0$ and $Q_1$ be finite sets, let $M$ be a $Q_0$-by-$Q_1$ matrix, let $X$ be a $Q_1$-matrix, and let $Y$ be a $Q_0$-matrix.

    Show that $M \cdot X = Y \cdot M$ implies $M \cdot X^* = Y^* \cdot M$.

    \emph{%
        Hint: if you're stuck, suppose $e, f, g \in \Exp$ with $e \cdot f \equiv g \cdot e$, and prove that $e \cdot f^* \equiv g^* \cdot e$ (cf.\ \Cref{\res{exercise:bisimulation-law}}).
        The structure of this proof can be adapted to your needs.
    }

    \item\label{exercise:matrix-fixpoint-vector}
    In this exercise, we will complete the proof of \Cref{\res{lemma:least-fixpoint-matrix}}.
    First, let us recall the setup there.
    We were given a finite $Q$-matrix $M$ and a $Q$-vector $b$, and set out to construct a $Q$-vector $s$ such that $b + M \cdot s \leq s$.
    Additionally, we needed to show that if $t$ is a $Q$-vector and $z \in \Exp$ such that $b \fatsemi z + M \cdot t \leq t$, then $s \fatsemi z \leq t$.

    In the inductive case, we assumed that this was possible for strictly smaller matrices and vectors.
    We then chose a $p \in Q$, and set $Q' = Q \setminus \{ p \}$, and constructed a $Q'$-matrix $M'$ and $Q'$-vector $b$:
    \begin{mathpar}
        M'(q, q') = M(q, q') + M(q, p) \cdot {M(p, p)}^* \cdot M(p, q')
        \and
        b'(q) = b(q) + M(q, p) \cdot {M(p, p)}^* \cdot b(p)
    \end{mathpar}
    By induction, we then obtained a $Q'$-vector $s'$ such that $b' + M' \cdot s'$, and moreover if $t'$ is a $Q'$-vector such that $b' \fatsemi z + M' \cdot t' \leq t'$, then $s' \fatsemi z \leq t'$.
    We then extended $s'$ to obtain a $Q$-vector $s$ by setting
    \[
        s(q) =
            \begin{cases}
            s'(q) & q \in Q' \\
            {M(p, p)}^* \cdot \left( b(p) + \sum_{q' \in Q'} M(p, q') \cdot s'(q') \right) & q = p
            \end{cases}
    \]
    \begin{exercises}
        \item
        Show that with this choice of $s$, we have that $b + M \cdot s \leq s$.
        \item
        Let $t$ be a $Q$-vector and $z \in \Exp$, with $b \fatsemi z + M \cdot t \leq t$.
        Show:
        \[
            {M(p, p)}^* \cdot \Bigl( b(p) \cdot z + \sum_{q' \in Q'} M(p, q') \cdot t(q') \Bigr) \leqq t(p)
        \]
        \item
        Use the fact above as well as the induction hypothesis applied to the $Q'$-vector $t'$ given by $t'(q) = t(q)$ to show that if $t$ is a $Q$-vector and $z \in \Exp$ with $b \fatsemi z + M \cdot t \leq t$, then $s \fatsemi z \leq t$.
    \end{exercises}

    \item\label{exercise:matrix-fixpoint-matrix}
    Use \Cref{\res{lemma:matrix-star}} to prove \Cref{\res{corollary:matrix-star-fixpoint}}.

    \item\label{exercise:other-star}
    In this exercise, we will prove the missing axioms about the Kleene star operator applied to a finite matrix $M$, i.e., that (1)~$M^* \cdot M + \mathbf{1} = M^*$ and (2)~if $B + S \cdot M \leq S$, then $B \cdot M^* \leq S$.
    \begin{exercises}
        \item
        Review the proofs of \Cref{\res{lemma:least-fixpoint-matrix},\res{lemma:matrix-star}} to confirm that (by analogy) we can construct for each finite matrix $M$ a matrix $M^\dagger$ such that (1)~$M^\dagger \cdot M + \mathbf{1} = M^\dagger$ and (2)~if $B + S \cdot M \leq S$, then $B \cdot M^\dagger \leq S$.
        \item
        Use these properties of $M^\dagger$, as well as \Cref{\res{lemma:matrix-star}} and \Cref{\res{corollary:matrix-star-fixpoint}} to prove that for finite matrices $M$, $M^* = M^\dagger$.
        \item
        Conclude that $M^*$ satisfies properties (1) and (2) above.
    \end{exercises}

    \item
    Let $R$ and $S$ be relations.
    Show that $M_{R \circ S} = M_R \cdot M_S$.

    \item
    Let $R$ be a relation.
    Show that $M_{R^*} = M_R^*$.

    \emph{%
        Hint: show that $M_R^* \leq M_{R^*}$ and $M_{R^*} \leq M_R^*$.
        For both inclusions, you need to use facts about the star of a matrix.
    }

    \item We consider some of the remaining cases of \Cref{\res{thm:ra_sound}}:
    \begin{exercises}
        \item
        Show that $\ore((e_1\cdot e_2)\cdot e_3)=\ore(e_1\cdot (e_2\cdot e_3))$ and $((e_1\cdot e_2)\cdot e_3)_{\ltr{a}}\equiv (e_1\cdot (e_2\cdot e_3))_{\ltr{a}}$ (one of the base cases).
        \item
        Show that if $\ore(e_1)=\ore( e_2)$ and $(e_1)_{\ltr{a}}\equiv (e_2)_{\ltr{a}}$, then $\ore(e_1\cdot f)= \ore(e_2\cdot f)$ and $(e_1\cdot f)_{\ltr{a}}\equiv (e_2\cdot f)_{\ltr{a}}$ (one of the inductive steps).
    \end{exercises}

    \item
    Prove the claim on page~\pageref{claim:locally-finite-unique-morphism}: \Cref{\res{lemma:exist_sol},\res{lemma:sol_hom},\res{lemma:sol_exp_id}} together imply that from any locally finite automaton $(S, <o_S, t_S>)$ there exists a \emph{unique} coalgebra morphism $\ceil{-}_S$ to $( \Exp /_\equiv, <\overline\ore,\overline\tre>)$.

    \item\label{exercise:ceil-monotone}
    In this exercise, we prove \Cref{\res{lemma:ceil-monotone}}: if $e \leqq f$, then $\ceil{[e]}_{\Exp/_\equiv} \leq \ceil{[f]}_{\Exp/_\equiv}$.
    First, we need the matrix $M_+$ over $\Exp/_\equiv$ given by $M_+([g], [h]) = 1$ if $g \equiv h + f$, and $M_+([g], [h]) = 0$ otherwise.
    We also define the right-multiplication of a vector $b$ by a matrix $M$ over $Q$ as another vector over $Q$, in the expected way:
    \[
        (b \cdot M)(q) = \sum_{q'} b(q') \cdot M(q', q)
    \]
    This multiplication interacts well with other matrix-matrix and matrix-vector multiplications, so we can drop parentheses.
    \begin{exercises}
        \item
        Prove that for any $b\colon \Exp/_\equiv \to \Exp/_\equiv$ and any $g \in \Exp$ we have that $(b \cdot M_+)([g]) = b([g + f])$.
        \item
        Use the fixpoint laws to show that $M_{\Exp/_\equiv}^* \cdot o \leq M_{\Exp/_\equiv}^* \cdot o \cdot M_+$.
        \item
        Use the two facts above to conclude $\ceil{[e]}_{\Exp/_\equiv} \leq \ceil{[f]}_{\Exp/_\equiv}$.
    \end{exercises}

    \item\label{exercise:check-bisim-up-to}
    Verify that the relation $R$ as given in \Cref{\res{example:bisim-up-to-involved}} is a bisimulation up-to congruence.
\end{exercises}

\onlyinsubfile{%
    \printbibliography%
}

\chapter{Kleene algebra with tests}%
\label{chapter:kat}
Previous chapters studied Kleene Algebra and equational reasoning over regular expressions.
As we mentioned in the introduction, regular expressions offer a useful algebraic perspective to reason about imperative programs.
However, to more faithfully model imperative programs we need to precisely capture \emph{control flow} structures such as conditionals and while loops.
Recall the two programs from the introduction.
\[
    \begin{tikzpicture}
        \node (p1) {
            \setlength{\algomargin}{0ex}
            \begin{minipage}{42mm}
                \begin{algorithm}[H]
                    \While{\testvar{a} \textnormal{\texttt{and}} \testvar{b}}{
                        $\progvar{p}$
                    }
                    \While{\testvar{a}}{
                        $\progvar{q}$\;
                        \While{\testvar{a} \textnormal{\texttt{and}} \testvar{b}}{
                           $\progvar{p}$\;
                        }
                    }
                \end{algorithm}
            \end{minipage}
        };
        \draw[densely dotted] ($(p1.north west) + (-0.15, 0)$) rectangle ($(p1.south east) + (-0.7, 0.1)$);
    \end{tikzpicture}
\qquad\qquad
    \begin{tikzpicture}
        \node (p2) {
            \setlength{\algomargin}{0ex}
            \begin{minipage}{30mm}
                \begin{algorithm}[H]
                    \While{\testvar{a}}{
                        \eIf{\testvar{b}}{
                            $\progvar{p}$\;
                        }{
                            $\progvar{q}$\;
                        }
                    }
                \end{algorithm}
            \end{minipage}
        };
        \draw[densely dotted] ($(p2.north west) + (-0.15, 0)$) rectangle ($(p2.south east) + (-0.7, 0.1)$);
    \end{tikzpicture}
\]
In this chapter we want to show how we can encode and reason about equivalence of these programs using regular expressions in a way that the effects of control flow are taken as first class citizens.
The key piece missing in regular expressions is a distinguished notion of \emph{test} that enables the conditional execution of program actions.
We will explain how one can develop a theory of regular expressions --- retaining all the elements from the previous chapters --- in which we can have both tests and program actions.
This will lead to the notion of \emph{Kleene algebra with tests} (\KAT), an equational system that combines Kleene and Boolean algebra originally proposed by Kozen~\cite{kozen-1996}.
Within \KAT one can model imperative programming constructs and assertions, which can then be used to analyze control flow~\cite{angusk01:kat:schemato,Koz08}, or data flow~\cite{kot-kozen-2005,fernandes-2007}, or even verify compiler optimizations~\cite{kozen-patron-2000}.
More recently, a variant of \KAT called \NetKAT was introduced to model forwarding behavior of networks, and verify important properties, such as reachability and access control~\cite{netkat}.

In \KAT, Boolean subexpressions called tests can be used to model assertions, i.e., properties that must hold at a certain point in the program before it can continue.
Using these assertions, we can encode conditionals and loops in a format that we encountered before~\cite{kozen-1996}:
\begin{mathpar}
   \whd{a}{e} \eqdef (ae)^* \overline{a}
   \and
   \ite{a}{e}{f} \eqdef ae + \overline{a}f
\end{mathpar}
Here, $a$ is a test (asserting $a$ is true), and $\overline{a}$ is its negation.

This chapter elides formal proofs of the claims made; in most cases, this is because they are analogous to their counterparts for \KA.
In particular, the claims about \KAT automata in \Cref{\res{sec:kat-automata}} can be proved by instantiating the coalgebraic machinery from \Cref{\res{chapter:coalgebra}}; we provide remarks with references to the appropriate results in some cases.

\section{Syntax and Semantics}
In this section we describe the syntax and semantics of \KAT, before treating both the language and relational interpretation of expressions.

In order to reason about common programming constructs such as if-then-else statements and while-loops, we need to add a notion of \emph{tests} to our syntax.
Given a set $\P$ of (primitive) actions and a set $\Bt$ of (primitive) tests, we define the set $\BExp$ of Boolean tests by:
\[
    \BExp\ni a,b \Coloneqq \testvar{a} \in \Bt \mid a \cdot b \mid a+b \mid \overline a \mid \0 \mid \1
\]
We intend for $\cdot$ to be read as conjunction (and), $+$ as conjunction (or), and $\overline{\phantom{a}}$ as negation (not); $\0$ and $\1$ represent the tests that are always false and always true, respectively.
The set of \KAT expressions is
\[
    \KExp \ni e,f\Coloneqq \p \in \P \mid b\in \BExp \mid e \cdot f \mid e+f \mid e^*
\]
For now, we can think of tests as ``assertions'' inside a program, which make the program fail in a way that undoes all earlier effects.
This explains, for example, the encoding of the \itebare statement above, which makes the program branch non-deterministically into two programs, $ae$ and $\overline{a}f$; among these two, either $a$ holds and the first branch survives, or $\overline{a}$ holds and the second branch does.

\subsection{Relational semantics}

As our first semantics of \KAT expressions, we extend the relational semantics given for \KA in \Cref{\res{section:ka-relational-semantics}}.
To this end, we must assign a relational semantics to tests; here, the idea is that each test is associated with a subset of the identity relation on states: if $(s, s)$ is in the semantics of $a$, it means that $a$ succeeds in state $s$, without changing the state.

We show that the language model is sound and complete for this interpretation.
Thus, \KAT equivalence implies program equivalence for any programming language with a suitable relational semantics.

\begin{definition}[Relational semantics]
A (\KAT) \emph{interpretation} (over a set $S$) consists of functions $\sigma\colon \P \to 2^{S \times S}$ and $\tau\colon \Bt \to 2^S$.
The \emph{relational semantics} of a \KAT expression $e$ with respect to $(\sigma, \tau)$, written $\sem{-}_{\sigma,\tau} \colon \KExp \to 2^{S \times S}$, is defined inductively, as follows:
\begin{align*}
\sem{\0}_{\sigma,\tau} &= \emptyset
    & \sem{\1}_{\sigma,\tau} &= \id_S \\
\sem{\p}_{\sigma,\tau} &= \sigma(\p)
    & \sem{\testvar{a}}_{\sigma,\tau} &= \id_{\tau(\testvar{a})} \\
\sem{e + f}_{\sigma,\tau} &= \sem{e}_{\sigma,\tau} \cup \sem{f}_{\sigma,\tau}
    & \sem{\overline{a}}_{\sigma,\tau} &= \At \setminus \sem{a}_{\sigma,\tau} \\
\sem{e \cdot f}_{\sigma,\tau} &= \sem{e}_{\sigma,\tau}\circ \sem{f}_{\sigma,\tau}
    & \sem{e^*}_{\sigma,\tau} &= \sem{e}_{\sigma,\tau}^*
\end{align*}
\end{definition}

Note how in the above, we make no distinction between tests (in $\BExp$) and full-fledged programs (in $\KExp$) when giving the semantics of $+$ and $\cdot$.
It is not too hard to check that for every $a \in \BExp$, it still holds that $\sem{a}_{\sigma, \tau} \subseteq \id_S$.
In fact, because relational composition coincides with intersection on subsets of the identity relation, we recover the expected interpretation of $\cdot$ as conjunction, i.e., $\sem{a \cdot b}_{\sigma, \tau} = \sem{a}_{\sigma, \tau} \cap \sem{b}_{\sigma, \tau}$.

\begin{example}\label{example:kat-rel-semantics}
Consider the expression ${\testvar{a}}+{\testvar{b}}\p$ and the interpretation over $S = \{s_1, s_2\}$ given by $\sigma(\p) = \{(s_1, s_2)\}$, $\tau(\testvar{a}) = \{s_2\}$, and $\tau(\testvar{b}) = \{s_1, s_2\}$.
It is easy to see that  $\sem{\testvar{a}}_{\sigma,\tau} = \{(s_2,s_2)\}$, $\sem{\testvar{b}}_{\sigma,\tau} = \{(s_1,s_1), (s_2,s_2)\}$ and $\sem{\p}_{\sigma,\tau} = \{(s_1,s_2)\}$.
In total, we obtain:
\[
 \sem{\testvar{a}+\testvar{b}\p}_{\sigma,\tau} = \{(s_2,s_2),(s_1,s_2)\}
 \]
\end{example}

\subsection{Language semantics}

Like \KA, \KAT comes equipped with a \emph{language model}~\cite{kozen-smith-1996}, which abstracts from the interpretation of the primitive programs and tests.
Instead of words, languages in this model are composed of \emph{guarded strings}~\cite{kaplan69}.
Like ordinary words, guarded strings model the possible traces of a program, but they also record which tests hold before the program starts, between actions, and when the program ends.

Formally, we define \emph{atoms} as subsets of $\Bt$, and write $\At$ for the set of atoms.
In the sequel it will be convenient to denote an atom in a condensed notation that makes explicit which tests are (not) part of the atom.
This is done by listing all of the primitive tests in some order, and putting a bar over the ones not included; for instance, when $\Bt = \{ \testvar{a}, \testvar{b} \}$, we will write $\overline{\testvar{a}}\testvar{b}$ for the atom $\{ \testvar{b} \}$ and $\testvar{a}\overline{\testvar{b}}$ for the atom $\{ \testvar{a} \}$.

We can then define the set \GSs of \emph{guarded strings} by $\GSs = (\At\cdot \P)^*\cdot \At$.
Given guarded strings $\alpha w \beta$ and $\beta x \gamma$, we define their \emph{fusion product} by $\alpha w \beta \diamond \beta x \gamma = \alpha w \beta x \gamma$.
Note that this is defined only when the inner atoms are the same.
We write $\GLs$ for $2^\GSs$, i.e., the set of \emph{guarded languages}.
Given $X, Y \subseteq \GSs$, we write $X\diamond Y$ for the set containing all fusion products $x\diamond y$ of $x\in X$ and $y\in Y$, whenever these are defined.
Furthermore, $X^{(n)}$ is defined inductively as $X^{(0)} = \At$ and $X^{(n+1)} = X \diamond X^{(n)}$; this allows us to define $X^{(*)}$ as $\bigcup_{n \geq 0} X^{(*)}$.

\begin{definition}[Language semantics]
Each \KAT expression $e$ denotes a set $\sem{e}$ of guarded strings
defined inductively on the structure of $e$:
\begin{align*}
\sem{\0} &= \emptyset
    & \sem{\1} &= \At \\
\sem{\p} &= \{\alpha\p\beta \mid \alpha,\beta \in\At\}
    & \sem{\testvar{a}} &= \{\alpha \in \At \mid \testvar{a} \in \alpha \} \\
\sem{e+f} &= \sem{e} \cup \sem{f}
    & \sem{\overline{a}} &= \At \setminus \sem{a} \\
\sem{e \cdot f} &= \sem{e} \diamond \sem{f}
    & \sem{e^*} &= \sem{e}^{(*)}
\end{align*}
A guarded language $L \subseteq \GSs$ is \emph{regular} if $L = \sem{e}$ for some $e \in \KExp$.
\end{definition}

The definition of the semantics of sequential composition in terms of fusion product has an intuitive explanation: a trace modeled by $e \cdot f$ is a trace of $e$ that ended in some machine state $\alpha$, which then continues with a trace of $f$ that starts in that same state $\alpha$.

\begin{example}\label{ex:kat_nd}
Consider the \KAT expression ${\testvar{a}}+{\testvar{b}}\p$ from \Cref{\res{example:kat-rel-semantics}}.
We compute its language semantics as follows:
\begin{align*}
    \sem{{\testvar{a}}+{\testvar{b}}\p}
        &= \sem{{\testvar{a}}}\cup (\sem{{\testvar{b}}}\diamond \sem{\p})\\
        &= \{\alpha \in \At \mid \testvar{a} \in \alpha\} \cup \left(\{\alpha \in \At \mid \testvar{b} \in \alpha \} \diamond  \{\alpha\p\beta \mid \alpha,\beta \in\At\}\right)\\
        &= \{\alpha \in \At \mid \testvar{a} \in \alpha \} \cup \{\alpha\p\beta \mid \alpha, \beta \in\At, \testvar{b} \in \alpha \}
\end{align*}
\end{example}

\begin{example}
Now consider the expression $\ite{a}{e}{e}$ and note that its semantics can be computed as follows:
\begin{align*}
    \sem{\ite{a}{e}{e}}
    &= \sem{a}\diamond \sem{e} \cup \sem{\overline{a}} \diamond \sem{e}\\
    &= (\sem{a}\cup \sem{\overline{a}})\diamond \sem{e}\\
    &= \At\diamond \sem{e}\\
    &= \sem{e}
\end{align*}
This suggests that the language semantics equates $\ite{a}{e}{e}$ with $e$ --- and this should be no surprise: $e$ is executed regardless of $a$.
\end{example}

Just like before, the language model abstracts the various relational interpretations in a sound and complete way~\cite{kozen-smith-1996}.
The proof is similar to that of \Cref{\res{thm:rel_equiv_lang}}, and is left as \Cref{\res{exercise:rel-vs-lang-kat}}.

\begin{theorem}\label{soundcompleteforrel}%
\label{thm:sound-complete-for-rel}
The language model is sound and complete for the relational model, in the following sense:
\[ \sem{e} = \sem{f} \quad \iff \quad\forall \sigma, \tau.\ \sem{e}_{\sigma,\tau} = \sem{f}_{\sigma,\tau}\]
\end{theorem}

\section{Reasoning}

Next, we extend the reasoning principles used for \KA to \KAT.
The basic idea is to include all of the laws of Boolean algebra for the tests.
\begin{definition}[Provable equivalence]\label{def:kat}
We define $\equiv_{\KAT}$ as the smallest congruence on $\KExp$ that includes the laws from \Cref{\res{def:ka}}, as well as the laws of \emph{Boolean algebra} --- that is, for all $a, b, c \in \BExp$:
\begin{mathpar}
    a+\0=a=a\cdot \1
    \and
    a\cdot (b\cdot c) = (a\cdot b)\cdot c
    \and
    a+b=b+a
    \and
    a\cdot b=b \cdot a
    \and
    a+(b+c) = (a+b)+c
    \and
    a+(a\cdot b) = a = a\cdot (a+b)
    \and
    \0=a \cdot\overline{a}
    \and
    \1= a+\overline{a}
    \and
    a+ (b\cdot c) = (a+ b)\cdot (a+ c)
    \and a\cdot (b+ c) = (a\cdot b)+ (a\cdot c)
\end{mathpar}
\end{definition}

To keep our notation simple, we will usually drop the subscript and write $\equiv$ for $\equiv_{\KAT}$.
Conversely, we may use $\equiv_{\KA}$ to signal that two \KAT expressions are equivalent by the laws of \KA alone.

As expected, the laws of \KAT are sound in the sense that if $e \equiv f$, then $\sem{e} = \sem{f}$.
We record this below; the proof is straightforward, by induction on the construction of $\equiv$, so we omit it.
\begin{theorem}\label{theorem:kat-soundness}
Let $e, f \in \KExp$.
If $e \equiv f$, then $\sem{e} = \sem{f}$.
\end{theorem}

\begin{example}[Basic conditional laws]\label{ex:basiccond}
We prove, using the axioms of Boolean and Kleene algebra, two familiar identities about conditionals.
\begin{mathpar}
    \ite{a}{e}{e} \equiv e
    \and
    \ite{a}{(\its{\overline{a}}{e})}{f} \equiv \its{\overline{a}}{f}
\end{mathpar}
Using our encodings, the above laws translate to the following claims:
\begin{mathpar}
    a{e} + \overline{a}e \equiv e \
    \and
    a(\overline{a}e + a) + \overline{a}f \equiv \overline{a}f + a
\end{mathpar}
These in turn can be proved as follows.
The first one is immediate:
\begin{align*}
    a{e} + \overline{a}e
        &\equiv (a + \overline{a}) e \tag{distributivity}\\
        &\equiv \1 e \tag{\ensuremath{a + \overline{a} \equiv \1}}\\
        &\equiv e    \tag{unit law}
\end{align*}
The second one requires a few more steps, but all are quite simple:
\begin{align*}
   a(\overline{a}e + a) + \overline{a}f
       &\equiv (a \overline{a}) e + aa + \overline{a}f \tag{distributivity, associativity}\\
       &\equiv (a \overline{a}) e + a + \overline{a}f \tag{\ensuremath{aa\equiv a}}\\
       &\equiv \0 e + a + \overline{a}f \tag{\ensuremath{a\overline{a} \equiv \0}}\\
       &\equiv a + \overline{a}f \tag{unit laws}\\
       &\equiv \overline{a}f + a \tag{commutativity} 
\end{align*}
\end{example}

As before, these laws can be formulated algebraically, allowing us to speak of ``a \KAT'' as an algebra that satisfies the laws from \Cref{\res{def:kat}}.

\begin{definition}[\KAT]\label{def:kat-algebra}
A Kleene algebra with tests (\KAT) is a two-sorted structure $(\alphabet, B, +, \cdot , -^*, \overline{\phantom{a}}, 0,1)$ where $(\alphabet, +, \cdot , -^*, 0, 1)$ is a Kleene algebra and
$(B, +, \cdot , \overline{\phantom{a}}, 0,1 )$ is a Boolean algebra.
In particular, the latter requirement tells us that $B$ is closed under $+$ and $\cdot$, and that the operators on $B$ satisfy the laws from \Cref{\res{def:kat}}.

Concrete examples of \KAT{}s include $(2^\GSs, \cup, \diamond, {-}^{(*)}, \overline{\phantom{a}}, \emptyset, \At)$, called the \emph{\KAT of guarded languages}, and $(2^{S \times S}, \cup, \circ, {-}^*, \overline{\phantom{a}}, \emptyset, 2^{\id_S})$ for any set $S$, called the \emph{\KAT of relations on $S$}. Note that in these two examples $\overline{\phantom{a}}$ is just set complement.
\end{definition}

\section{Automata and Bisimulation}\label{sec:kat-automata}

Now that we familiarized ourselves with how the syntax and semantics of \KA can be extended with tests, it is time to talk about the accompanying automata model.
Here too, the tweaks are relatively minor: instead of accepting words consisting of just letters, we need to recognize guarded words.
To achieve that, the \KAT automata will move from state to state while reading both an atom $\alpha \in \At$ and a primitive letter $\p \in \P$; acceptance of a word in each state is determined by the final atom.
If we formally encode these transitions and acceptance conditions, the model below arises naturally; it is a deterministic version of automata on guarded strings as introduced by \citet{kozen03}.

\begin{definition}[\KAT automata~\cite{kozen08}]
A \KAT automaton is a pair $(S,<o_S,t_S>)$ where $o_S\colon S \to 2^\At$ and $t_S\colon S \to S^{\At\times \P}$.
\end{definition}

\begin{remarkblue}
Blue remarks such as this one make a return in this section, for those who read \Cref{\res{chapter:coalgebra}} and are interested in the coalgebraic background.
In the case of the definition above, it should be clear that \KAT automata are $F$-coalgebras for
\(
    F(X) = 2^\At \times X^{\At\times P}
\).
\end{remarkblue}

\begin{figure}
    \centering
    \begin{tikzpicture}
        \node[state] (s0) {$s_0$};
        \node[state,right=2cm of s0] (s1) {$s_1$};
        \draw[->] (s0) edge node[above] {$\overline{\testvar{a}}\q$} (s1);
        \draw[->] (s0) edge[loop above] node[above] {$\1\p$, $\testvar{a}\q$} (s0);
        \node[left=4mm of s0] (s0out) {$\overline{\testvar{a}}\testvar{b}$};
        \draw[->] (s0) edge[double,double distance=0.7mm,-implies] (s0out);
        \draw[->] (s1) edge[loop above] node[above] {$\1\p$, $\1\q$} (s1);
    \end{tikzpicture}
    \caption{An example \KAT automaton on $\P = \{ \p, \q \}$ and $\Bt = \{ \testvar{a}, \testvar{b} \}$.}\label{figure:example-kat-automaton}
\end{figure}

Just like classical automata, we can draw \KAT automata as graphs, as in \Cref{\res{figure:example-kat-automaton}}.
Rather than labeling the transitions with single letters, we endow them with pairs of atoms and letters --- i.e., for a pair of states $s$ and $s'$, we attach $\alpha\p$ to the arrow from $s$ to $s'$ when $t_S(s)(\alpha, \p) = s'$.
In lieu of accepting states the acceptance function is depicted by drawing double arrows ($\Rightarrow$) from each state to atoms that are accepted in that state --- i.e., if $\alpha$ is such that $o_S(\alpha) = 1$, then $s$ has a double arrow to $\alpha$.
To keep these drawings compact, we cluster atoms --- e.g., a transition label $\testvar{b}\p$ is shorthand for the labels $\testvar{a}\testvar{b}\p$ and $\overline{\testvar{a}}\testvar{b}\p$; more formally, a label $b\p$ is shorthand for $\alpha_1\p, \dots, \alpha_n\p$ where $\alpha_1, \dots, \alpha_n \leqq b$.

\medskip
Given a \KAT automaton $(S, <o_S, t_S>)$, the language of a state $s \in S$, also denoted $L(s)$, is guarded language defined inductively:
\begin{mathpar}
    L(s)(\alpha) = o_S(\alpha)
    \and
    L(s)(\alpha\p w) = L(t_S(\alpha, \p))(w)
\end{mathpar}
Informally, acceptance of a guarded string consisting of just an atom is determined by $o_S$, while for longer guarded strings this depends on whether the remainder of the string is accepted by the state reached after reading the first atom and letter.

\begin{remarkblue}
The set of guarded languages $\GLs = 2^\GSs$ has an $F$-coalgebra structure $<o_\GLs, t_\GLs>$ given by
\begin{mathpar}
    o_\GLs(L)(\alpha) = L(\alpha)
    \and
    t_\GLs(L)(\alpha, \p) = \{ w \mid \alpha\p w \in L \}
\end{mathpar}
In fact, $(\GLs, <o_\GLs, t_\GLs>)$ is the final $F$-coalgebra, and the map $L$ discussed above, which sends each state of a given $F$-coalgebra to a guarded language in $\GLs$, is the unique $F$-coalgebra homomorphism from that $F$-coalgebra into the final $F$-coalgebra; one can prove this claim by replaying the arguments from \Cref{\res{thm:languages-final}}.
\end{remarkblue}

\subsection{Expressions to automata}

We now have two formalisms to represent guarded languages: \KAT expressions and \KAT automata.
Unsurprisingly, there is a two-way correspondence between these, in the style of Kleene's theorem: a guarded language is represented by a \KAT expression if and only if it is represented by a \KAT automaton.
For the purpose of this chapter, we limit ourselves to discussing the conversion from expressions to automata using Brzozowski derivatives; for the other direction, a matrix-based approach similar to the one in \Cref{\res{subsection:a2e}} suffices.

\begin{definition}[Brzozowski derivatives for \KAT expressions]
Given a \KAT expression $e\in \KExp$, we define $E \colon  \KExp \to  2^\At$ and $D\colon  \KExp \to \KExp^{\At\times \P}$ by induction on the structure of $e$.
First, $E(e)$ is given by:
\begin{align*}
E(\0) &= \emptyset
    & E(\1) &= \At \\
E(\p) &= \emptyset
    & E(\testvar{a}) &= \{\alpha\in\At \mid \testvar{a} \in \alpha\} \\
E(e+f) &= E(e) \cup E(f)
    & E(\overline{b}) &= \At \setminus E(b) \\
E(ef) &= E(e) \cap E(f)
    & E(e^*) &= \At
\end{align*}
Next, we define $e_{\alpha\q} = D(e)(\alpha,\q)$ by
\[
\begin{array}{llll}
\p_{\alpha\q} =\begin{cases} \1 & \text{if }\p=\q\\ \0 & \text{if }\p\neq \q\end{cases}  &
\b_{\alpha\q} = \0 &
(ef)_{\alpha\q}=\begin{cases} e_{\alpha\q}f +f_{\alpha\q} & \text{if }\alpha\in E(e)\\ e_{\alpha\q}f &
\text{if }\alpha\not\in E(e)\end{cases}   \\[4ex]
(e+f)_{\alpha\q}=  e_{\alpha\q}+f_{\alpha\q}&& (e^*)_{\alpha\q} = e_{\alpha\q}e^*
\end{array}
\]
\end{definition}

\begin{example}\label{example:kat-derivatives}
Let $\P = \{ \p \}$ and $\Bt = \{{\testvar{a}},{\testvar{b}}\}$.
If $e = {\testvar{a}}\q + {\testvar{b}}\p$, then
\begin{align*}
    E(e) &= (E(\testvar{a}) \cap E(\p)) \cup (E(\testvar{b}) \cap E(\q)) \\
         &= (\{\alpha \mid \testvar{a} \in \alpha \} \cap \emptyset) \cup (\{\alpha \mid \testvar{b} \in \alpha \} \cap \emptyset)
          = \emptyset
\end{align*}
Moreover, we can compute the values of $D(e)$ as follows:
\begin{align*}
    e_{{\testvar{a}}{\testvar{b}}\p}
        &= {\testvar{a}}_{{\testvar{a}}{\testvar{b}}\p} + ({\testvar{b}}\p)_{{\testvar{a}}{\testvar{b}}\p}
         = \0 + ({\testvar{b}}_{{\testvar{a}}{\testvar{b}}\p} \cdot \p + \p_{{\testvar{a}}{\testvar{b}}\p})
         = \0 + (\0 \cdot \p + \1)
    \\
    e_{\overline {\testvar{a}}{\testvar{b}}\p}
        &= {\testvar{a}}_{\overline {\testvar{a}}{\testvar{b}}\p} +({\testvar{b}}\p)_{\overline{\testvar{a}}{\testvar{b}}\p}
         = \0 + ({\testvar{b}}_{{\overline{\testvar{a}}}{\testvar{b}}\p} \cdot \p + \p_{{\overline{\testvar{a}}}{\testvar{b}}\p})
         = \0 + (\0 \cdot \p + \1)
    \\
    e_{{\testvar{a}}\overline {\testvar{b}}\p}
        &= {\testvar{a}}_{{\testvar{a}}\overline {\testvar{b}}\p} +({\testvar{b}}\p)_{{\testvar{a}}\overline {\testvar{b}}\p}
         = \0 + {\testvar{b}}_{{\testvar{a}}\overline {\testvar{b}}\p} \cdot \p
         = \0 + \0 \cdot \p
    \\
    e_{\overline {\testvar{a}}\overline {\testvar{b}}\p}
        &= {\testvar{a}}_{\overline {\testvar{a}}\overline {\testvar{b}}\p} +({\testvar{b}}\p)_{\overline {\testvar{a}}\overline {\testvar{b}}\p}
         = \0 + {\testvar{b}}_{\overline {\testvar{a}}\overline {\testvar{b}}\p} \cdot \p
         = \0 + \0 \cdot \p
\end{align*}
\end{example}

These derivatives make $(\KExp, <E, D>)$ a \KAT automaton.
The semantics of \KAT automata then gives us a map $L\colon \KExp \to \GLs$, associating with each expression the language it accepts as a state of this automaton.
As expected, this map coincides with $\sem{-}\colon \KExp \to \GLs$.

\begin{theorem}\label{theorem:kat-bialgebra}
For all $e \in \KExp$, it holds that $L(e) = \sem{e}$.
\end{theorem}

\begin{remarkblue}
Although the theorem above can be proved directly, a somewhat more elegant proof arises coalgebraically.
Note that, by finality, $L$ is the unique $F$-coalgebra homomorphism that makes the following diagram commute:
\[
    \begin{tikzpicture}
        \node (KExp) {$\KExp$};
        \node[right=4cm of KExp] (GLs) {$2^\GSs$};
        \node[below=7mm of KExp] (FKExp) {$2^\At \times \KExp^{\At\times\P}$};
        \node at (FKExp -| GLs) (FGLs) {$2^\At \times {(2^\GSs)}^{\At\times\P}$};

        \draw[dashed,->] (KExp) edge node[above] {\footnotesize $L$} (GLs);
        \draw[->] (KExp) edge node[left] {\footnotesize $<E,D>$} (FKExp);
        \draw[dashed,->] (FKExp) edge node[below] {\footnotesize $F(L)$} (FGLs);
        \draw[->] (GLs) edge node[right] {\footnotesize $<o_\GLs, t_\GLs>$} (FGLs);
    \end{tikzpicture}
\]
Thus, if we can show that $\sem{-}$ is also an $F$-coalgebra homomorphism that fits in the same place, the claim follows.
In essence, this latter condition comes down proving that:
\[
    \sem{e} = E(e) \cup \bigcup_{\alpha \in \At,\ \p \in \P} \{ \alpha \p w \mid w \in \sem{e_{\alpha\p}} \}
\]
This follows from a version of \Cref{\res{thm:fundamental-re}} for \KAT.
\end{remarkblue}

Of course, this only shows that there is a \KAT automaton for each expression; for this correspondence to be useful, we need to obtain a \emph{finite} \KAT automaton.
Fortunately, the same technique as before lets us tame $(\KExp, <E, D>)$ to obtain a finite automaton for each expression.

\begin{lemma}\label{lemma:kat-aci-finite}
Let $\cong$ be the congruence on $\KExp$ that equates terms up to associativity, commutativity and idempotence of $+$ (see \Cref{\res{definition:congruence-aci}}).
There exists, for every $e \in \KExp$, a finite set $\rho(e) \subseteq \KExp$ such that, in $(\KExp, <E, D>)$, $e$ reaches only states $\cong$-related to some element of $\rho(e)$.
\end{lemma}

\begin{lemma}\label{lemma:kat-aci-bisimulation}
$E$ and $D$ are well-defined up to $\cong$, i.e., if $e \cong f$ then $E(e) = E(f)$, and for all $\alpha \in \At$ and $\p \in \P$ we have that $e_{\alpha\p} \cong f_{\alpha\p}$.
\end{lemma}

\begin{remarkblue}
The last lemma is tantamount to claiming that $\cong$ is an $F$-bisimulation on the $F$-coalgebra $(\KExp, <E, D>)$, and can be shown quite easily by induction on the construction of $\cong$.
\end{remarkblue}

Together, these two lemmas tell us that we can consider states in the Brzozowski automaton $(\KExp, <E, D>)$ up to $\cong$; with this convention, each state can reach only a finite number of other states that differ under $\cong$.
Thus, for a given \KAT expression $e \in \KExp$, we can (lazily) generate a finite automaton just like we did for regular expressions.

\begin{example}\label{ex:kat-automaton}
Let $\P$, $\Bt$ and $e$ be as in \Cref{\res{example:kat-derivatives}}.
Because for all $\alpha \in \At$, we have that ${\0 + (\0 \cdot \p + \1)}_{\alpha \p} = \0 + (\0 \cdot \p + \0) \cong \0 + \0 \cdot \p$ and ${(\0 + \0 \cdot \p)}_{\alpha \p} = \0 + \0 \cdot \p$, we obtain the following automaton for $e$:
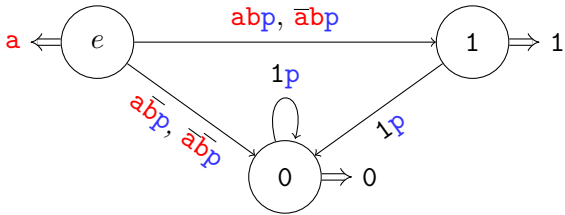
\[
    \begin{tikzpicture}
        \node[state] (e) {$e$};
        \node[state,right=4cm of e] (one) {$\1$};
        \node[state,below=13mm of $(e)!0.5!(one)$] (zero) {$\0$};
        \draw[->] (e) edge node[above] {$\testvar{a}\testvar{b}\p$, $\overline{\testvar{a}}\testvar{b}\p$} (one);
        \draw[->] (e) edge node[sloped,below] {$\testvar{a}\overline{\testvar{b}}\p$, $\overline{\testvar{a}}\overline{\testvar{b}}\p$} (zero);
        \draw[->] (one) edge node[sloped,below] {$\1\p$} (zero);
        \draw[->] (zero) edge[loop above] node[above] {$\1\p$} (zero);
        \node[left=4mm of e] (eout) {$\testvar{a}$};
        \draw[->] (e) edge[double,double distance=0.7mm,-implies] (eout);
        \node[right=4mm of one] (oneout) {$\1$};
        \draw[->] (one) edge[double,double distance=0.7mm,-implies] (oneout);
        \node[right=4mm of zero] (zeroout) {$\0$};
        \draw[->] (zero) edge[double,double distance=0.7mm,-implies] (zeroout);
    \end{tikzpicture}
\]
\end{example}

\subsection{Decision procedure}

It is not too hard to transpose the notion of bisimulation for automata (\Cref{\res{def:automatabis}}) to \KAT automata; this yields the following.

\begin{definition}[\KAT bisimulation]\label{def:kat-bisimulation}
Given \KAT automata $(S,<o_X,t_X>)$ and $(Y,<o_Y,t_Y>)$ we call $R\subseteq X\times Y$ is a \emph{bisimulation} if for all $(x,y)\in R$:
\begin{enumerate}
\item $o_X(x)(\alpha) = o_Y(y)(\alpha)$ for all $\alpha \in \At$, and
\item for all $\alpha \in\At$ and $\p\in \P$, $(t_X(x)(\alpha, \p), t_Y(y)(\alpha, \p))\in R$.
\end{enumerate}
\end{definition}

We denote the largest bisimulation on $(\KExp, <E, D>)$ by $\bisim$, and we can use it as a sound and complete proof principle for equivalence.
\begin{theorem}[\KAT coinduction]\label{thm:katcoinduction}
Let $e, f$ be \KAT expressions.
\[
e \bisim f \iff \sem{e} = \sem{f}
\]
\end{theorem}

\begin{remarkblue}
It is not too hard to see that, in analogy with \Cref{\res{remark:bisimulation-coalgebra}}, bisimulations the sense of \Cref{\res{def:kat-bisimulation}} are precisely $F$-bisimulations when \KAT automata are regarded as coalgebras.
With that in mind, the forward direction of the claim above follows directly from \Cref{\res{lemma:bisim-implies-behavioral-equiv}}, and the backward direction follows from \Cref{\res{corollary:behavioral-equiv-implies-bisim}} provided we can show that $F$ is kernel-compatible; the latter is not too hard to show.
\end{remarkblue}

\Cref{\res{thm:katcoinduction}} provides an alternative way to prove that two expressions are equivalent: build the automaton corresponding to each expression using Brzozowski derivatives and then check if they are bisimilar.
To make the latter decidable, we have to leverage \Cref{\res{lemma:kat-aci-finite,lemma:kat-aci-bisimulation}}, and work with bisimulations up-to $\cong$.

Indeed, we can take this idea one step further, and work with bisimulations up-to $\equiv$, like we did in \Cref{\res{sec:leveraging-completeness}}, but for \KAT.
Given a relation $R \subseteq \KExp$, we can write $\equiv_R$ or the smallest congruence on $\KExp$ containing both $R$ and $\equiv$.
We now say that $R$ is a \emph{bisimulation up-to congruence} if it satisfies the following for all $e \mathrel{R} f$:
\begin{enumerate}
\item $E(e)(\alpha) = E(f)(\alpha)$ for all $\alpha \in \At$, and
\item for all $\alpha \in\At$ and $\p\in \P$, $D(e)(\alpha, \p) \equiv_R D(f)(\alpha, \p)$.
\end{enumerate}
The following generalization of \Cref{\res{\res{lemma:bisimulation-up-to-bisimulation}}} then follows by analogy:
\begin{lemma}\label{lemma:kat-bisimulation-up-to-bisimulation}
Let $R \subseteq \KExp^2$.
If $R$ is a bisimulation up-to congruence, then $\equiv_R$ is a bisimulation on $(\KExp, <E, D>)$.
\end{lemma}

Thus, to show that $e \equiv f$, it suffices find a bisimulation up-to congruence $R$ such that $e \mathrel{R} f$.
This is also necessary --- it is not too hard to show that $\equiv$ is a bisimulation on $(\KExp, <E, D>)$, which makes it a bisimulation up-to congruence (note that ${\equiv_\equiv}$ coincides with $\equiv$); hence, if $e \equiv f$, then there exists a bisimulation up-to congruence relating $e$ to $f$.
To demonstrate this technique, we will momentarily apply it to
\[
(\testvar a(\testvar b\progvar p + \overline{\testvar b}\progvar q))^*\overline{\testvar a}\qquad\text{ and }\qquad(\testvar a\testvar b \progvar p)^*(\overline{\testvar a}+\overline{\testvar b}) (\testvar a \progvar q (\testvar a\testvar b\progvar p)^*(\overline{\testvar a}+\overline{\testvar b}))^*\overline{\testvar a}
\]
which encode the second and the first program from the introduction.
By way of contrast, we first consider a purely algebraic derivation.

\begin{description}
\item[Algebraic derivation]
We start by proving three intermediate results.
First, we know that for all $e \in \KExp$:
\begin{equation}\label{aux1}
    \overline{\testvar a} (\testvar a e)^* \equiv \overline{\testvar a}  (\1+(\testvar a e)(\testvar a e)^*) \equiv \overline{\testvar a} + \overline{\testvar a} \testvar a e(\testvar a e)^* \equiv \overline{\testvar a} +\0 \equiv\overline{\testvar a}
\end{equation}
Second, using \Cref{\res{lem:slide}} we can show that for all $e \in \KExp$:
\begin{equation}\label{aux2}
 (\testvar a e \overline{\testvar a })^*\equiv \1 + \testvar a e \overline{\testvar a}(\testvar a e \overline{\testvar a })^* \equiv \1+ \testvar a e (\overline{\testvar a } \testvar a e)^* \overline{\testvar a } \equiv \1+\testvar a e \overline{\testvar a}
\end{equation}
And third, if we abbreviate $(\testvar a\testvar b\progvar p)^*$ with $x$, then:
\begin{equation}\label{aux3}
    \overline{\testvar b} (\testvar a \progvar q x(\overline{\testvar a}+\overline{\testvar b}))^*\overline{\testvar a} \equiv (\overline{\testvar b} \testvar a \progvar q x)^*(\overline{\testvar b}\overline{\testvar a} + \overline{\testvar b}\testvar a \progvar q x \overline{\testvar a})
\end{equation}
The latter can be derived as follows:
\begin{align*}
&
\overline{\testvar b} (\testvar a \progvar q x(\overline{\testvar a}+\overline{\testvar b}))^*\overline{\testvar a}  \\
&\equiv
\overline{\testvar b} (\testvar a \progvar q x\overline{\testvar a}+ \testvar a \progvar q x\overline{\testvar b})^*\overline{\testvar a} \tag{distributivity}\\
&\equiv
\overline{\testvar b} ((\testvar a \progvar q x\overline{\testvar a})^*\testvar a \progvar q x\overline{\testvar b})^*(\testvar a \progvar q x\overline{\testvar a})^*\overline{\testvar a} \tag{\Cref{\res{exercise:denesting}}}\\
&\equiv
(\overline{\testvar b} (\testvar a \progvar q x\overline{\testvar a})^*\testvar a \progvar q x)^*\overline{\testvar b}(\testvar a \progvar q x\overline{\testvar a})^*\overline{\testvar a} \tag{\Cref{\res{lem:slide}}}\\
&\equiv
(\overline{\testvar b} (\1+\testvar a \progvar q x\overline{\testvar a})\testvar a \progvar q x)^*\overline{\testvar b}(\1+\testvar a \progvar q x\overline{\testvar a})\overline{\testvar a} \tag{\ref{aux2}}\\
&\equiv
(\overline{\testvar b} \testvar a \progvar q x+\overline{\testvar b}\testvar a \progvar q x\overline{\testvar a}\testvar a \progvar q x)^*(\overline{\testvar b}\overline{\testvar a} + \overline{\testvar b}\testvar a \progvar q x\overline{\testvar a}) \tag{distributivity}\\
&\equiv
(\overline{\testvar b} \testvar a \progvar q x)^*(\overline{\testvar b}\overline{\testvar a} + \overline{\testvar b}\testvar a \progvar q x \overline{\testvar a}) \tag{\ensuremath{\overline{\testvar a}\testvar a\equiv \0}}
\end{align*}
We can now rewrite the second program to obtain the first:
\begin{align*}
& (\testvar a\testvar b\progvar p)^*(\overline{\testvar a}+\overline{\testvar b}) (\testvar a \progvar q (\testvar a\testvar b\progvar p)^*(\overline{\testvar a}+\overline{\testvar b}))^*\overline{\testvar a} \\
    &\equiv  x(\overline{\testvar a}+\overline{\testvar b}) (\testvar a \progvar q x(\overline{\testvar a}+\overline{\testvar b}))^*\overline{\testvar a} \tag{def.\ $x$} \\
    &\equiv x\overline{\testvar a} (\testvar a \progvar q x(\overline{\testvar a}+\overline{\testvar b}))^*\overline{\testvar a}  +
    x\overline{\testvar b} (\testvar a \progvar q x(\overline{\testvar a}+\overline{\testvar b}))^*\overline{\testvar a} \tag{distributivity} \\
    &\equiv x\overline{\testvar a}   +x(\overline{\testvar b} \testvar a \progvar q x)^*(\overline{\testvar b}\overline{\testvar a} + \overline{\testvar b}\testvar a \progvar q x\overline{\testvar a}) \tag{\ref{aux1},~\ref{aux3}}\\
    &\equiv x(\1  + (\overline{\testvar b} \testvar a \progvar q x)^*\overline{\testvar b}\testvar a \progvar q x)\overline{\testvar a} + x(\overline{\testvar b}  \testvar a \progvar q x)^*\overline{\testvar b}\overline{\testvar a}\tag{distributivity}\\
    &\equiv x(\overline{\testvar b} \testvar a \progvar q x)^*\overline{\testvar a} +x(\overline{\testvar b} \testvar a \progvar q x)^*\overline{\testvar b}\overline{\testvar a}\tag{star axiom}\\
    &\equiv x(\overline{\testvar b} \testvar a \progvar q x)^*(\overline{\testvar a} +\overline{\testvar b}\overline{\testvar a}) \tag{distributivity} \\
    &\equiv x(\overline{\testvar b} \testvar a \progvar q x)^*\overline{\testvar{a}} \tag{\ensuremath{\overline{\testvar a} +\overline{\testvar b}\overline{\testvar a}\equiv \overline{\testvar a}}} \\
    &\equiv (x\overline{\testvar b} \testvar a \progvar q)^* x\overline{\testvar{a}} \tag{\Cref{\res{lem:slide}}} \\
    &\equiv  ((\testvar a\testvar b \progvar p)^* \testvar a \overline{ \testvar b }\progvar q)^* (\testvar a\testvar b \progvar p)^*\overline{\testvar a} \tag{def.\ $x$} \\
    &\equiv  (\testvar a\testvar b \progvar p + \testvar a \overline{\testvar b}\progvar q)^* \overline{\testvar a} \tag{\Cref{\res{exercise:denesting}}} \\
    &\equiv (\testvar a(\testvar b \progvar p +\overline{\testvar b}\progvar q))^* \overline{\testvar a} \tag{distributivity}
\end{align*}

\item[Bisimulation proof]
We now repeat the same proof, but using a bisimulation up-to congruence.
First, we analyze the derivatives of the two expressions.
We start with $(\testvar a(\testvar b\progvar p + \overline{\testvar b}\progvar q))^*\overline{\testvar a}$ and observe that $E((\testvar a(\testvar b\progvar p + \overline{\testvar b}\progvar q))^*\overline{\testvar a}) = \{ \overline{\testvar a}\testvar{b}, \overline{\testvar a}\overline{\testvar b} \}$.
Moreover, many derivatives are equivalent to $\0$; for instance, in the case of $\overline{\testvar{a}}\testvar{b}$ and $\p$, we calculate:
\begin{align*}
((\testvar a(\testvar b\progvar p + \overline{\testvar b}\progvar q))^*\overline{\testvar a})_{\overline{\testvar a}\testvar b\p}
    &= {(\testvar a(\testvar b\progvar p + \overline{\testvar b}\progvar q))^*}_{\overline{\testvar a}\testvar b\p}\overline{\testvar a} + \0 \\
    &= {(\testvar a(\testvar b\progvar p + \overline{\testvar b}\progvar q))}_{\overline{\testvar a}\testvar b\p} {(\testvar a(\testvar b\progvar p + \overline{\testvar b}\progvar q))^*}\overline{\testvar a} + \0 \\
    &= {({\testvar a}_{\overline{\testvar a}\testvar b\p}(\testvar b\progvar p + \overline{\testvar b}\progvar q))} {(\testvar a(\testvar b\progvar p + \overline{\testvar b}\progvar q))^*}\overline{\testvar a} + \0 \\
    &= \0 {(\testvar a(\testvar b\progvar p + \overline{\testvar b}\progvar q))^*}\overline{\testvar a} + \0 \equiv \0
\end{align*}

Similar derivations exist for $\alpha\p$ with $\alpha \in \{ \overline{\testvar a}\overline{\testvar b}, \testvar{a}\overline{\testvar b} \}$, and $\beta\q$ with $\beta \in \{ \overline{\testvar a}\testvar{b}, \overline{\testvar a}\overline{\testvar b}, \testvar{a}\testvar{b} \}$.
The derivatives for $\testvar a\testvar b\p$ and $\testvar a\overline{\testvar b}\q$ remain to be analyzed.
For the first of these, we can derive as follows:
\begin{align*}
((\testvar a(\testvar b\progvar p + \overline{\testvar b}\progvar q))^*\overline{\testvar a})_{\testvar a\testvar b\p}
&= ((\testvar a (\testvar b\progvar p + \overline{\testvar b}\progvar q))^*)_{\testvar a\testvar b\p} \overline{\testvar a} +  (\overline{\testvar a})_{\testvar a\testvar b\p}\\
&=(\testvar a (\testvar b\progvar p + \overline{\testvar b}\progvar q))_{\testvar a\testvar b\p}(\testvar a (\testvar b\progvar p + \overline{\testvar b}\progvar q))^*\overline{\testvar a} + \0\\
&= (\0+ (\testvar b\progvar p + \overline{\testvar b}\progvar q)_{\testvar a\testvar b\p}) (\testvar a (\testvar b\progvar p + \overline{\testvar b}\progvar q))^*\overline{\testvar a}+
\0\\
&=(\0+ ((\testvar b\progvar p)_{\testvar a\testvar b\p} + (\overline{\testvar b}\progvar q)_{\testvar a\testvar b\p})) (\testvar a (\testvar b\progvar p + \overline{\testvar b}\progvar q))^*\overline{\testvar a}+
\0\\
&= (\0+ ((\0+\1) + (\0+\0))) (\testvar a (\testvar b\progvar p + \overline{\testvar b}\progvar q))^*\overline{\testvar a}+ \0 \\
&\equiv (\testvar a (\testvar b\progvar p + \overline{\testvar b}\progvar q))^*\overline{\testvar a}
\intertext{The final derivative of $(\testvar a(\testvar b\progvar p + \overline{\testvar b}\progvar q))^*\overline{\testvar a}$ is obtained as follows:}
((\testvar a(\testvar b\progvar p + \overline{\testvar b}\progvar q))^*\overline{\testvar a})_{\testvar a\overline{\testvar b}\q} &= ((\testvar a (\testvar b\progvar p + \overline{\testvar b}\progvar q))^*)_{\testvar a\overline{\testvar b}\q}  \overline{\testvar a} +  (\overline{\testvar a})_{\testvar a\overline{\testvar b}\q} \\
&=(\testvar a (\testvar b\progvar p + \overline{\testvar b}\progvar q))_{\testvar a\overline{\testvar b}\q} (\testvar a (\testvar b\progvar p + \overline{\testvar b}\progvar q))^*\overline{\testvar a} + \0\\
&= (\0+ (\testvar b\progvar p + \overline{\testvar b}\progvar q)_{\testvar a\overline{\testvar b}\q}) (\testvar a (\testvar b\progvar p + \overline{\testvar b}\progvar q))^*\overline{\testvar a}+
\0\\
&=(\0+ ((\testvar b\progvar p)_{\testvar a\overline{\testvar b}\q} + (\overline{\testvar b}\progvar q)_{\testvar a\overline{\testvar b}\q})) (\testvar a (\testvar b\progvar p + \overline{\testvar b}\progvar q))^*\overline{\testvar a}+
\0\\
&=(\0+ ((\0+\0) + (\0+\1))) (\testvar a (\testvar b\progvar p + \overline{\testvar b}\progvar q))^*\overline{\testvar a}+ \0 \\
&\equiv (\testvar a (\testvar b\progvar p + \overline{\testvar b}\progvar q))^*\overline{\testvar a}
\end{align*}

In summary, in the two cases where the derivative of $(\testvar a (\testvar b\progvar p + \overline{\testvar b}\progvar q))^*\overline{\testvar a}$ is not $\0$, it is equivalent to $(\testvar a (\testvar b\progvar p + \overline{\testvar b}\progvar q))^*\overline{\testvar a}$.

Let us now consider $(\testvar a\testvar b \progvar p)^*(\overline{\testvar a}+\overline{\testvar b}) (\testvar a \progvar q (\testvar a\testvar b\progvar p)^*(\overline{\testvar a}+\overline{\testvar b}))^*\overline{\testvar a}$.
Note that $E((\testvar a\testvar b \progvar p)^*(\overline{\testvar a}+\overline{\testvar b}) (\testvar a \progvar q (\testvar a\testvar b\progvar p)^*(\overline{\testvar a}+\overline{\testvar b}))^*\overline{\testvar a}) = \{ \overline{\testvar a}\testvar{b}, \overline{\testvar a}\overline{\testvar b} \}$.
As before, many derivatives are equivalent to $\0$; indeed, routine calculation will show that this is the case for $\alpha \p$ with $\alpha=\overline{\testvar a}\testvar b$ or $\alpha=\overline{\testvar a}\overline{\testvar b}$ or $\alpha=\testvar a \overline{\testvar b}$, and for $\beta\q$ with $\beta=\overline{\testvar a}\testvar b$ or $\beta=\overline{\testvar a}\overline{\testvar b}$ or $\beta=\testvar a \testvar b$.

We calculate the remaining derivatives; first, for $\testvar{a}\testvar{b}\p$:
\begin{align*}
&\phantom{=}((\testvar a\testvar b \progvar p)^*(\overline{\testvar a}+\overline{\testvar b}) (\testvar a \progvar q (\testvar a\testvar b\progvar p)^*(\overline{\testvar a}+\overline{\testvar b}))^*\overline{\testvar a})_{\testvar a\testvar b\p} \\
&= ((\testvar a\testvar b \progvar p)^*)_{\testvar a\testvar b\p}(\overline{\testvar a}+\overline{\testvar b}) (\testvar a \progvar q (\testvar a\testvar b\progvar p)^*(\overline{\testvar a}+\overline{\testvar b}))^*\overline{\testvar a} \\
&\phantom{=} {} + ((\overline{\testvar a}+\overline{\testvar b}) (\testvar a \progvar q (\testvar a\testvar b\progvar p)^*(\overline{\testvar a}+\overline{\testvar b}))^*\overline{\testvar a})_{\testvar a\testvar b\p} \\
&=(\testvar a\testvar b \progvar p)_{\testvar a\testvar b\p}(\testvar a\testvar b \progvar p)^*(\overline{\testvar a}+\overline{\testvar b}) (\testvar a \progvar q (\testvar a\testvar b\progvar p)^*(\overline{\testvar a}+\overline{\testvar b}))^*\overline{\testvar a} + \0\\
&=\1 (\testvar a\testvar b \progvar p)^*(\overline{\testvar a}+\overline{\testvar b}) (\testvar a \progvar q (\testvar a\testvar b\progvar p)^*(\overline{\testvar a}+\overline{\testvar b}))^*\overline{\testvar a} + \0 \\
&\equiv (\testvar a\testvar b \progvar p)^*(\overline{\testvar a}+\overline{\testvar b}) (\testvar a \progvar q (\testvar a\testvar b\progvar p)^*(\overline{\testvar a}+\overline{\testvar b}))^*\overline{\testvar a}
\intertext{%
    Finally, the derivative for $\testvar{a}\overline{\testvar b}\q$ works out as follows:
}
&\phantom{=}((\testvar a\testvar b \progvar p)^*(\overline{\testvar a}+\overline{\testvar b}) (\testvar a \progvar q (\testvar a\testvar b\progvar p)^*(\overline{\testvar a}+\overline{\testvar b}))^*\overline{\testvar a})_{\testvar a\overline{\testvar b}\q}\\
&= ((\testvar a\testvar b \progvar p)^*)_{\testvar a\overline{\testvar b}\q}(\overline{\testvar a}+\overline{\testvar b}) (\testvar a \progvar q (\testvar a\testvar b\progvar p)^*(\overline{\testvar a}+\overline{\testvar b}))^*\overline{\testvar a} \\
&\phantom{=} {} + ((\overline{\testvar a}+\overline{\testvar b}) (\testvar a \progvar q (\testvar a\testvar b\progvar p)^*(\overline{\testvar a}+\overline{\testvar b}))^*\overline{\testvar a})_{\testvar a\overline{\testvar b}\q} \\
&=(\testvar a\testvar b \progvar p)_{\testvar a\overline{\testvar b}\q}(\testvar a\testvar b \progvar p)^*(\overline{\testvar a}+\overline{\testvar b}) (\testvar a \progvar q (\testvar a\testvar b\progvar p)^*(\overline{\testvar a} +\overline{\testvar b}))^*\overline{\testvar a} \\
&\phantom{=} {}  + (\0 + ((\testvar a \progvar q (\testvar a\testvar b\progvar p)^*(\overline{\testvar a}+\overline{\testvar b}))^*\overline{\testvar a})_{\testvar a\overline{\testvar b}\q})\\
&= \0 + (\testvar a \progvar q (\testvar a\testvar b\progvar p)^*(\overline{\testvar a}+\overline{\testvar b}))_{\testvar a\overline{\testvar b}\q}(\testvar a \progvar q (\testvar a\testvar b\progvar p)^*(\overline{\testvar a}+\overline{\testvar b}))^*\overline{\testvar a} + (\overline{\testvar a})_{\testvar a\overline{\testvar b}\q}\\
&= \0 + \1(\testvar a\testvar b\progvar p)^*(\overline{\testvar a}+\overline{\testvar b})(\testvar a \progvar q (\testvar a\testvar b\progvar p)^*(\overline{\testvar a}+\overline{\testvar b}))^*\overline{\testvar a} + \0 \\
&\equiv (\testvar a\testvar b\progvar p)^*(\overline{\testvar a}+\overline{\testvar b})(\testvar a \progvar q (\testvar a\testvar b\progvar p)^*(\overline{\testvar a}+\overline{\testvar b}))^*\overline{\testvar a}
\end{align*}

Thus, $(\testvar a\testvar b\progvar p)^*(\overline{\testvar a}+\overline{\testvar b})(\testvar a \progvar q (\testvar a\testvar b\progvar p)^*(\overline{\testvar a}+\overline{\testvar b}))^*\overline{\testvar a}$ and $(\testvar a (\testvar b\progvar p + \overline{\testvar b}\progvar q))^*\overline{\testvar a}$ are alike, in that (1)~they are assigned the same value by $E$, (2)~the same six derivatives are equivalent to $\0$, and (3)~they are equivalent to their remaining two derivatives.
All of this is sufficient to verify that the following is a bisimulation up-to congruence:
\[R = \{ (\testvar a(\testvar b\progvar p + \overline{\testvar b}\progvar q)^*\overline{\testvar a}, (\testvar a\testvar b \progvar p)^*(\overline{\testvar a}+\overline{\testvar b}) (\testvar a \progvar q (\testvar a\testvar b\progvar p)^*(\overline{\testvar a}+\overline{\testvar b}))^*\overline{\testvar a}) \}
\]
\end{description}

\noindent
Among the above, the second (bisimulation-based) approach required more text to describe, in spite of us eliding several of the derivations.
Still, whereas the first (algebraic) proof required some ingenuity in which lemmas to apply, it should be noted that all steps in the bisimulation-based approach could be mechanized, from the calculation of $E$ and $D$ to the constant-folding that occurs at the end of each derivation.
More generally, if one of the expressions has a derivation that simplifies to an expression not already seen, we can continue by analyzing the derivatives of that expression, and so on, until no new expressions remain; \Cref{\res{lemma:kat-aci-finite}} guarantees that this process terminates.
For automated reasoning, the bisimulation-based approach is therefore preferable.

\section{Completeness, a sketch}
For \KAT it was shown in~\cite{kozen-smith-1996} that there exists a similar result as we proved for \KA in the previous chapter --- i.e., that semantic equivalence and provable equivalence coincide.
More formally, we have
\begin{theorem}[Soundness and Completeness for \KAT]\label{theorem:completeness-KAT}
For $e, f \in \KExp$:
\[e\equiv f \Leftrightarrow \sem e = \sem f\]
\end{theorem}

The forward direction of this theorem (soundness) we had already seen in \Cref{\res{theorem:kat-soundness}}.
We finish the technical contents of this booklet by sketching the proof of the backwards claim (completeness).
However, rather than replaying the arguments from \Cref{\res{chapter:completeness}}, we rely on a technique called \emph{reduction}, which lets the completeness of \KA do the heavy lifting for us.
This idea relies the insight that expressions in \KAT are simply regular expressions over the alphabet $\alphabet = \P \cup \BExp$.\footnote{%
    Granted, this is an infinite alphabet, whereas we assumed beforehand that $\alphabet$ was finite.
    However, it should be clear that expressions over any alphabet will only use finitely many letters; hence, we can restrict our attention to the part of $\alphabet$ that \emph{is} used, and instantiate our earlier results to that finite subalphabet.
}
Thus, in the special case that $e$ and $f$ have the same regular language, completeness tells us that $e \equiv_{\KA} f$, and hence that $e \equiv_{\KAT} f$ (since laws of \KA are laws of \KAT).
Obviously, this is not true for all $e$ and $f$, and so the game becomes to massage them to obtain $\KAT$-equivalent expressions where we can apply this insight.

To further illustrate this idea, suppose for instance that $e = \p\overline{\overline{\testvar{a}}\testvar{b}}$ and $f = \p\testvar{a} + \p\overline{\testvar{b}}$.
In this case, $\sem{e}_{\KAT} = \sem{f}_{\KAT}$, but
\[
    \sem{e}_{\KA} = \{ \p\overline{\overline{\testvar{a}}\testvar{b}} \} \neq \{ \p\testvar{a}, \p\overline{\testvar{b}} \} = \sem{f}_{\KA}
\]
Nevertheless, we can show that
\begin{align*}
    e &\equiv_{\KAT} \p(\testvar{a}\testvar{b} + \testvar{a}\overline{\testvar b} + \overline{\testvar a}\overline{\testvar b}) \\
    f &\equiv_\KAT \p(\testvar{a}\testvar{b} + \testvar{a}\overline{\testvar b}) + \p(\testvar{a}\overline{\testvar b} + \overline{\testvar a}\overline{\testvar b})
\end{align*}
Furthermore, the expressions on the right-hand side above have the same regular language.
By completeness of \KA, they are therefore provably equivalent using $\equiv_\KA$ (in fact, a proof is not too hard to work out manually in this case).
We conclude that $e \equiv_\KAT f$.

The approach for the two expressions above relies on substituting a test (e.g., $\overline{\overline{\testvar{a}}\testvar{b}}$) with the sum of atoms obtained through $E$ (e.g., $\testvar{a}\testvar{b} + \testvar{a}\overline{\testvar b} + \overline{\testvar a}\overline{\testvar b}$).
In general, this transformation is sound by the laws of Boolean algebra, and hence also by the laws of $\KAT$: we have that for $b \in \BExp$, $b \equiv_{\KAT} \sum_{\alpha \leq b} \alpha$.
Such a substitution is therefore sound.

Unfortunately, this does not suffice in general for our larger aim.
Consider for instance $e = (\p + \testvar{a})\overline{\testvar{a}}$ and $f = \p\overline{\testvar a}$.
Simply substituting $\testvar{a}$ with $\testvar{a}\testvar{b} + \testvar{a}\overline{\testvar b}$ and $\overline{\testvar a}$ with $\overline{\testvar a}\testvar{b} + \overline{\testvar a}\overline{\testvar b}$ does not get us to our goal: the regular language semantics of $(\p + (\testvar{a}\testvar{b} + \testvar{a}\overline{\testvar b}))(\overline{\testvar a}\testvar{b} + \overline{\testvar a}\overline{\testvar b})$ contains the word $\testvar{a}\testvar{b} \cdot \overline{\testvar a}\testvar{b}$ (the $\cdot$ marks the boundary between the ``letters'' $\testvar{a}\testvar{b}$ and $\overline{\testvar a}\testvar{b}$), which is not found in the regular language obtained from $\p(\overline{\testvar a}\testvar{b} + \overline{\testvar a}\overline{\testvar b})$.
However, if we distribute the multiplication in $e$, we get $\p\overline{\testvar a} + \testvar{a}\overline{\testvar a}$.
Here, the subexpression $\testvar{a}\overline{\testvar a}$ is now a single test, which does not have any atoms below it and can therefore be replaced with $\0$, the empty sum.
For the substitution to be effective, therefore, tests that are executed in sequence must also appear as a single (conjunctive) test.

Fortunately, this obstacle can also be overcome, and with some care it is possible to derive a translation $r\colon \KExp \to \Exp$~\cite{kozen-smith-1996} which satisfies the following two properties for all $e \in \KExp$:
\begin{mathpar}
    \sem{e}_{\KAT} = \sem{r(e)}_{\KA}
    \and
    e \equiv_{\KAT} r(e)
\end{mathpar}
These two properties lead to a proof of \Cref{\res{theorem:completeness-KAT}} that goes as follows.
First, take $e, f \in \KExp$, and suppose $\sem{e}_\KAT = \sem{f}_\KAT$.
The first property above tells us that $\sem{r(e)}_\KA = \sem{r(f)}_\KA$.
From the latter, it follows that $r(e) \equiv_\KA r(f)$ by completeness, and hence also $r(e) \equiv_{\KAT} r(f)$.
Finally, since $e \equiv_\KAT r(e)$ and $r(f) \equiv_\KAT f$ by the second property, we can chain these equivalences to conclude that $e \equiv_\KAT f$.

This strategy for proving completeness is an instance of a more general technique, captured in a framework called \emph{Kleene algebra with hypotheses}~\cite{cohen,DoumaneKPP19,kozen02,km14,fossacs2020,kappethesis, DBLP:journals/lmcs/PousRW24}.
Intuitively, additional axioms are added to \KA, and completeness is established through completeness of \KA via a function satisfying similar properties to $r$.
The reduction is used to capture syntactically (i.e., on the level of expressions) the behaviour of the newly added axioms, such that the regular language semantics of the reduced terms ($r(e)$ and $r(f)$ above) coincides with the semantics of the \KA with additional axioms.
Indeed, in some cases it is possible to combine a reduction for one extension of \KA with a reduction for another extension, and thus obtain a reduction (and hence a completeness proof) for the combination of those extensions.
More pointers to further reading on this topic can be found in Chapter~\ref{\res{chapter:further-reading}}.

\section{Exercises}%
\label{section:exerciseskat}

\begin{exercises}
    \item
    Let $b \in \BExp$.
    Prove that $b + b \equiv b$ and $b \cdot b \equiv b$.

    \item
    In the following, let $b, c \in \BExp$.

    \begin{exercises}
        \item
        Suppose $b + c \equiv \1$ and $b \cdot c \equiv \0$.
        Prove $b \equiv \overline{c}$.

        \item
        Use the fact above to prove that $\overline{\overline{b}} \equiv b$.

        \item
        Prove DeMorgan's first law: $\overline{b + c} \equiv \overline{b} \cdot \overline{c}$.

        \item
        Prove DeMorgan's second law: $\overline{b \cdot c} \equiv \overline{b} + \overline{c}$.
    \end{exercises}

    \item
    The axioms that we presented for Boolean algebra are somewhat redundant, in that we can prove some using the other laws.
    In this exercise, we will explore this a bit more.
    Below, let $b, c \in \BExp$.
    \begin{exercises}
        \item
        Prove that $b + \1 \equiv \1$ and $b \cdot \0 \equiv \0$, without using the absorption axioms $b + (b \cdot c) = b$ or $b \cdot (b+c) = b$.

        \item
        Use the facts derived above to prove the absorption axioms.
    \end{exercises}

    \item
    We now look at some laws of Boolean algebra that can be proved \emph{without the associativity axioms $b \cdot (c \cdot d) = (b \cdot c) \cdot d$ and $b + (c + d) = (b + c) + d$}; you are forbidden from using these laws in this exercise.
    \begin{exercises}
        \item
        Prove that $b + (c \cdot d) \equiv b + (b + c)\cdot d$.

        \item
        Prove that $b + (c + (b \cdot d)) \equiv b + c$ using the above.

        \item
        Conclude that $b + (\overline{b} + c) \equiv \1$.
    \end{exercises}

    \item
    Let $b, c \in \BExp$ and $e, f, g \in \KExp$.

    \begin{exercises}
        \item  Show that it holds that $ (b + e)^* \equiv e^*$.
        \item
        Prove that \itebare is ``skew-associative'', i.e.:
        \begin{align*}
            &\ite{b}{(\ite{c}{e}{f})}{g} \\
                &\quad {}\equiv \ite{b \cdot c}{e}{(\ite{b}{f}{g})}
        \end{align*}

        \item
        Suppose $\ite{b}{e \cdot f}{g} \leqq f$.
        Show that
        \[
            (\whd{b}{e}) \cdot g \leqq f
        \]

        \emph{Hint: You need to use one of the star axioms.}
    \end{exercises}

    \item
    \begin{exercises}
        \item Show, using the axioms of \KAT, that the following programs are equivalent: $(\whd{(\testvar{a} + \testvar{b})}{\progvar{p}}) \cdot (\ite{\testvar{b}}{\progvar{q}}{\progvar{r}})$ and $(\whd{(\testvar{a}+\testvar{b})}{\progvar{p}}) \cdot \progvar{r}$.

        \item Show the above equivalence using an argument based on bisimulation up-to congruence.
    \end{exercises}

    \item\label{exercise:rel-vs-lang-kat}
    Prove \Cref{\res{thm:sound-complete-for-rel}} by generalizing the strategy from \Cref{\res{theorem:lang-implies-rel},\res{theorem:rel-implies-lang}}, which were used to prove \Cref{\res{thm:rel_equiv_lang}}.

    \item\label{exercise:kat-hoare}
    \emph{Hoare logic} is a formalism for reasoning about the correctness of programs~\cite{hoare-1969,floyd-1967}.
    Its statements are \emph{triples} denoted $\htr{P}{C}{Q}$, where $P$ and $Q$ are logical formulas, and $C$ is a program.
    Such a triple \emph{holds} when, if a machine starts in a state satisfying $P$, it reaches a state satisfying $Q$ after executing $C$.

    We can encode Hoare logic using \KAT~\cite{kozen-2000}, by writing
    \[
        \htr{b}{e}{c}
        \quad\quad
        \text{as a shorthand for}
        \quad\quad
        b \cdot e \cdot \overline{c} \equiv \0
    \]
    The idea is that, if $b \cdot e \cdot \overline{c} \equiv \0$, then there is no valid way to execute $b \cdot e \cdot \overline{c}$, i.e., to assert that $b$ holds, execute $e$, and then assert that $\overline{c}$ holds.
    Thus, necessarily, if $e$ runs to completion after asserting $b$, then $c$ holds.

    In this exercise, you will verify that the rules of inference in Hoare logic are actually compatible with this encoding in \KAT.

    \begin{exercises}
        \item
        Hoare logic contains a rule for sequential composition:
        \[
            \inferrule{%
                \htr{P}{C_1}{Q} \\
                \htr{Q}{C_2}{R} \\
            }{%
                \htr{P}{C_1; C_2}{R}
            }
        \]

        Verify the compatibility of this rule: given $b, c, d \in \BExp$ and $e, f \in \KExp$ such that $\htr{b}{e}{c}$ and $\htr{c}{f}{d}$ hold, prove that $\htr{b}{e \cdot f}{d}$ holds.

        \item
         The following rule governs \whdbare constructs.
        \[
            \inferrule{%
                \htr{P \wedge Q}{C}{P}
            }{%
                \htr{P}{\whd{Q}{C}}{\neg Q \wedge P}
            }
        \]
        Verify this rule as well: given $b, c \in \BExp$ and $e \in \KExp$ such that $\htr{b \cdot c}{e}{b}$ holds, prove that $\htr{b}{\whd{c}{e}}{\overline{c} \cdot b}$ holds.

        \emph{Hint: use the slide rule (\Cref{\res{lem:slide}}).}

        \item
        We can alternatively encode a Hoare triple using either of the following equivalences:
        \begin{mathpar}
        b \cdot e \leqq e \cdot c
        \and
        b \cdot e \equiv b \cdot e \cdot c
        \end{mathpar}
        Show that these three encodings are equivalent --- i.e., if any of the three equivalences hold, then so do the others.
    \end{exercises}
\end{exercises}

\onlyinsubfile{%
    \printbibliography%
}

\chapter{Further Reading}%
\label{chapter:further-reading}

In this booklet, we barely scratched the surface of the vast literature out there relating to \KA.
This chapter gives an (inevitably incomplete) list of pointers for further reading, organized along several themes.

\paragraph{Axiomatizing \KA}
Regular expressions were proposed by Kleene~\cite{kleene-1956}, and the idea of encoding programs can be traced back to Pratt~\cite{pratt-1976}.
There exists a variety of axioms for Kleene algebra; possibilities include the ones proposed by Salomaa~\cite{salomaa-1966}, Conway~\cite{conway-1971}, Krob~\cite{krob-1990}, Boffa~\cite{boffa-1990}, and Kozen~\cite{kozen-1994}.
The axioms presented in this booklet are due to Kozen~\cite{kozen-1994}.
Each of these works also includes a completeness proof; as mentioned, the proof presented in this booklet is based on that of Jacobs~\cite{jacobs-2006}, who based his work on an earlier article by Kozen~\cite{kozen-2001}.

We can also drop the right-handed fixpoint laws to obtain \emph{left-handed \KA}, which is still powerful enough to axiomatize equivalence of regular expressions~\cite{kozen-silva-2020,das-doumane-pous-2018,krob-1990}.
For an alternative proof of this we refer to~\cite{kappe-fmp}.
The latter work also offers some thoughts on a \emph{finite model property} for Kleene algebra, which says that the equations that are true in all Kleene algebras are precisely those satisfied by \emph{finite} Kleene algebras.
Palka first studied this property~\cite{palka-2005}, and related it to completeness.

\paragraph{Decision procedures}
Equivalence in Kleene algebra is efficiently decidable, by converting regular expressions into automata, and checking bisimilarity.
There are various techniques to obtain automata from regular expressions, with equally various characteristics in terms of efficiency and complexity~\cite{thompson-1968,antimirov-1996,glushkov-1961,berry-sethi-1986,mcnaughton-yamada-1960}; see~\cite{watson-1993} for an overview.

There are also different strategies for establishing whether two automata are bisimilar.
A rather efficient algorithm was proposed in~\cite{DBLP:conf/popl/BonchiP13}, where one can also find references to other procedures for checking bisimilarity.
A decision procedure for Kleene algebra that was proven correct in a proof assistant was provided in~\cite{bp:itp10:kacoq}.

\paragraph{Extensions}
Kleene algebra captures reasoning about programs in a general setting.
However, for many applications, it is interesting to impose additional structure on the algebraic theory or add extra primitives to \KA.
Such extensions can be useful to express additional equivalences for programs that are not captured by the (rather general) \KA axioms, but are required in a specific domain.

The most prominent \KA extension is \KAT~\cite{kozen96}, which we reviewed in \Cref{\res{chapter:kat}}.
\KAT was proven to be complete and decidable in~\cite{kozen-smith-1996}.
A more efficient decision procedure was developed and certified in~\cite{pous:itp13:ra}.
\KAT subsumes \emph{propositional Hoare logic}, in the sense that any valid rule of Hoare logic can be proved in KAT via the encoding from~\Cref{\res{exercise:kat-hoare}}~\cite{kozen-2000}.

\emph{Kleene algebra with observations} (\KAO)~\cite{kao} is a variant of \KAT that does not identify concatenation of tests with conjunction --- that is, the test $a \wedge b$ is different from the test $a \cdot b$.
Instead, the axiom $a\wedge b\leq a\cdot b$ is used, which semantically can be understood as ``the behaviour of $a\wedge b$ is included in the behaviour of $a\cdot b$''.
In a concurrent setting this is necessary, as $a\wedge b$ is one event, but between $a$ and $b$ in $a\cdot b$, other events can take place in a parallel thread.
Hence, $a\wedge b$ and $a\cdot b$ do not necessarily have the same behaviour in a concurrent setting.

\KAT has also been extended with mutable state in~\cite{GrathwohlKM14} and with networking primitives; the latter extension is known as \NetKAT~\cite{netkat}.
\NetKAT can be used to analyse packet behaviour in networks, and comes with an efficient (coalgebraic) decision procedure~\cite{netkat2}.

\emph{Guarded Kleene Algebra with Tests} (\GKAT) was introduced~\cite{DBLP:journals/pacmpl/SmolkaFHKKS20} to formalise the fragment of \KAT that corresponds to programs formed using \itebare and \whdbare.
Deciding equivalence in \GKAT is easier than deciding equivalence in \KAT: whereas equivalence in \KAT is PSPACE-complete, in \GKAT it can be done in nearly linear time.
This result assumes that the number of tests is fixed; otherwise, the problem is co-NP hard~\cite{DBLP:journals/pacmpl/SmolkaFHKKS20}.
Right now, the completeness problem for \GKAT is not fully settled yet, though some progress has been made~\cite{schmid-etal-2021,DBLP:conf/esop/SchmidKS23,schmid-kappe-2025}.
An introduction to and survey of \GKAT can be found in~\cite{kappe-2025}.

\emph{Kleene algebra with domain}~\cite{DBLP:journals/tocl/DesharnaisMS06} extends \KA with axioms for a domain and codomain operation, which allows for the definition of modal operators.
Completeness of this extension has been settled, and several results about its complexity are also known~\cite{sedlar-2023}
\emph{Modal Kleene algebra} is Kleene algebra enriched by forward and backward box and diamond operators, defined via domain and codomain operations~\cite{DesharnaisMoellerStruth2004,DBLP:conf/amast/MollerS04}.
Both extensions can be thought of as a bridge between Kleene Algebra and Propositional Dynamic Logic (PDL)~\cite{fischer-ladner-1979,harel-etal-2000}.

\emph{Kleene algebra with a top element}, an extension relevant for reasoning about abnormal termination~\cite{DBLP:conf/fossacs/Mamouras17}, has been studied in~\cite{DBLP:conf/concur/PousW22,DBLP:journals/corr/abs-2304-07190}.
Completeness results are provided for language and relational models of \KA with top, and \KAT with top.
Kleene algebra with tests and top can be used to model \emph{incorrectness logic}~\cite{incorrect}.\footnote{%
    The completeness proof for \KAT with top given in this paper is incorrect, as discussed and fixed in~\cite{DBLP:conf/concur/PousW22}.
}

\paragraph{Hypotheses}
Many of the aforementioned extensions of \KA come with proofs of completeness w.r.t.\ the appropriate language or relational models and decidability.
Such results are desirable for program verification --- e.g., in a proof assistant, such structures make it possible to write algebraic proofs of correctness, and to mechanize some of the steps.
In particular, when two expressions $e$ and $f$ representing two programs happen to be provably equivalent in for instance \KAT, it suffices to call a certified decision procedure~\cite{bp:itp10:kacoq,pous:itp13:ra}.
Obtaining such decidability and completeness results is often non-trivial.

To develop a unifying perspective on extensions of \KA, and a more modular approach to proving completeness and decidability for new \KA extensions, a general framework called \emph{Kleene algebra with hypotheses}~\cite{cohen,DoumaneKPP19,kozen02,km14,fossacs2020,kappethesis, DBLP:journals/lmcs/PousRW24,wagemaker-thesis} was proposed, which encompasses for instance \KAT, \KAO and \NetKAT~\cite{DBLP:journals/lmcs/PousRW24}.
Kleene algebra with hypotheses comes with a canonical language model constructed from a set of hypotheses, in terms of \emph{closed} languages~\cite{DoumaneKPP19}.
In the case of \KAT, the canonical language model obtained from hypotheses corresponds to the familiar interpretation of expressions as languages of guarded strings.
The hypotheses framework cannot be used when the resulting languages are not regular, which is the case for instance with commutative \KA~\cite{DBLP:journals/corr/abs-1910-14381,pilling-1970,conway-1971}.
We refer to~\cite{DBLP:journals/lmcs/PousRW24} for an overview.

From a logical perspective, KA with hypotheses is about axiomatizing the \emph{Horn theory} of regular expressions, which is about proving valid conditional equivalences between regular expressions, like the one from \Cref{\res{exercise:bisimulation-law}}.
The challenge is that such implications are undecidable in general~\cite{kozen-1996,kuznetsov-2023,amorim-etal-2025}; the focus has therefore been in dealing with certain kinds of hypotheses, such as those of the form $e \equiv 0$~\cite{cohen,hardin-2005}.
Crucially, there are implications that are true in relational semantics, but not in all KAs, such as $e \leq 1 \implies e^2 = e$; for more on this, see~\cite{hardin-thesis}.

\paragraph{Bisimilarity}
Kleene Algebra assumes left-distributivity, i.e., the program $e \cdot (f + g)$ is equivalent to the program $e \cdot f + e \cdot g$ --- even though the latter program ``chooses'' whether or not to execute $f$ or $g$ before $e$ is run.
Similarly, in \KA $e \cdot 0$ is equivalent to $0$ --- even though the former may perform some actions before ultimately failing.
To distinguish these, one can drop these laws from \KA by interpreting a regular expression up to bisimilarity instead.
In~\cite{DBLP:journals/jcss/Milner84} a sound axiomatisation of bisimilarity for regular expressions is provided.
A completeness proof for a fragment of this syntax can be found in~\cite{grabmayer-fokkink-2020}; a proof of completeness for the entire syntax has also been announced~\cite{DBLP:conf/lics/Grabmayer22}.
Part of what makes this version of \KA so hard is that not every automaton can be translated back to a bisimilar expression~\cite{DBLP:journals/jcss/Milner84}, and thus one half of Kleene's theorem, which was pivotal in the completeness proof, is missing.

There is limited work on bisimulation semantics for variants of \KA.
Simulation rather than bisimulation is characterised in probabilistic versions of Kleene algebra~\cite{DBLP:conf/lpar/McIverW05,DBLP:journals/jlp/McIverGCM08}.
\emph{Probabilistic \KA} was introduced for resolving non-deterministic choices as they occur.
Completeness of probabilistic \KA w.r.t.\ an appropriate class of automata is considered in~\cite{DBLP:conf/RelMiCS/McIverRS11}.
\emph{(Probabilistic) \GKAT modulo bisimilarity} is studied in~\cite{DBLP:conf/icalp/RozowskiKKS023,schmid-etal-2021}.

\paragraph{Concurrency}
The aforementioned extensions of Kleene algebra all consider programs in a sequential setting.
Kleene algebras that model programs with parallel behaviour have also been studied widely.
\emph{Concurrent Kleene algebra} was proposed in~\cite{DBLP:conf/RelMiCS/HoareMSW09,DBLP:journals/jlp/HoareMSW11}, and models concurrency via an axiom called the exchange law as a combination of interleaving and certain parts of programs taking place ``truly concurrently'' (i.e., there is no causal relation between these events).
An overview of the history of \CKA and the exchange law can be found in~\cite{DBLP:books/mc/21/Struth21}.
Completeness of \CKA is established in~\cite{DBLP:journals/corr/LaurenceS17,DBLP:conf/esop/KappeB0Z18}, both of which are based on~\cite{DBLP:conf/RelMiCS/LaurenceS14}.

Adding tests to \CKA is hard --- indeed, curious inconsistencies may crop up~\cite{kao}.
\KAO circumvents these, which allows an extension to \emph{concurrent Kleene algebra with observations} (\CKAO), which in turn gives rise to further extensions to analyse concurrent programs with state (\POCKA)~\cite{DBLP:conf/concur/WagemakerBDKR020} and concurrent behaviours in networks (\CNetKAT)~\cite{DBLP:conf/esop/WagemakerFKKRS22}; these are all developed using \KA with hypotheses.
\DyNetKAT~\cite{DBLP:conf/fossacs/CaltaisHMT22} is a different concurrent extension of \NetKAT that does not use \CKA but instead uses a parallel operator in the style of process algebra.

\emph{Synchronous Kleene algebra} (\SKA)~\cite{DBLP:journals/jlp/Prisacariu10} was proposed as a way to model lock-step concurrency using \KA.
The appropriate language model for \SKA is much simpler than for \CKA, but less expressive because arbitrary interleaving between programs is not possible.
In fact, \SKA is an instance of Kleene algebra with hypotheses, and not of concurrent Kleene algebra with hypotheses~\cite{wagemaker-thesis}.
Completeness is established in~\cite{skadoorons}.

\emph{Probabilistic \CKA}~\cite{DBLP:journals/corr/McIverRS13,DBLP:conf/lpar/McIverRS13} can capture non-determinism, concurrency and probability using a ``truly concurrent'' model.
Its algebraic theory is shown to be sound for a set of probabilistic automata modulo probabilistic simulation; completeness appears to be an open question.

Finally, there is work on \emph{higher-dimensional automata} (HDAs)~\cite{pratt-1991}, which model the beginning and end of concurrent events using higher-dimensional structure.
This allows them to describe concurrent behavior like message passing.
While equational reasoning has been less of a concern in this line of work, HDAs do enjoy many of the formal results that one expects from plain automata~\cite{fahrenberg-ziemianski-2024,fahrenberg-etal-2024}, so it is conceivable that the results around \KA can one day be extended to this setting.

\onlyinsubfile{%
    \printbibliography%
}

\backmatter%
\printbibliography%

\end{document}